\documentstyle[12pt,progopt,epsf]{article}
\input epsf

\def\theequation{\thesection.\arabic{equation}}
\newcommand{\us}{{^{(s)}}}
\newcommand{\qpd}{{\cal W}^{(s)}(\alpha)}
\newcommand{\spd}{{P^{(s)}(\theta)}}
\newcommand{\pbd}{{P(\theta)}}
\newcommand{\I}{{\rm i}}
\newcommand{\Hs}{${\cal H}^{(\sigma)}$}
\newcommand{\e}{{\rm e}}
\newcommand{\plus}{{^{\dagger}}}
\newcommand{\phaseop}{{\hat\Phi_{\theta}}}
\newcommand{\cs}  {{$|\alpha\rangle_{(\sigma)}$}}
\newcommand{\ba}{{\overline\alpha}}
\newcommand{\tcs} {{$|\ba\rangle_{(\sigma)}$ }}

\def\mref#1{(\ref{#1})}

\def\ket#1{{\,|#1\rangle}}
\def\bra#1{{\langle#1|\,}}
\def\var#1{{\langle(\Delta{#1})^2\rangle}}

\def\citeY#1{{[\citeyearNP{#1}]}}
\def\citeAY#1{\citeNP{#1}}

\begin{document}
\begin{center}
\setcounter{page}{355} {\sc PROGRESS IN OPTICS, VOL. XXXV\\
EDITED BY E. WOLF\\
ELSEVIER, AMSTERDAM, 1996\\
PAGES 355--446}

\vspace*{2cm} VI

\vspace*{2cm}
{\bf QUANTUM PHASE PROPERTIES OF \\
NONLINEAR OPTICAL  PHENOMENA}

\vspace*{2cm} BY

\vspace*{2cm}
{\sc R. TANA\'S and A. MIRANOWICZ}\\

\vspace*{5mm}
{\em Nonlinear Optics Division, \\
Institute of Physics, \\
Adam Mickiewicz University, \\
60-780 Pozna\'n, Poland}\\[12pt]
and\\[12pt]

{\sc Ts. GANTSOG}\\

\vspace*{5mm}
{\em Max-Planck-Institut f\"ur Quantenoptik, \\
D-85748 Garching bei M\"unchen, Germany and \\
Department of Theoretical Physics, \\
National University of Mongolia, \\
210646 Ulaanbaatar, Mongolia } \end{center}

\newpage

\tableofcontents

\newpage
% \setcounter{page}{1}
%%%%%%%%%%%%%%%%%%%%%%%%%%%%%%%%%%%%%%%%%%%%%%%%%%%%%%%%%%%%%%%%%%%%%%%
%                   Introduction                                      %
%%%%%%%%%%%%%%%%%%%%%%%%%%%%%%%%%%%%%%%%%%%%%%%%%%%%%%%%%%%%%%%%%%%%%%%
\section{Introduction}

In classical optics, the concepts of intensity and phase of
optical fields have a well-defined meaning. The oscillating
real electromagnetic field associated with single mode,
$E=A\cos(\Phi)$, has a well-defined amplitude and phase.
Apart from a constant factor, the squared real amplitude,
$A^2$, is the intensity of the field. In classical
electrodynamics, contrary to quantum electrodynamics, there
is no real need to use complex numbers to describe the field.
However, it is convenient to work with exponentials rather
than cosine and sine functions, and complex amplitudes of the
field, $E=A\exp(-\I \Phi)$, are commonly used to describe the
field. The modulus squared of such an amplitude is the
intensity of the field and the argument is the phase. Both
the intensity and the phase can be measured simultaneously in
classical optics. In quantum optics, it was quite natural to
associate the photon number operator with the intensity of
the field and somehow construct the phase operator conjugate
to the number operator. The latter task, however, turned out
not to be easy.

The first attempts to construct explicitly a quantum phase
operator as a quantity conjugate to the number operator were
made by \citeAY{Dir27}. His idea was to perform a polar
decomposition of the annihilation operator, similar to the
polar decomposition of the complex amplitude performed for
classical fields. It turned out later that such a
decomposition suffers from serious drawbacks, and the phase
operator introduced in this way cannot be considered as a
properly defined Hermitian phase operator. \citeAY{SG64}
exposed the contradictions inherent in Dirac's polar
decomposition and introduced, instead of the phase operator
that appeared to be non-Hermitian, the operators
$\widehat{\cos}\Phi_{\rm SG}$ and $\widehat{\sin}\Phi_{\rm
  SG}$ corresponding to the cosine and sine of the phase.
Unfortunately, these $\widehat{\cos}\Phi_{\rm SG}$ and
$\widehat{\sin}\Phi_{\rm SG}$ operators, although being
Hermitian, do not commute, so that they cannot represent a
single phase angle. Historically, the idea to use
$\widehat{\cos}\Phi$ and $\widehat{\sin}\Phi$ as Hermitian
operators describing the phase, was first raised by
\citeAY{Lou63} in his short Letter, but he did not construct
the explicit form of these operators. \citeAY{CN68}, in their
review paper,  gave a detailed record of the problems
encountered on the way to constructing of the Hermitian phase
operator and discussed thoroughly the properties of
$\widehat{\cos}\Phi_{\rm SG}$ and $\widehat{\sin}\Phi_{\rm
SG}$ operators. From their analysis it became clear that it
is the boundedness of the photon number spectrum which does
not allow for negative values, and which precludes the
existence of a properly defined Hermitian phase operator in
the infinite-dimensional Hilbert space.  The difficulty in
finding the form of the Hermitian phase operator led to the
widespread belief that no such operator exists, although
there were a number of ingenious attempts to construct a
suitable operator within the infinite-dimensional Hilbert
space (\citeAY{GW70,Tur72}, \citeA{PY73}~\citeY{PY73,PY92},
\citeAY{Pau74,DY78}, \citeA{Gal84} \citeY{Gal84a,Gal84}).  It
was known \cite{New80,BP86,LP91,LPK92} that extension of the
oscillator energy spectrum to negative values allows for the
mathematical construction of the Hermitian phase operator,
but this solution was unsatisfactory because of its recourse
to unphysical states.

Recently, \citeA{PB88} \citeY{PB88,PB89} (see also
\citeA{BP89} \citeY{BP89,BP90,BP92,BP93},
\citeAY{PBV89,BPV90,PVB90}, \citeA{VP90a},
\citeY{VP90a,VP90b,VP93}, \citeNP{VBP92}) have found a way
out of the difficulties with construction of a Hermitian
phase operator. The key idea in the development of the
Hermitian optical phase operator was abandonment of the
conventional infinite-dimensional Hilbert space for the
description of quantum states of a single field mode. They
introduced, instead, a state space ${\cal H}^{(\sigma)}$ of
formally finite dimension together with a prescription for
taking the infinite-dimensional limit only after c-number
expectation values and moments have been calculated. This
idea reintroduced a symmetry to the photon number spectrum,
which became bounded not only from below but also from above,
and removed the main obstacle in constructing a Hermitian
phase operator. An essential and indispensable ingredient of
the Pegg-Barnett construction is the way the
infinite-dimensional limit is taken, which distinguishes it
from the quantum-mechanical constructions based on
finite-dimensional spaces that have been studied before
(\citeA{Lev73} \citeY{Lev73,Lev76,Lev77},
\citeAY{ST76},~\citeA{San76}~\citeY{San76,San77,San77a},
\citeAY{SS78,Gol80}), but in which, when the limiting
procedure has been applied for the phase operator, the
original problems reappeared. The consequences of the
limiting procedure in the Pegg-Barnett approach have been
discussed by \citeNP{BP92}, and \citeNP{GMT92}. The proposal
of the Pegg-Barnett approach has renewed interest in the
problem of the proper description of the quantum-optical
phase.

Almost at the same time, \citeAY{SSW89} (see also
\citeAY{SS91}) used an alternative approach based on the
quantum estimation theory and probability operator
measures~\cite{Hel76} to describe the phase properties of
optical fields.  This approach does not rely on the existence
of the Hermitian phase operator but rather on the existence
of the eigenstates of the Susskind-Glogower nonunitary
exponential phase operator \cite{SG64}. The eigenstates of
the Susskind-Glogower phase operators form a basis for the
probability operator measures. The \citeAY{SSW89} idea has
gained some popularity (\citeA{Hal91}
\citeY{Hal93},~\citeAY{HF91,SDH91,BLC92,Bra92},
\citeA{Hra92}~\citeY{Hra92,Hra92a},
\citeAY{HS92,LBC93,Jon93,Sha93}). It turned out, however,
that it gives for physical states [i.e., states with a finite
mean number of photons (finite energy and its higher
moments)], the same results as the Pegg-Barnett approach
after the limit transition to the infinite-dimensional space.
The eigenkets of the Susskind-Glogower exponential phase
operators can be used in a similar fashion as coherent states
(eigenkets of the annihilation operator) to define the
resolution of the system operators; e.g., the phase operator
(\citeA{LP91} \citeY{LP91,LP93a},~\citeA{BB94}
\citeY{BB94,BB94b}, \citeAY{VB95}. In this case the ordering
of the phase exponentials is relevant, and, if the antinormal
ordering is taken, the results agree with those obtained from
the Pegg-Barnett formalism.

Another way to describe the phase properties of the field is
to use quasiprobability distribution functions. The idea
behind this approach is relatively simple: to integrate the
suitable quasiprobability distribution function, such as the
Glauber-Sudarshan $P$-function, Wigner function, or Husimi
$Q$-function over the ``radial'' variable and getting in this
way a corresponding phase distribution, which can then be
applied in calculations of the mean values of the
phase-dependent quantities.  Since the quasiprobability
description of the quantum state of the system can be in some
cases associated with realistic measurements performed on the
system, this approach to the phase problem has focused the
attention of many authors (see \S 2.2).  The phase
distribution functions obtained by integrating the
quasidistributions are different for different
quasidistributions, and they are all different from the
Pegg-Barnett phase distribution. The situation is even worse
because for some states of the field the $P$-function and the
Wigner function take on negative values, and the
corresponding phase distribution can also be negative. This
means that such phase distributions must be used with some
care, but in many cases this approach gives results
describing properly the phase properties of the field.

\citeA{NFM91}
\citeY{NFM91,NFM92,NFM92a,NFM93,NFM93a,NFM93b,NFM93c,NFM93d}
(see
also~\citeAY{FMM94},~\citeAY{BP93},~\citeA{Hra93a}~\citeY{Hra93a,Hra95},
\citeAY{HB93}) presented an operational approach to the
quantum phase problem. Their idea is to define phase in terms
of what actually is, or can be, measured. They do not search
for a phase operator which would satisfy some mathematical
criteria, but start their considerations from the
experimental schemes usual in classical phase measurements.
Examining various measuring schemes, they identify certain
operators, $\hat{C}_M$ and $\hat{S}_M$, corresponding to the
measured cosine and sine of the phase difference between two
fields. As a result, \citeA{NFM91} came to the conclusion
that there is no unique phase operator, and that different
measuring schemes correspond to different operators.
Nevertheless, recent theoretical studies
\cite{RW94,EW95,EWR95} show that the intrinsic Hermitian
phase operator associated with the \citeA{NFM91} apparatus
can be found. The phase distribution measured in this
experiment, under some conditions, is the radius-integrated
Q-function (\citeAY{FS93,LP93b},
\citeA{FVS93}~\citeY{FVS93,FVS93a},~\citeAY{Ban93,KC94a}).
The measurements of \citeA{NFM91} are important since until
then only isolated phase measurements were available
performed by \citeAY{GBL74} (\citeAY{GWF73}, and for
theoretical analysis see also
\citeAY{Nie77,Lev77},~\citeA{Lyn87}~\citeY{Lyn87,Lyn90,Lyn93,Lyn95},~\citeAY{GU90,Ban91,TR92},~\citeAY{MJ80},~\citeA{WC84}~\citeY{WC84,WC86}).

Quite recently, another experimental technique, {\em optical
homodyne tomography}, was invented and applied to
measurements of the quantum state of the field
\cite{SBFR93,BSR93,BSCR93,SBCR93,SBCRF93} allowing quantum
phase mean values to be calculated from the measured field
density matrix. This technique opens new possibilities for
quantum measurements. Overviews of various techniques for
measuring phase distributions are presented by, e.g.,
\citeAY{LP93c,PL94},~\citeAY{PHJ94} and~\citeAY{Lyn95}.

Another interesting approach to the phase problem was
presented by \citeNP{BE91}. They introduced the idea of
phasors and phasor bases that can be used for studying
possible candidates for the quantum phase operators.
Different phasor bases lead to different phase operators, and
according to \citeNP{BE91} extrapolation of the classical
concept of phase to the quantum regime is not unique.

Since the absolute phase of the single-mode field is not
accessible for measurements, and it is always the difference
with respect to a reference phase that we are forced to deal
with in real measurements, it is tempting to define the
phase-difference operator as a fundamental quantity
describing the optical phase. \citeA{LS93c}
\citeY{LS93c,LS94} have defined a Hermitian phase-difference
operator, which is in fact a polar decomposition of the
Stokes operators for the two-mode field, and it is not the
same as the difference of the two Pegg-Barnett operators. The
difference between the two is most pronounced for weak
fields.  \citeA{Ban91a} \citeY{Ban91a,Ban91b,Ban91c,Ban92}
has introduced yet another phase operator, based on the
two-mode description of the field.  In recent years, many
different aspects of the quantum phase problem have been
studied (\citeAY{BSP89},~\citeA{NK89}
\citeY{NK89,NK90,NK91b}, \citeAY{CE90,LS90,SP90},
\citeA{Hra90}~\citeY{Hra90,Hra93b},
\citeA{LP90}~\citeY{LP90,LP92,LP93,LP94},
\citeA{Vou90}~\citeY{Vou90,Vou92,Vou93},
\citeAY{AJV91,CCMW91,Dow91},
\citeA{Ell91}~\citeY{Ell91,Ell91a,Ell92},
\citeAY{GT91d,NK91a}, \citeA{OS91a}~\citeY{OS91a,OS91b},
\citeAY{Pau91,Tan91b,WBK91,ACTS92,Ban92c,JABS92},
\citeA{LPK92a}~\citeY{LPK92a,LPK94},
\citeAY{Rit92,SDH92,TR92}
\citeAY{Aga93,ASW93,Ban93a,BFS93,CGM93,CM93,DMRS93},
\citeA{DP93}~\citeY{DP93,DP94},
\citeAY{JD93,LS93a,Ste93,SM93,TG93,AGOSW94,BB94a},
\citeA{VP94a}~\citeY{VP94a,VP94b,VP94c},
\citeA{BBB94}~\citeY{BBB94,BBB95},
\citeAY{Das94,Fra94,GLLVV94,Opa94,SL94,SFS94}), and the
number of publications on the subject is growing steadily.
Various conceptions of the quantum-optical phase have been
described by \citeAY{BD93} in a special issue of Physica
Scripta devoted to ``Quantum phase and phase dependent
measurements'', and in the same issue one can find very
interesting historical facts, given by \citeAY{Nie93},
concerning the development of our knowledge on quantum phase.

Nowadays, although the quantum phase is still a subject of
some controversy, significant progress has been achieved in
clarifying the status of the quantum-mechanical phase
operator, describing the phase properties of optical fields
in terms of various phase distribution functions, and
measuring phase-dependent physical quantities. We can now
risk the statement that, despite the existence of various
different conceptions of phase, we are en route to unified
view and understanding of the quantum-optical phase.

It is not our aim in this review article to give a detailed
account of different descriptions of the quantum phase
showing their similarities and differences. We will not focus
our attention on the quantum phase formalisms, which are
interesting on their own right and deserve separate
treatment. We shall instead concentrate on the description of
the quantum properties of real-field states which are
generated in various nonlinear optical processes. Nonlinear
optical phenomena are sources of optical fields, the
statistical properties of which have been changed in a
nontrivial way as a result of nonlinear transformation.
Quantum phase properties are among those statistical
properties which undergo nonlinear changes, and fields
generated in different nonlinear processes have different
phase properties. With the existing phase formalisms, the
quantum-phase properties of such fields can be studied in a
systematic way, and quantitative comparisons between
different quantum-field states can be made. Using the
Pegg-Barnett phase formalism and the phase formalism based on
the $s$-parametrized quasidistribution functions, we will
study a number of both one- and two-mode field states from
the point of view of their phase properties.

%%%%%%%%%%%%%%%%%%%%%%%%%%%%%%%%%%%%%%%%%%%%%%%%%%%%%%%%%%%%%%%%%%%%%%%
%                   Phase formalisms                                  %
%%%%%%%%%%%%%%%%%%%%%%%%%%%%%%%%%%%%%%%%%%%%%%%%%%%%%%%%%%%%%%%%%%%%%%%
\section{Phase formalisms}

At the very beginning of quantum electrodynamics,
\citeNP{Dir27} raised the idea that the optical phase should
be described by a Hermitian phase operator canonically
conjugate to the number operator; that is, for the
single-mode field the two operators should obey the canonical
commutation relation
%----------------------------------------------------------------------
\begin{eqnarray}
  [\hat{\Phi},\hat{n}] &=& - \I.
\label{N01}
\end{eqnarray}
This commutation relation implies directly the
``traditional'' Heisenberg uncertainty relation
%----------------------------------------------------------------------
\begin{eqnarray}
  \var{\hat{n}} \var{\hat{\Phi}} \geq {1 \over 4},
\label{N02}
\end{eqnarray}
which appeared to be wrong \cite{SG64,CN68}. Closer
investigation of the commutator~\mref{N01} showed that the
matrix elements of the phase operator $\hat{\Phi}$ in a
number-state basis are undefined \cite{Lou63}:
%----------------------------------------------------------------------
\begin{eqnarray}
  (n-n^{'}) \bra{n^{'}}\hat{\Phi}\ket{n}\;=\;-\I\delta_{nn^{'}}.
\label{N03}
\end{eqnarray}
Since it was suggested that the problems in eq.~\mref{N03}
are due to the multivalued nature of $\hat{\Phi}$,
\citeNP{Lou63} introduced the operators $\widehat{\cos}\Phi$
and $\widehat{\sin}\Phi$, which should, as he suggested,
satisfy the commutation relations:
%----------------------------------------------------------------------
\begin{eqnarray}
  \left[\widehat{\cos}\Phi, \hat{n}\right] &=& \I \widehat{\sin}\Phi,
  \nonumber\\ \left[\widehat{\sin}\Phi, \hat{n}\right] &=& -\I
  \widehat{\cos}\Phi.
\label{N04}
\end{eqnarray}
However, \citeNP{Lou63} did not give the explicit form of the
cosine and sine operators; thus, his idea did not help much
in solving the phase problem. Moreover, it turned out that
the problem expressed in eq.~\mref{N03} was not due to the
multivalued nature of $\hat{\Phi}$, but rather to the
improper application of the correspondence between the
Poisson bracket and the commutator. \citeNP{SG64} returned to
the original Dirac idea of polar decomposition of the
creation and annihilation operators and introduced the
exponential phase operators:
%---------------------------------------------------------------------------
\begin{eqnarray}
  \widehat{\exp}(\I \Phi_{\rm SG})
  &\equiv&\sum_{n=0}^{\infty}|n\rangle\langle n+1|,
\label{N05}
\end{eqnarray}
%---------------------------------------------------------------------------
\begin{eqnarray}
  \widehat{\exp}(-\I \Phi_{\rm SG})&\equiv&\bigr[\widehat{\exp} (\I
  \Phi_{\rm SG})\bigl]^{\dagger}.
\label{N06}
\end{eqnarray}
From eqs.~\mref{N05} and~\mref{N06} one obtains:
%---------------------------------------------------------------------------
\begin{eqnarray}
  \widehat{\exp}(\I \Phi_{\rm SG})\widehat{\exp}(-\I \Phi_{\rm
    SG})&=&1, \nonumber\\ \widehat{\exp}(-\I \Phi_{\rm
    SG})\widehat{\exp}(\I \Phi_{\rm SG})&=&
    1-|0\rangle\langle0|,
\label{N07}
\end{eqnarray}
which explicitly shows the non-unitarity of the
Susskind-Glogower phase operator.

The Susskind-Glogower exponential operators~\mref{N05}
and~\mref{N06} allow for constructing two Hermitian
combinations corresponding to cosine and sine of the phase.
However, the two combinations do not commute, so they cannot
be considered as describing a single phase angle. Despite
this deficiency, the Susskind-Glogower phase operators were
widely used in description of optical fields until recently.
The eigenkets of the Susskind-Glogower operator~\mref{N05},
given by
%----------------------------------------------------------------------
\begin{eqnarray}
  \ket{\theta} &\equiv& \frac{1}{\sqrt{2\pi}}
  \sum_{n=0}^{\infty}\exp(\I n\theta)\ket{n},
\label{N08}
\end{eqnarray}
generate the resolution of the identity, and, despite their
nonorthogonality, they can be used to form the probability
operator measure applied to the phase description by
\citeNP{SS91}.

\citeNP{GW70} made an attempt to construct a Hermitian phase
operator in the infinite-dimensional Hilbert space which, as
they demanded, should satisfy the canonical commutation
relation~\mref{N01}. Their work was almost completely
forgotten.  \citeNP{BE91} have made a comparison of the
Garrison-Wong and Pegg-Barnett phase operators, pointing out
that if the limit to infinite-dimensional space is performed
on the latter operator (but not on the expectation values),
the former operator is obtained. In their view the
Pegg-Barnett phase formalism is only an approximation to the
``correct'' phase formalism. \citeA{BE91} introduced their
own quantum-phase description, which has not gained broader
acceptance. Nevertheless, their paper is an essential
contribution which clarifies a number of points in the
quantum-phase problem. The Garrison-Wong approach turned out
to lead to phase distributions which are asymmetric and
difficult to accept on physical grounds \cite{GMT92}; e.g.,
even the vacuum is phase-anisotropic. For reference, we give
a sketch of their approach in Appendix~A.

The renewed interest in quantum-phase problems has resulted
in a reexamination of some of the earlier approaches and the
creation of other, completely new descriptions.  From a
number of different phase formalisms that are now available,
we choose only two, which we shall apply for the description
of the phase properties of fields generated in various
nonlinear optical processes. These are: the Pegg-Barnett
phase formalism, which represents the canonical phase
formalism based on the idea of finding a Hermitian phase
operator, and another formalism based on the description of
the optical phase through $s$-parametrized phase
distributions, which for some values of $s$ represents the
experimentally measured phase probability distributions.

%%%%%%%%%%%%%%%%%%%%%%%%%%%%%%%%%%%%%%%%%%%%%%%%%%%%%%%%%%%%%%%%%%%%%%%%%%%%
\subsection{Pegg-Barnett phase formalism}
%%%%%%%%%%%%%%%%%%%%%%%%%%%%%%%%%%%%%%%%%%%%%%%%%%%%%%%%%%%%%%%%%%%%%%%%%%%%

\citeA{PB88}~\citeY{PB88,PB89} (and~\citeAY{BP89}) introduced
the Hermitian phase formalism, which is based on the
observation that, in a finite-dimensional state space, the
states with well-defined phase exist~\cite{Lou73}.  Thus,
they restrict the state space to a finite
$(\sigma+1)$-dimensional Hilbert space ${\cal H}^{(\sigma)}$
spanned by the number states $|0\rangle$, $|1\rangle$,\ldots,
$|\sigma\rangle$. In this space they define a complete
orthonormal set of phase states by:
%----------------------------------------------------------------------
\begin{eqnarray}
  |\theta_{m}\rangle &=& \frac{1}{\sqrt{\sigma+1}} \:
  \sum_{n=0}^{\sigma} \exp(\I n\theta_{m}) \, |n\rangle, \hspace{1cm}
  m=0,1,\ldots,\sigma \, ,
\label{N09}
\end{eqnarray}
where the values of $\theta_m$ are given by
%----------------------------------------------------------------------
\begin{eqnarray}
  \theta_{m} &=& \theta_{0} + \frac{2\pi m}{\sigma+1}.
\label{N10}
\end{eqnarray}
The value of $\theta_0$ is arbitrary and defines a particular
basis set of ($\sigma+1$) mutually orthogonal phase states.
The number state $|n\rangle$ can be expanded in terms of the
$|\theta_m\rangle$ phase state basis as
%---------------------------------------------------------------------------
\begin{eqnarray}
  |n\rangle&=&\sum_{m=0}^{\sigma}|\theta_m\rangle\langle\theta_m|n\rangle
  \;=\;\frac{1}{\sqrt{\sigma+1}}\sum_{m=0}^{\sigma}\exp(-\I
  n\theta_m)|\theta_m\rangle.
\label{N11}
\end{eqnarray}
From~eqs.~\mref{N09} and~\mref{N11} we see that a system in a
number state is equally likely to be found in any state
$|\theta_m\rangle$, and a system in a phase state is equally
likely to be found in any number state $|n\rangle$.

The Pegg-Barnett (PB) Hermitian phase operator is defined as
%----------------------------------------------------------------------
\begin{eqnarray}
  \phaseop \:\equiv\:
  \left(\hat{\Phi}_{\theta_0}^{(\sigma)}\right)_{\rm PB} &=&
  \sum_{m=0}^{\sigma} \theta_{m} \,
  |\theta_{m}\rangle\langle\theta_{m}|.
\label{N12}
\end{eqnarray}
Of course, the phase states~\mref{N09} are eigenstates of the
phase operator~\mref{N12} with the eigenvalues $\theta_m$
restricted to lie within a phase window between $\theta_0$
and $\theta_0+2\pi\sigma/(\sigma+1)$. The Pegg-Barnett
prescription is to evaluate any observable of interest in the
finite basis~\mref{N09}, and only after that to take the
limit $\sigma\rightarrow\infty$.

Since the phase states~\mref{N09} are orthonormal,
$\langle\theta_m|\theta_{m'}\rangle =\delta_{mm'}$, the $k$th
power of the Pegg-Barnett phase operator~\mref{N12} can be
written as
%----------------------------------------------------------------------
\begin{eqnarray} \phaseop^{k} &=& \sum_{m=0}^{\sigma}
  \theta_{m}^{k} \, |\theta_{m}\rangle\langle\theta_{m}|.
\label{N13}
\end{eqnarray}
Substituting eqs.~\mref{N09} and~\mref{N10} into
eq.~\mref{N12} and performing summation over $m$ yields
explicitly the phase operator in the Fock basis:
%----------------------------------------------------------------------
\begin{eqnarray}
  \phaseop &=& \theta_{0} \:+\: \frac{\sigma\pi}{\sigma +1} \:+\:
  \frac{2\pi}{\sigma+1}\sum_{n\neq n'}
  \frac{\exp[\I\,(n-n')\theta_{0}]\, |n\rangle \langle
    n'|}{\exp[\I\,(n-n')2\pi/(\sigma+1)]-1}\:.
\label{N14}
\end{eqnarray}
It is readily apparent that the Hermitian phase operator
$\phaseop$ has well-defined matrix elements in the
number-state basis and does not suffer from such problems as
the original Dirac phase operator. A detailed analysis of the
properties of the Hermitian phase operator was given
by~\citeAY{PB89} and~\citeAY{BP89}.

The unitary phase operator $\exp(\I\phaseop)$ can be defined
as the exponential function of the Hermitian operator
$\phaseop$. This operator acting on the eigenstate
$|\theta_m\rangle$ gives the eigenvalue $\exp(\I \theta_m)$,
and can be written as (\citeA{PB88}~\citeY{PB88,PB89}):
%---------------------------------------------------------------------------
\begin{eqnarray}
  \exp(\I\phaseop) &\equiv& \sum_{n=0}^{\sigma-1} |n \rangle \langle
  n+1 | + \exp\bigl[\I(\sigma+1)\theta_0\bigr] |\sigma \rangle \langle
  0 |,
\label{N15}
\end{eqnarray}
Its Hermitian conjugate is
%---------------------------------------------------------------------------
\begin{eqnarray}
  [\exp(\I\phaseop)]^{\dagger}=\exp(-\I\phaseop)
\label{N16}
\end{eqnarray}
with the same set of eigenstates $|\theta_m\rangle$ but with
eigenvalues $\exp(-\I \theta_m)$. This is the last term in
eq.~\mref{N15} that distinguishes the unitary phase operator
from the Susskind-Glogower phase operator~\cite{SG64}. The
first sum in eq.~\mref{N15} reproduces the Susskind-Glogower
phase operator if the limit $\sigma\rightarrow\infty$ is
taken.  In contrast to the Pegg-Barnett unitary phase
operator, the Susskind-Glogower exponential
operator~\mref{N05} is defined as a whole and is not unitary,
as is apparent from eq.~\mref{N07}.  The sine and cosine
operators in the Pegg-Barnett formalism are the sine and
cosine functions of the Hermitian phase operator $\phaseop$.
They are more consistent with the classical notion of phase
than their counterparts in the Susskind-Glogower phase
formalism. In particular, they satisfy the ``natural''
relations:
%---------------------------------------------------------------------------
\begin{eqnarray}
  \cos^2\phaseop+\sin^2\phaseop&=&1,
\label{N17} \\
\bigl[\cos\phaseop,\sin\phaseop\bigr]&=&0,
\label{N18}   \\
\langle n |\cos^2\phaseop|n\rangle=\langle n|
\sin^2\phaseop|n\rangle&=&{1\over2}. \label{N19}
\end{eqnarray}
The last relation is also true for the vacuum state, in sharp
contrast to the Susskind-Glogower phase operators. This is
consistent with the phase of vacuum being random, as well as
for any other number state.

The Pegg-Barnett phase operator~\mref{N14}, expressed in the
Fock basis, readily gives the phase-number
commutator~\cite{PB89}:
%----------------------------------------------------------------------
\begin{eqnarray}
  [\phaseop,\hat{n}] &=& - \frac{2\pi}{\sigma+1} \sum_{n\neq n'}
  \frac{(n-n')\exp[\I (n-n') \theta_0]} {\exp[\I (n-n')
    2\pi/(\sigma+1) ]-1} |n\rangle\langle n'|.
\label{N20}
\end{eqnarray}
Equation~\mref{N20} looks very different from the famous
Dirac postulate of the phase-number commutator~\mref{N01}.
\citeAY{San76} and \citeAY{PVB90} examined canonically
conjugate operators in the finite-dimensional Hilbert space.
According to the generalized definition by~\citeA{PVB90}, the
photon number and phase are indeed canonically conjugate
variables, similar to momentum and position or angular
momentum and azimuthal phase angle.

As the Hermitian phase operator is defined, one can calculate
the expectation value and variance of this operator for a
given state of the field $|f\rangle$. Moreover, the
Pegg-Barnett phase formalism allows the introduction of the
continuous-phase probability distribution which is a
representation of the quantum state of the field and
describes the phase properties of the field in a very
spectacular fashion. Examples of such phase distributions for
particular states of the field will be given later.

A general pure state of the field mode with a decomposition
%---------------------------------------------------------------------------
\begin{eqnarray}
  |{f}\rangle=\sum_{n=0}^{\sigma}c_n|n\rangle,
\label{N21}
\end{eqnarray}
can be re-expressed in the phase state basis, according to
eq.~\mref{N11}, as
%---------------------------------------------------------------------------
\begin{eqnarray}
  |{f}\rangle={1\over{\sqrt{\sigma+1}}}\sum_{n}\sum_{m}c_n\exp (-\I
  n\theta_m)|\theta_m\rangle.
\label{N22}
\end{eqnarray}
We should remark here that the coefficients $c_{n}$ in the
decomposition~\mref{N21} in a finite-dimensional space should
differ from the coefficients in the infinite-dimensional
space if the state $\ket{f}$ is to be normalized. A short
discussion of this problem is given in Appendix~B.

The phase probability distribution is given by~\cite{PB89}:
%---------------------------------------------------------------------------
\begin{eqnarray}
  \big |\langle\theta_m|{f}\rangle\big |^2 &=&
  {1\over{\sigma+1}}{\left |\sum_{n}c_n\exp (-\I n\theta_m)\right
    |}^2,
\label{N23}
\end{eqnarray}
which leads to the expectation value and variance of the
phase operator:
%---------------------------------------------------------------------------
\begin{eqnarray}
  \langle{f}|\phaseop|{f}\rangle&=&\sum_m\theta_m \big
  |\langle\theta_m|{f}\rangle\big |^2,
\label{N24}\\
\var{\phaseop}&=&\sum_m(\theta_m-\langle \phaseop\rangle)^2
\big |\langle\theta_m|{f}\rangle\big |^2. \label{N25}
\end{eqnarray}
If the field state $|{f}\rangle$ is a partial phase state,
i.e., if the amplitude has the form
%---------------------------------------------------------------------------
\begin{eqnarray}
  c_n=b_n \mbox{e}^{\I n\varphi},
\label{N26}
\end{eqnarray}
the phase probability distribution~\mref{N23} becomes
%---------------------------------------------------------------------------
\begin{eqnarray}
  \big |\langle\theta_m|{f}\rangle\big |^2&=&{1\over{\sigma+1}}{\left
      |\sum_{n}b_n\exp [\I n(\varphi-\theta_m)]\right |}^2 \nonumber\\
  &=&{1\over{\sigma+1}}+{2\over{\sigma+1}}\sum_{n>k}
  b_nb_k\cos{\left[(n-k)(\varphi-\theta_m)\right]}.
\label{N27}
\end{eqnarray}
The mean and variance of $\phaseop$ will depend on the chosen
value of $\theta_0$.~\citeAY{Jud64}, in his description of
the uncertainty relation for angle variables, suggested that
the choice of phase window which minimizes the variance can
be used to specify uniquely the mean or variance.  For the
partial phase state, the most convenient and physically
transparent way of choosing $\theta_0$ is to symmetrize the
phase window with respect to the phase $\varphi$.  This means
the choice
%---------------------------------------------------------------------------
\begin{eqnarray}
  \theta_0=\varphi - {{\pi \sigma}\over{\sigma+1}},
\label{N28}
\end{eqnarray}
and after introducing a new phase label
%---------------------------------------------------------------------------
\begin{eqnarray}
  \mu =m-{\sigma\over2},
\label{N29}
\end{eqnarray}
the phase probability distribution~\mref{N27} becomes
%---------------------------------------------------------------------------
\begin{eqnarray}
  \big |\langle\theta_{\mu}|{f}\rangle\big |^2 &=&
  {1\over{\sigma+1}}+{2\over{\sigma+1}}\sum_{n>k}b_nb_k\cos{\left [
      (n-k){{2\pi\mu}\over{\sigma+1}}\right ]}
\label{N30}
\end{eqnarray}
with $\mu$, which goes in integer steps from $-\sigma/2$ to
$\sigma/2$.  Since the distribution~\mref{N30} is symmetrical
in $\mu$, we immediately get, according to
eqs.~\mref{N28}--\mref{N30}:
%---------------------------------------------------------------------------
\begin{eqnarray}
  \langle{f}|\phaseop|{f}\rangle=\varphi.
\label{N31}
\end{eqnarray}
This result means that for a partial phase state with phase
$\varphi$, the choice of $\theta_0$ as in eq.~\mref{N28}
relates directly the expectation value of the phase operator
with the phase $\varphi$. With this choice of $\theta_0$ the
variance of the phase operator has a particularly simple
form:
%---------------------------------------------------------------------------
\begin{eqnarray}
  \var{\phaseop} &=&
  {{4\pi^2}\over{(\sigma+1)^2}}\sum_{\mu=-\sigma/2}^{\sigma/2}\mu^2
  \big |\langle\theta_{\mu}|{f}\rangle\big |^2.
\label{N32}
\end{eqnarray}
So-called physical states play a significant role in the
Pegg-Barnett formalism. Physical states $|p\rangle$ are
defined by~\citeAY{PB89} as the states of finite energy. Most
of the expressions in the Pegg-Barnett formalism take a much
simpler form for physical states. For example, the
commutator~\mref{N20} reduces to
%----------------------------------------------------------------------
\begin{eqnarray}
\label{N33} [\phaseop,\hat{n}]_p &=& -
\I\;[1-(\sigma+1)|\theta_0\rangle\langle\theta_0|],
\end{eqnarray}
a form more similar to the standard canonical commutation
relation~\mref{N01}. On the other hand, when physical states
are considered, we can simplify the calculation of the sum in
eqs.~\mref{N24} and~\mref{N25} by replacing it by the
integral in the limit as $\sigma$ tends to infinity.  Since
the density of states is $(\sigma+1)/2\pi$, we can write the
expectation value of the $k$th power of the phase operator as
%----------------------------------------------------------------------
\begin{eqnarray}
  \langle f|\phaseop^{k}|f\rangle &=&
  \int\limits_{\theta_{0}}^{\theta_{0}+2\pi} {\rm d}\theta \,
  \theta^{k} \pbd,
\label{N34}
\end{eqnarray}
where the continuous-phase distribution $\pbd$ is introduced
by
%----------------------------------------------------------------------
\begin{eqnarray}
  P(\theta) \;\equiv\; P_{\rm PB}(\theta) &=& \lim_{\sigma \rightarrow
    \infty} \frac{\sigma+1}{2\pi} |\langle\theta_{m}|f\rangle|^{2},
\label{N35}
\end{eqnarray}
and $\theta_m$ has been replaced by the continuous-phase
variable $\theta$. If the state $|f\rangle$ has the
number-state decomposition~\mref{N21}, then the Pegg-Barnett
phase distribution is ~\cite{PB89}:
%---------------------------------------------------------------------------
\begin{eqnarray}
  \pbd &=& \frac{1}{2\pi} \left\{ 1+2 {\rm Re} \sum_{n,m \atop m>n}
    c_m c_n^* \exp[-\I (m-n) \theta] \right\}.
\label{N36}
\end{eqnarray}
In the case of fields being in mixed states described by the
density matrix $\hat{\rho}$, formula~\mref{N36} generalizes
to
%---------------------------------------------------------------------------
\begin{eqnarray}
  \pbd &=& \frac{1}{2\pi} \left\{ 1+2 {\rm Re} \sum_{m>n} \rho_{mn}
    \exp[-\I (m-n) \theta] \right\},
\label{N37}
\end{eqnarray}
where $\rho_{mn}=\langle m|\hat{\rho}|n\rangle$ are the
density matrix elements in the number-state basis.
Equations~\mref{N36} or~\mref{N37} can be used for
calculations of the Pegg-Barnett phase distribution for any
state with known amplitudes $c_n$ or matrix elements
$\rho_{mn}$. Despite the fact that they are exact, they can
rarely be summed up into a closed form, and usually numerical
summation must be performed to find the phase distribution.
Such numerical summations have been widely applied in
studying the phase properties of optical fields.  The
Pegg-Barnett phase distribution, eqs.~\mref{N36} or
\mref{N37}, is obviously $2\pi$-periodic, and for all states
with the density matrix diagonal in the number states, the
phase distribution is uniform over the $2\pi$-wide phase
window.  These are nondiagonal elements of the density matrix
that lead to the structure of the phase distribution. The
Pegg-Barnett distribution is positive definite and
normalized.

After introducing the continuous-phase distribution $\pbd$,
formula~\mref{N32} for the phase variance, if the
symmetrization is performed, can be rewritten into the form:
%---------------------------------------------------------------------------
\begin{eqnarray}
  \var{\phaseop} &=& \int\limits_{-\pi}^{\pi}\theta^2 \pbd \,{\rm
    d}\theta.
\label{N38}
\end{eqnarray}
This means that as the phase distribution function $\pbd$ is
known, all quantum-mechanical phase characteristics can be
calculated with this function in a classical-like manner. The
result for the variance~\mref{N38} is~\cite{BP89}:
%---------------------------------------------------------------------------
\begin{eqnarray}
  \var{\phaseop}= {{\pi^2}\over
    3}+4\sum_{n>k}b_nb_k{{(-1)^{n-k}}\over{(n-k)^2}}.
\label{N39}
\end{eqnarray}
The value $\pi^2/3$ is the variance for the uniformly
distributed phase, as in the case of a single number state.

For physical states there are some additional useful
relations between the expectation values of the Pegg-Barnett
phase operators and of the Susskind-Glogower phase operators.
For example, the following relations hold~\cite{VP89}:
%---------------------------------------------------------------------------
\begin{eqnarray}
  \langle \exp(\I m\phaseop)\rangle_p&=& \langle\widehat{\exp}(\I
  m\Phi_{\rm SG})\rangle_p,
\label{N40}\\
\langle
\cos\phaseop\rangle_p&=&\langle\widehat{\cos}\Phi_{\rm
  SG}\rangle_p,
\label{N41}\\
\langle
\sin\phaseop\rangle_p&=&\langle\widehat{\sin}\Phi_{\rm
  SG}\rangle_p,
\label{N42}\\
\langle \cos^2\phaseop\rangle_p&=&
\langle\widehat{\cos}^2\Phi_{\rm
  SG}\rangle_p+{1\over4}\langle(|0\rangle \langle 0|)\rangle_p,
\label{N43}\\
\langle\sin^2\phaseop\rangle_p&=&\langle\widehat{\sin}^2\Phi_{\rm
SG} \rangle_p+{1\over4}\langle(|0\rangle\langle 0|)\rangle_p.
\label{N44}
\end{eqnarray}
where the subscript $p$ refers to a physical state
expectation value.

%%%%%%%%%%%%%%%%%%%%%%%%%%%%%%%%%%%%%%%%%%%%%%%%%%%%%%%%%%%%%%%%%%%%%%%%%%%%
\subsection{Phase distributions from $s$-parametrized
  quasidistributions}
%%%%%%%%%%%%%%%%%%%%%%%%%%%%%%%%%%%%%%%%%%%%%%%%%%%%%%%%%%%%%%%%%%%%%%%%%%%%

According to \citeANP{CG69a}~\citeY{CG69a,CG69b}, the
$s$-parametrized quasidistribution function $\qpd$ describing
a field state, can be derived from the formula
%---------------------------------------------------------------------------
\begin{eqnarray}
  \qpd &=& \frac{1}{\pi} \,{\rm Tr}\,\{ \hat{\rho}\,
  \hat{T}\us(\alpha)\},
\label{N45}
\end{eqnarray}
where the operator $\hat{T}\us(\alpha)$ is given by
%---------------------------------------------------------------------------
\begin{eqnarray}
  \hat{T}\us(\alpha) &=& \frac{1}{\pi} \int
  \exp(\alpha\xi^*-\alpha^*\xi) \hat{D}\us(\xi) \,{\rm d}^2 \xi,
\label{N46}
\end{eqnarray}
%---------------------------------------------------------------------------
\begin{eqnarray}
  \hat{D}\us(\xi) &=& {\rm e}^{s|\xi|^2/2} \hat{D}(\xi)
\label{N47}
\end{eqnarray}
with $\hat{D}(\xi)$ being the displacement operator;
$\hat{\rho}$ is the density matrix of the field, and we have
introduced the extra $1/\pi$ factor with respect to the
original definition of~\citeAY{CG69b}. The operator
$\hat{T}\us(\alpha)$ can be rewritten in the
form~\cite{CG69a}:
%---------------------------------------------------------------------------
\begin{eqnarray}
  \hat{T}\us(\alpha) &=& \frac{2}{1-s} \sum_{n=0}^{\infty}
  \hat{D}(\alpha) |n\rangle \left( \frac{s+1}{s-1}\right)^n \langle n|
  \hat{D}^{\dagger}(\alpha),
\label{N48}
\end{eqnarray}
which gives explicitly its $s$-dependence. So, the
$s$-parametrized quasidistribution function $\qpd$ has the
following form in the number-state basis:
%---------------------------------------------------------------------------
\begin{eqnarray}
  \qpd &=& \frac{1}{\pi}\sum_{m,n} \rho_{mn} \langle
  n|\hat{T}\us(\alpha) |m\rangle,
\label{N49}
\end{eqnarray}
where the matrix elements of the operator~\mref{N46} are
given by~\cite{CG69a}:
%----------------------------------------------------------------------
\begin{eqnarray}
\label{N50} \langle n|\hat{T}^{(s)}(\alpha)|m\rangle &=&
\left(\frac{n!}{m!} \right)^{1/2} \left(
\frac{2}{1-s}\right)^{m-n+1} \left(
  \frac{s+1}{s-1} \right)^n \nonumber \\ &&\times\; {\rm e}^{-\I
  (m-n)\theta} |\alpha|^{m-n} \exp \left(-\frac{2|\alpha|^2}{1-s}
\right) L_{n}^{m-n}\left( \frac{4|\alpha|^2}{1-s^2}\right),
\end{eqnarray}
in terms of the associate Laguerre polynomials
$L_n^{m-n}(x)$. In eq.~\mref{N50} we have also separated
explicitly the phase of the complex number $\alpha$ by
writing
%---------------------------------------------------------------------------
\begin{eqnarray}
  \alpha&=& |\alpha| {\rm e}^{\I\theta}.
\label{N51}
\end{eqnarray}
In the following, the phase $\theta$ will be treated as the
quantity representing the field phase.

With the quasiprobability distributions $\qpd$ the
expectation values of the $s$-ordered products of the
creation and annihilation operators can be obtained by proper
integrations in the complex $\alpha$ plane. In particular,
for $s=1,0,-1$, the $s$-ordered products are normal,
symmetrical, and antinormal ordered products of the creation
and annihilation operators, and the corresponding
quasiprobability distributions are the Glauber-Sudarshan
$P$-function, Wigner function, and Husimi $Q$-function.

By virtue of the relation inverse to eq.~\mref{N49}, given
by~\cite{CG69b}:
%----------------------------------------------------------------------
\begin{eqnarray}
\label{N52} \hat{\rho} &=& \int {\rm d}^2 \alpha\:
\hat{T}^{(-s)}(\alpha) W\us(\alpha),
\end{eqnarray}
the Pegg-Barnett phase distribution~\mref{N37} can be related
to the $s$-parametrized quasidistribution function~\mref{N45}
as follows \cite{ER91}:
%----------------------------------------------------------------------
\begin{eqnarray}
\label{N53} P(\theta) &=& \int {\rm d}^2 \alpha\:
K\us(\alpha,\theta) W\us(\alpha),
\end{eqnarray}
where the kernel is given by
%----------------------------------------------------------------------
\begin{eqnarray}
\label{N54} K\us(\alpha,\theta) &=& \frac{1}{2\pi} \sum_{m,n}
\langle n|\hat{T}^{(-s)}(\alpha)|m\rangle \exp[\I
(m-n)\theta]
\end{eqnarray}
in terms of the matrix elements~\mref{N50} for $(-s)$. The
kernel~\mref{N54} is convergent for $s>-1$ only.
Nevertheless, the remaining relation between the Husimi
$Q$-function and the Pegg-Barnett distribution can also be
expressed by eq.~\mref{N53}, albeit with the following
kernel~\cite{Mir94}:
%----------------------------------------------------------------------
\begin{eqnarray}
\label{N55} K^{(-1)}(\alpha,\theta) &=& \frac{1}{2\pi}
\sum_{n,m}\int {\rm d}^2 \beta\:
\frac{(\alpha+\beta)^m(\alpha^*-\beta^*)^{n}}{\sqrt{n!m!}}
\nonumber\\ &&\times\;
\exp[\I(n-m)\theta-|\alpha|^2+\alpha\beta^*-\alpha^*\beta].
\end{eqnarray}

On integrating the quasiprobability distribution $\qpd$ over
the ``radial'' variable $|\alpha|$, we get the ``phase
distribution'' associated with this quasiprobability
distribution. The $s$-parametrized phase distribution is thus
given by
%---------------------------------------------------------------------------
\begin{eqnarray}
\label{N56} \spd &=& \int\limits_{0}^{\infty} \qpd |\alpha|
\:{\rm d}|\alpha|
\end{eqnarray}
or equivalently by
%---------------------------------------------------------------------------
\begin{eqnarray}
\label{N57} \spd &=& \frac{1}{2}\int\limits_{0}^{\infty}
{\cal W}^{(s)}(W,\theta) \:{\rm d} W,
\end{eqnarray}
where integration is performed over the intensity
$W=|\alpha|^2$. Inserting eq.~\mref{N49} into eq.~\mref{N56}
yields
%---------------------------------------------------------------------------
\begin{eqnarray}
  \spd &=& \frac{1}{\pi} \sum_{m,n} \rho_{mn} \left(
    \frac{n!}{m!}\right)^{1/2} \left( \frac{2}{1-s}\right)^{m-n+1}
  \left( \frac{s+1}{s-1}\right)^{n} {\rm e}^{-\I (m-n)\theta}
  \nonumber \\ &&\times\; \int\limits_{0}^{\infty} |\alpha|^{m-n}
  \exp\left(-\frac{2|\alpha|^2}{1-s}\right)
  L_n^{m-n}\left(\frac{4|\alpha|^2}{1-s^2}\right) |\alpha| \:{\rm
    d}|\alpha|.
\label{N58}
\end{eqnarray}
If the definition of the Laguerre polynomial is invoked, the
integrations in eq.~\mref{N58} can be performed explicitly,
and we get for the $s$-parametrized phase distribution a
formula similar to the Pegg-Barnett phase
distribution~\mref{N37}, which reads
%---------------------------------------------------------------------------
\begin{eqnarray}
  \spd &=& \frac{1}{2\pi} \left\{ 1+2{\rm Re} \sum_{m>n} \rho_{mn}
    {\rm e}^{-\I (m-n)\theta}G^{(s)}(m,n) \right\}.
\label{N59}
\end{eqnarray}
The difference between eqs.~\mref{N37} and~\mref{N59} lies in
the coefficients $G^{(s)}(m,n)$, which are given by
%--------------------------------------------------------------------------
\begin{eqnarray}
  G^{(s)}(m,n) &=& \left( \frac{2}{1-s}\right)^{\frac{m+n}{2}}\:
  \sum_{l=0}^{{\rm min}(m,n)} (-1)^l \left(\frac{1+s}{2}\right)^{l}
  \nonumber \\ &&\times\; \sqrt{\left( n\atop l \right)\left( m\atop l
    \right)}
  \frac{\Gamma\left(\frac{m+n}{2}-l+1\right)}{\sqrt{(m-l)!(n-l)!}}.
\label{N60}
\end{eqnarray}
The $s$-parametrized coefficients $G^{(s)}(m,n)$,
[eq.~\mref{N60}], can be rewritten in a compact
form~\cite{Mir94,LVBP95} ($m\ge n$):
%----------------------------------------------------------------------
\begin{eqnarray}
\label{N61} G^{(s)}(m,n) &=& \sqrt{\frac{n!}{m!}} \left(
\frac{s+1}{s-1}\right)^n \left(
\frac{2}{1-s}\right)^{(m-n)/2} \nonumber \\ &&\times\;\;
\Gamma\left(\frac{m-n}{2}+1\right)
P_n^{\left(m-n,-\frac{1}{2}(m+n)\right)}
\left(\frac{s-3}{s+1}\right)
\end{eqnarray}
in terms of the Jacobi polynomials $P_n^{(\nu,\mu)}(x)$, or
equivalently as
%----------------------------------------------------------------------
\begin{eqnarray}
\label{N62} G^{(s)}(m,n) &=& \sqrt{\frac{m!}{n!}}
\frac{1}{(m-n)!} \left(
  \frac{s+1}{s-1}\right)^n \left( \frac{2}{1-s}\right)^{(m-n)/2}
\nonumber \\ &&\times\;\;
\Gamma\left(\frac{m-n}{2}+1\right)\:
_2F_1\left(-n,\frac{m-n}{2}+1,m-n+1,\frac{2}{1+s}\right)
\end{eqnarray}
using the hypergeometric (Gauss) function $_2F_1(a,b,c,x)$.

For s=0, we have the coefficients for the {\em Wigner phase
  distribution} $P^{(0)}(\theta)$, i.e., the phase distribution
associated with the Wigner function. In this special case of
$s=0$, eq.~\mref{N60} reduces to the expression obtained
by~\citeAY{TMGC92}, whereas eq.~\mref{N62} goes over into the
expression of~\citeA{GK92}~\citeY{GK92,GK93}:
%----------------------------------------------------------------------
\begin{eqnarray}
\label{N63} G^{(0)}(m,n) &=& \left\{
\begin{array}{ll}
  2^{(m-n)/2}\sqrt{\frac{n!}{m!}} \frac{\Gamma(m/2+1)}{(n/2)!} & n \:
  {\rm even}\\ 2^{(m-n)/2}\sqrt{\frac{n!}{m!}}
  \frac{\Gamma((m+1)/2)}{[(n-1)/2]!} & n \: {\rm odd.}
\end{array}
\right.
\end{eqnarray}
Equations~\mref{N61}--\mref{N63} are given for $m\ge n$.
Otherwise the indices $m$ and $n$ should be interchanged.

For $s=-1$, only the term with $l=0$ survives in
eq.~\mref{N60}, and we get the coefficients for the {\em
Husimi phase distribution} $P^{(-1)}(\theta)$, i.e., the
phase distribution associated with the $Q$-function.  Now,
eq.~\mref{N60} reduces to~\cite{Pau74,TGMK91,TG92b}:
%----------------------------------------------------------------------
\begin{eqnarray}
  G^{(-1)}(m,n) &=& \frac{\Gamma\left(
      \frac{n+m}{2}+1\right)}{\sqrt{n!m!}}.
\label{N64}
\end{eqnarray}
It is apparent from eqs.~\mref{N59}--\mref{N62} that for the
phase distribution associated with the $P$-function ($s$=1)
the coefficients $G^{(s)}(m,n)$ become infinity, and one can
conclude that the phase distribution $P^{(1)}(\theta)$ is
indeterminate. However, at least for a special class of
states, summation can be performed numerically or even
analytically for $P^{(1)}(\theta)$. For instance, for the
states described by the density matrix $\hat{\rho}$ of the
form
%----------------------------------------------------------------------
\begin{eqnarray}
  \rho_{mn}&=&|\rho_{mn}| \exp[-\I (m-n)\vartheta_0],
\label{N65}
\end{eqnarray}
the $s$-parametrized phase distribution $\spd$ can be
rewritten as~\cite{Mir94}:
%----------------------------------------------------------------------
\begin{eqnarray}
  \spd &=& \frac{1}{2\pi} \left\{ 1+2 \sum_{m=1}^{\infty} a_m^{(s)}
    \cos[m(\theta-\vartheta_0)] \right\}
\label{N66}
\end{eqnarray}
with the coefficients
%----------------------------------------------------------------------
\begin{eqnarray}
  a_m^{(s)} &=& \sum_{n=0}^{\infty} |\rho_{m+n,n}| G^{(s)} (m+n,n).
\label{N67}
\end{eqnarray}
Equations~\mref{N66} and~\mref{N67}, for $s=0$ and
$\vartheta_0=0$, go over into expressions obtained
by~\citeNP{BR93}.  Numerical calculation of
$\lim\limits_{s\rightarrow 1} a_m^{(s)}$ is usually
straightforward. For coherent states, the coefficients
$a_m^{(1)}$ are equal to unity. Hence, $P^{(1)}_{\rm
coh}(\theta)$, given by eq.~\mref{N66}, is the Dirac delta
function $\delta(\theta-\vartheta_0)$ [see \S 3.1].

Formulas~\mref{N59}--\mref{N62} allow calculation of the
$s$-parametrized phase distributions for any state with known
$\rho_{mn}$ and their comparison with the Pegg-Barnett phase
distribution, for which $G^{(s)}(m,n)\equiv 1$. The phase
distributions associated with particular quasiprobability
distributions have been used widely in the literature to
describe the phase properties of field states. For example,
the Husimi phase distribution $P^{(-1)}(\theta)$ was used
by~\citeAY{BP69,Pau74,FS93},~\citeA{FVS93}~\citeY{FVS93,FVS93a},
\citeAY{LP93b,Ban93} and {KC94a} in their schemes for phase
measurement. \citeAY{BC90} applied $P^{(-1)}(\theta)$ to
describe the phase properties of generalized squeezed states.
The Wigner phase distribution $P^{(0)}(\theta)$ was used
by~\citeA{SHV89a}~\citeY{SHV89,SHV89a} in their description
of the phase probability distribution for highly squeezed
states.~\citeAY{HPR93} showed in general that the Wigner
phase distribution can be interpreted as an approximation of
the Pegg-Barnett distribution. To estimate the difference
between the $P^{(0)}(\theta)$ and $P(\theta)$ they analyzed
the deviation of the Wigner function ${\cal W}^{(0}(\alpha)$
for a phase state from Dirac's delta function.
Recently,~\citeAY{HFS95} have compared the Pegg-Barnett,
Husimi, and Wigner phase distributions for large-amplitude
classical states.~\citeAY{ER91} applied the $s$-parametrized
quasiprobability distributions to study properties of the
Jaynes-Cummings model with cavity damping.

%%%%%%%%%%%%%%%%%%%%%%%%%%%%%%%%%%%%%%%%%%%%%%%%%%%%%%%%%%%%%%%%%%%%%%%%%%%%
\begin{figure} % Fig. 1
\vspace*{0cm} \hspace*{0mm} \epsfbox{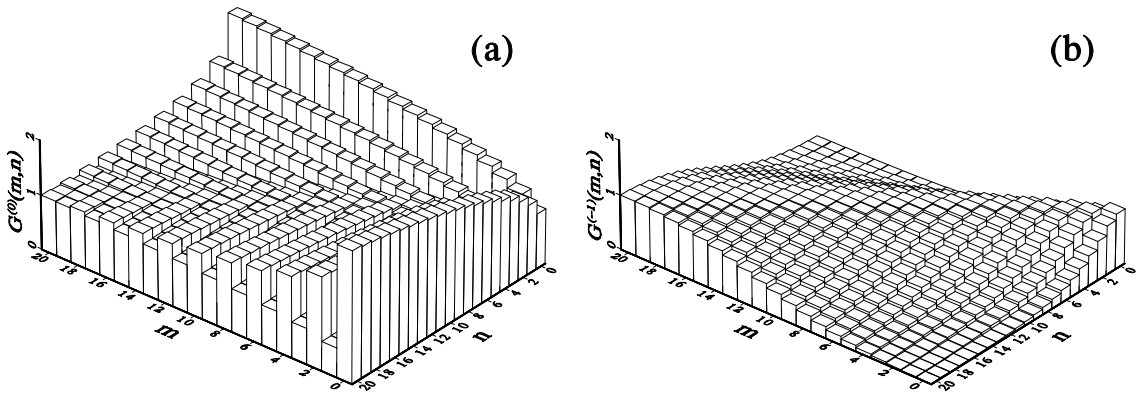}
\vspace*{-0cm} \caption{The coefficients $G^{(s)}(m,n)$ for
(a) $s=0$, and (b) $s=-1$}
\end{figure} \vspace*{-0.cm}
For some field states the quasiprobability distribution
functions $\qpd$ can be found in a closed form via direct
integrations according to the definitions
\mref{N45}--\mref{N47}, and sometimes the next integration
leading to the $s$-parametrized phase distributions can also
be performed according to definition~\mref{N56}. In the next
Sections, we shall illustrate the differences between the
Pegg-Barnett phase distribution and the $s$-parametrized
phase distributions obtained by integrating the
$s$-parametrized quasiprobability distribution functions. For
any field with known number-state matrix elements $\rho_{mn}$
of the density matrix, the $s$-parametrized phase
distribution can be calculated according to eq.~\mref{N59}
with the coefficients $G^{(s)}(m,n)$ given by eq.~\mref{N60}.
The distribution of the coefficients $G^{(s)}(m,n)$, for
$s=0,-1$, is illustrated in fig.~1. It is apparent that for
the Husimi phase distribution the coefficients decrease
monotonically as we go further away from the diagonal. This
means that all nondiagonal elements $\rho_{mn}$ are weighted
with numbers that are less than unity, and the phase
distribution for $s=-1$ is always broader than the
Pegg-Barnett phase distribution (for which
$G^{(s)}(m,n)\equiv 1$). For $s=0$ the situation is not so
simple, because the coefficients $G^{(0)}(m,n)$ show even-odd
oscillations with values that are both smaller and larger
than unity. This usually leads to a phase structure sharper
than the Pegg-Barnett distribution. Moreover, since the
Wigner function ($s=0$) can take negative values, the
positive definiteness of the Wigner phase distribution is not
guaranteed. Also, the oscillatory behavior of the coefficient
$G^{(0)}(m,n)$ suggests that, at least for some states, the
Wigner phase distribution $P^{(0)}(\theta)$ can exhibit
negative values. This nonclassical feature of
$P^{(0)}(\theta)$ was shown explicitly by
\citeA{GK92}~\citeY{GK92,GK93} for the simple example of the
number-state superposition (only for convenience, we assume
that $m>n$):
%----------------------------------------------------------------------
\begin{eqnarray}
\label{N68} |\Psi\rangle &=& 2^{-1/2} (|n\rangle +
|m\rangle).
\end{eqnarray}
In a straightforward manner, the general expressions for the
phase distributions $\pbd$, [eq.~\mref{N37}] and
$\spd$,[eq.~\mref{N59}] reduce to
%----------------------------------------------------------------------
\begin{eqnarray}
\label{N69} \pbd &=& \frac{1}{2\pi} (1+\cos [(m-n)\theta]),
\end{eqnarray}
and
%----------------------------------------------------------------------
\begin{eqnarray}
\label{N70} \spd &=& \frac{1}{2\pi} \left(1+
G\us(m,n)\cos[(m-n)\theta]\right).
\end{eqnarray}
%%%%%%%%%%%%%%%%%%%%%%%%%%%%%%%%%%%%%%%%%%%%%%%%%%%%%%%%%%%%%%%%%%%%%%%%%%%%
\begin{figure} % Fig. 2
\vspace*{-1.5cm} \hspace*{15mm} \epsfxsize=12cm
\epsfbox{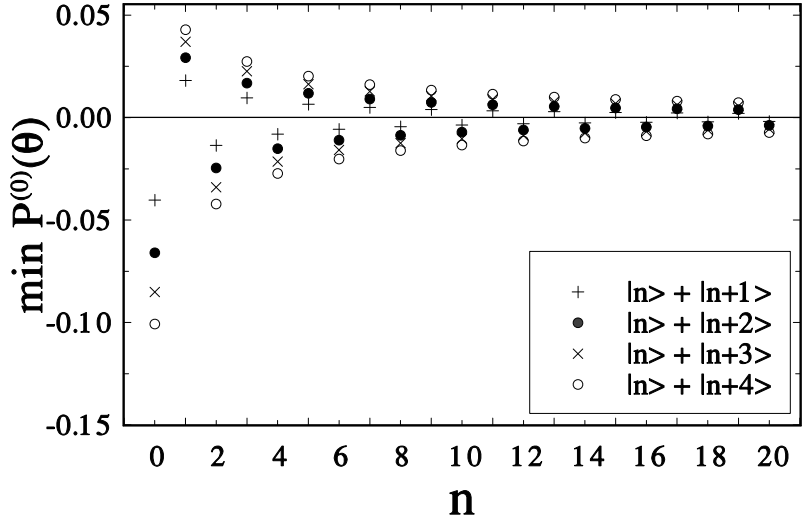} \vspace*{0cm} \caption{ The minima of the
Wigner phase distributions $P^{(0)}(\theta) $, eq. (2.70),
for the superpositions of two number states (2.68) for
various values of $n$ and $m-n$=1, 2, 3, 4. }
\end{figure}
respectively, The Pegg-Barnett, $\pbd$, and Husimi,
$P^{(-1)}(\theta)$, phase distributions are obviously
positive definite for any superposition~\mref{N68}. As seen
in fig.~2, the Wigner phase distribution $P^{(0)}(\theta)$ is
positive for superpositions with odd $m-n$. However, it takes
negative values for even $m-n$. In this case, the smaller is
$n$ for fixed $m-n$, or the higher is the value of $m-n$ for
a given $n$, the minimum of the Wigner phase distribution is
more strongly negative. Hence, one obtains the greatest
negativity for the superposition
$(|0\rangle+|2m\rangle)/\sqrt{2}$ in the limit of
$m\rightarrow\infty$.  As was emphasized by~\citeAY{GK93}
(see also fig.~2), for large values of $n$ the Pegg-Barnett
distribution is approached for both even and odd $m$.

It is highly illustrative to consider analytically the
special case of eq.~\mref{N68} when $m-n=2$
(\citeA{GK92}~\citeY{GK92,GK93}). These results will be
helpful in the analysis presented in \S 3.2 for even and odd
coherent states. Now, the coefficients $G\us(m,n)$, given by
eqs.~\mref{N60}--\mref{N62}, can be rewritten in a much
simpler form:
%----------------------------------------------------------------------
\begin{eqnarray}
\label{N71} G\us(n+2,n) &=& \frac{1-s}{2\sqrt{(n+1)(n+2)}}
\left[
  \left(\frac{s+1}{s-1}\right)^{n+2} - 1 \right] +
\left(\frac{n+2}{n+1}\right)^{1/2}.
\end{eqnarray}
For $s=0$, eq.~\mref{N71} goes over into
(\citeA{GK92}~\citeY{GK92,GK93}):
%----------------------------------------------------------------------
\begin{eqnarray}
\label{N72} G^{(0)}(n+2,n) &=&
\left(\frac{n+2}{n+1}\right)^{(-1)^n/2}
\end{eqnarray}
and for $s=-1$ one obtains
%----------------------------------------------------------------------
\begin{eqnarray}
\label{N73} G^{(-1)}(n+2,n) &=&
\left(\frac{n+1}{n+2}\right)^{1/2}.
\end{eqnarray}
Equation~\mref{N72} provides direct proof of the oscillatory
behavior of $G^{(0)}(n+2,n)$ with increasing $n$. For even
$n$, the right-hand-side of eq.~\mref{N72} is greater than
unity, which implies a negative minimum of the Wigner phase
distribution~\mref{N70} [solid circles in fig.~2]. However,
for odd $n$, the coefficients $G^{(0)}(n+2,n)$ are less than
unity and equal to $G^{(-1)}(n+2,n)$. Hence the Husimi and
Wigner phase distributions for such states (with odd $n$) are
equal and positive definite.

From the form of the coefficients $G^{(s)}(m,n)$ it is
evident that there is no $s$ such that $G^{(s)}(m,n)=1$ for
all $m$, $n$. This means that there is no ``phase ordering''
of the field operators; that is, the ordering for which
$P\us(\theta)$ would be equal to $\pbd$. However, for a given
state of the field one can find $s$ such that the two
distributions are ``almost identical''.  Formula~\mref{N59}
is quite general, and it was used in earlier studies of the
phase properties of the anharmonic oscillator~\cite{TGMK91},
parametric down conversion~\cite{TG92b}), and
displaced-number states~\cite{TMGC92}. A disadvantage of
formula~\mref{N59} is the fact that the numerical summations
can be time consuming and even difficult to perform for field
states with slowly converging number-state expansions. This,
for example, is the case for highly squeezed states. In some
cases, instead of using the number-state expansions, we can
find analytical formulas for $\spd$ via direct integrations,
as shown in \S 3. In many cases such formulas can be treated
as good approximations to the Pegg-Barnett phase
distribution.

%%%%%%%%%%%%%%%%%%%%%%%%%%%%%%%%%%%%%%%%%%%%%%%%%%%%%%%%%%%%%%%%%%%%%%%%%%%%
%           Phase properties of single-mode optical fields
%%%%%%%%%%%%%%%%%%%%%%%%%%%%%%%%%%%%%%%%%%%%%%%%%%%%%%%%%%%%%%%%%%%%%%%%%%%%
\section{Phase properties of single-mode optical fields}
\setcounter{equation}{0}

Optical fields produced in nonlinear optical processes have
specific phase properties which depend on the nonlinear
process in which the field is produced and on the state of
the field before it undergoes the nonlinear transformation.
Since there is a large variety of nonlinear optical
processes, there is the possibility to generate fields with
different phase properties. Here, some examples of such field
states will be given and their phase properties discussed
briefly.

%%%%%%%%%%%%%%%%%%%%%%%%%%%%%%%%%%%%%%%%%%%%%%%%%%%%%%%%%%%%%%%%%%%%%%%%
\subsection{Coherent states}
%%%%%%%%%%%%%%%%%%%%%%%%%%%%%%%%%%%%%%%%%%%%%%%%%%%%%%%%%%%%%%%%%%%%%%%%

The most common single-mode field in quantum optics is a
Glauber coherent state. Its phase properties have probably
been analyzed within each known phase formalism. We shall
focus our attention on two of them only.

The $s$-parametrized quasiprobability distribution function
for a coherent state
%--------------------------------------------------------------------------
\begin{eqnarray}
  |\alpha_0\rangle &=& \hat{D}(\alpha_0) |0\rangle
\label{N74}
\end{eqnarray}
can be calculated from eqs.~\mref{N45}--\mref{N47} as
%---------------------------------------------------------------------------
\begin{eqnarray}
  \qpd &=& \frac{1}{\pi^2} \int
  \exp(\alpha\xi^*-\alpha^*\xi+s|\xi|^2/2) \langle
  0|\hat{D}^{\dagger}(\alpha_0)\hat{D}(\xi)\hat{D}(\alpha_0)|0\rangle
  \, {\rm d}^2\xi \nonumber \\ &=& \frac{1}{\pi^2} \int
  \exp[(\alpha-\alpha_0)\xi^*-(\alpha^*-\alpha_0^*)\xi+s|\xi|^2/2]
  \langle 0|\hat{D}(\xi)|0\rangle \, {\rm d}^2\xi \nonumber \\ &=&
  \frac{1}{\pi^2}\int
  \exp[(\alpha-\alpha_0)\xi^*-(\alpha^*-\alpha_0^*)\xi
  +s|\xi|^2/2-|\alpha_0|^2/2] \, {\rm d}^2\xi \nonumber \\ &=&
  \frac{1}{\pi} \frac{2}{1-s} \exp \left\{ -\frac{2}{1-s}
    |\alpha-\alpha_0|^2 \right\}.
\label{N75}
\end{eqnarray}
The corresponding $s$-parametrized phase distribution is
(\citeAY{TMG93}; for $s=0$ see also~\citeAY{GK93} and
{BR93}):
%--------------------------------------------------------------------------
\begin{eqnarray}
  \spd &=& \int\limits_{0}^{\infty} \qpd |\alpha| \: {\rm d}|\alpha|
  \nonumber \\ &=& \frac{1}{2\pi}\exp[-(X_0^2-X^2)] \left\{
    \exp(-X^2)+\sqrt{\pi}X\,(1+{\rm erf}(X))\right\},
\label{N76}
\end{eqnarray}
where
%--------------------------------------------------------------------------
\begin{eqnarray}
  X &=& X\us(\theta) \:=\: \sqrt{\frac{2}{1-s}} |\alpha_0|
  \cos(\theta-\vartheta_0),
\label{N77}
\end{eqnarray}
and $X_0=X\us(\vartheta_0)$; $\vartheta_0$ is the phase of
$\alpha_0$. The phase distribution $P^{(1)}(\theta)$
associated with the $P$-function can be obtained from
eqs.~\mref{N76} and~\mref{N77} in the limit of $s\rightarrow
1$:
%----------------------------------------------------------------------
\begin{eqnarray}
\label{N78} P^{(1)}(\theta) &=&\delta (\theta-\vartheta_0),
\end{eqnarray}
which is the Dirac delta function. This result can also be
achieved from eq.~\mref{N66}. As was shown by~\citeAY{Mir94},
the coefficients $a_m^{(1)}$ are unity for arbitrary $m$.
Hence, eq.~\mref{N66} reduces to
%----------------------------------------------------------------------
\begin{eqnarray}
\label{N79} P^{(1)}(\theta) &=& \frac{1}{2\pi}\, \left\{ 1+2
  \sum_{m=1}^{\infty}\cos[m(\theta-\vartheta_0)] \right\} \nonumber\\
&\equiv&\delta (\theta-\vartheta_0),
\end{eqnarray}
which is the desired function~\mref{N78}. This example shows
that the general expression~\mref{N59} for the
$s$-parametrized phase distributions is also valid in the
special case of $s$=1.

%%%%%%%%%%%%%%%%%%%%%%%%%%%%%%%%%%%%%%%%%%%%%%%%%%%%%%%%%%%%%%%%%%%%%%%%%%%%
\begin{figure} % Fig. 3
\vspace*{-4cm} \hspace*{-5mm} \epsfbox{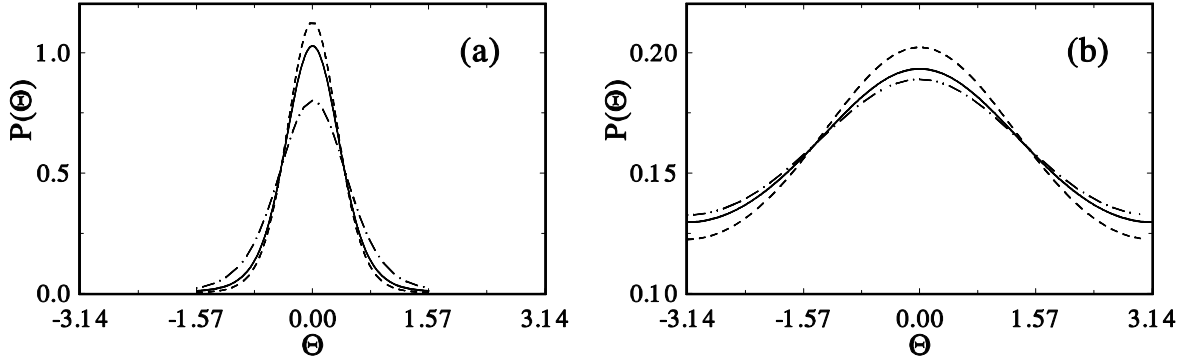}
\vspace*{-4.3cm} \caption{{ Phase distributions for the
coherent states with the mean number of photons: (a)
$|\alpha_0|^2=2$, and (b) $|\alpha_0|^2=0.01$; the
Pegg-Barnett distribution (solid line), the Wigner phase
function $P^{(0)}(\theta)$ (dashed line), and the Husimi
phase distribution $P^{(-1)}(\theta)$ (dotted-dashed line).
}}
\end{figure}
Formula~\mref{N76} is exact, it is $2\pi$-periodic, positive
definite and normalized, so it satisfies all requirements for
the phase distribution. Moreover, formula~\mref{N76} has a
quite simple and transparent structure. For small
$|\alpha_0|$, the first term in braces plays an essential
role, and for $|\alpha_0|\rightarrow 0$ we get a uniform
phase distribution. For large $|\alpha_0|$, the second term
in the braces predominates, and if we replace ${\rm erf}(X)$
by unity, we obtain the approximate asymptotic formula given
by \citeAY{SDHV90} (for $s=0$):
%--------------------------------------------------------------------------
\begin{eqnarray}
  P^{(0)}(\theta) &\approx& \sqrt{\frac{2}{\pi}} |\alpha_0|
  \cos(\theta-\vartheta_0)\exp[-2|\alpha_0|^2 \sin^2(\theta-
  \vartheta_0)],
\label{N80}
\end{eqnarray}
which however, can be applied only to
$-\pi/2\leq(\theta-\vartheta_0)\leq\pi/2$.  After
linearization of formula~\mref{N80} with respect to $\theta$,
the approximate formula for coherent states with large mean
number of photons obtained by~\citeAY{BP89} is recovered. The
presence of the error function in eq.~\mref{N76} handles
properly the phase behavior in the whole range of phase
values $-\pi\leq(\theta-\vartheta_0)<\pi$.

The Pegg-Barnett distribution $\pbd$ for the coherent state
$|\alpha_0\rangle$ can be calculated from eq.~\mref{N36} with
the superposition coefficients
%----------------------------------------------------------------------
\begin{eqnarray}
\label{N81} c_n &=& b_n \exp(\I n\vartheta_0), \ \ \ \ b_n
\;=\; \exp(-|\alpha_0|^2/2) \frac{|\alpha_0|^n}{\sqrt{n!}}.
\end{eqnarray}
The exact formula for the $s$-parametrized phase
distributions $\spd$ for coherent states is given by
eqs.~\mref{N76} and~\mref{N77}. Alternatively, the $\spd$ are
given by eq.~\mref{N59} after insertion of $c_n$ given by
eq.~\mref{N81}. In fig.~3 we show the phase distributions
$\pbd$, $P^{(0)}(\theta)$, and $P^{(-1)}(\theta)$ for a
coherent state with the mean number of photons
$|\alpha_0|^2=2$ (a), and $|\alpha_0|^2=0.01$ (b).  It is
seen that the Pegg-Barnett phase distribution is located
somewhere between the Wigner and Husimi phase distributions.
It becomes closer to $P^{(0)}(\theta)$ for $|\alpha_0|^2\gg
1$, and closer to $P^{(-1)}(\theta)$ for $|\alpha_0|^2\ll 1$.
For $|\alpha_0|^2\rightarrow\infty$, the Pegg-Barnett
distribution tends to the Wigner phase
distribution~\cite{SHV89a,BP89}, and for $|\alpha_0|^2
\rightarrow 0$ all the distributions tend to the uniform
distribution, but the Pegg-Barnett distribution in this
region tends to the Husimi phase distribution.  This means
that for coherent states with large mean numbers of photons,
$P^{(0)}(\theta)$ is a good approximation to the Pegg-Barnett
phase distribution, while for small numbers of photons
$P^{(-1)}(\theta)$ becomes a good approximation to the
Pegg-Barnett distribution.

%%%%%%%%%%%%%%%%%%%%%%%%%%%%%%%%%%%%%%%%%%%%%%%%%%%%%%%%%%%%%%%%%%%%%%%%
\subsection{Superpositions of coherent states}
%%%%%%%%%%%%%%%%%%%%%%%%%%%%%%%%%%%%%%%%%%%%%%%%%%%%%%%%%%%%%%%%%%%%%%%%

Superpositions of macroscopically distinguishable coherent
states have attracted much interest (see, for example,
\citeAY{BK95} and references therein) due to their property
of being prototypes for the Schr\"odinger cats, and important
nonclassical properties, such as sub-Poissonian photon
statistics, quadrature squeezing, higher-order squeezing,
etc.. Their phase properties have also been a subject of
interest.

Let us consider the normalized superposition $|\psi\rangle$
of coherent states defined as
%----------------------------------------------------------------------
\begin{eqnarray}
  |\psi \rangle &=& \sum_{k=1}^{N} c_k |\exp(\I
  \phi_k)\alpha_0\rangle.
\label{N82}
\end{eqnarray}
This superposition of two well-separated components is called
a {\em Schr\"odinger cat}, whereas for $N>2$ the notions {\em
  Schr\"odinger cat-like state} or {\em kitten states} are often used.
The phase distributions $\pbd$, [eq.~\mref{N37}] and $\spd$,
[eq.~\mref{N59}] for the state~\mref{N82} can be rewritten in
a form showing explicitly the superposition structure
(\citeAY{TGMK91},~\citeA{GK92}~\citeY{GK92,GK93},
\citeAY{BGK93,BKG93,TAC93,HG93,Buz93,Mir94,BK95}).

The Pegg-Barnett phase distribution $\pbd$ splits into two
terms~\cite{TGMK91}:
%----------------------------------------------------------------------
\begin{eqnarray}
  \pbd &=& P_0(\theta) \,+\, P_{\rm int}(\theta),
\label{N83}
\end{eqnarray}
where
%----------------------------------------------------------------------
\begin{eqnarray}
  P_0(\theta) &=& \sum_{k=1}^{N} |c_k|^2\, P_k(\theta)
\label{N84}
\end{eqnarray}
is the sum of phase distributions
%----------------------------------------------------------------------
\begin{eqnarray}
  P_k(\theta) &=& \frac{1}{2\pi} \left\{ 1+2 \sum_{m>n} b_m b_n \cos
    \left[ (m-n)(\phi_k+\vartheta_0-\theta) \right] \right\}
\label{N85}
\end{eqnarray}
for the coherent states of the superposition, and the second
distribution
%----------------------------------------------------------------------
\begin{eqnarray}
  P_{\rm int}(\theta) &=& \sum_{k,l=1 \atop k\neq l}^{N} c_k c_l^*\,
  P_{kl}(\theta),
\label{N86}
\end{eqnarray}
%----------------------------------------------------------------------
\begin{eqnarray}
  P_{kl}(\theta) &=& \frac{1}{2\pi} \sum_{m,n} b_m b_{n} \exp\left[ \I
    m\,(\phi_k+\vartheta_0-\theta) -\I n\,(\phi_l+\vartheta_0-\theta)
  \right]
\label{N87}
\end{eqnarray}
represents interference terms emerging from the quantum
interference between the component states of the
superposition. In fig.~4, the phase distributions~\mref{N83}
and~\mref{N84} are presented in polar coordinates for the
discrete superpositions of coherent states in the anharmonic
oscillator model [see \S 3.5,eq.~\mref{N132}].  It is evident
from fig.~4 that as the number of components in the
superposition becomes larger, the interference terms play an
increasing role and the symmetry of the Pegg-Barnett
distribution [eq.~\mref{N83}] is destroyed. These terms are
negligible for well-separated components of the superposition
only \cite{TGMK91}.

Analogously, the $s$-parametrized quasidistribution $\qpd$
for the superposition state~\mref{N82} is represented as
(\citeAY{Mir94}; for $s=-1$ see~\citeAY{MTK90}, and for $s=0$
see also~\citeAY{BK95}):
%----------------------------------------------------------------------
\begin{eqnarray}
  \qpd &=& {\cal W}_0^{(s)}(\alpha) \,+\, {\cal W}_{\rm
    int}^{(s)}(\alpha),
\label{N88}
\end{eqnarray}
where the sum of coherent terms is
%----------------------------------------------------------------------
\begin{eqnarray}
  {\cal W}_0^{(s)}(\alpha) &=& \sum_{k=1}^{N} |c_k|^2 \;{\cal
    W}_k^{(s)}(\alpha)
\label{N89}
\end{eqnarray}
with
%----------------------------------------------------------------------
\begin{eqnarray}
  {\cal W}_k^{(s)}(\alpha) &=& \frac{1}{\pi}\frac{2}{1-s} \exp \left\{
    -\frac{2}{1-s}\left|\alpha-{\rm e}^{{\rm
          i}\phi_k}\alpha_0\right|^2 \right\}
\label{N90}
\end{eqnarray}
The interference part is given by:
%----------------------------------------------------------------------
\begin{eqnarray}
\label{N91} {\cal W}_{\rm int}^{(s)}(\alpha) &=& 2
\sum_{k,l=1\atop k>l}^{N} |c_k|\;|c_l| \;{\cal
W}_{kl}^{(s)}(\alpha)
\end{eqnarray}
with
%----------------------------------------------------------------------
\begin{eqnarray}
\label{N92} {\cal W}_{kl}^{(s)}(\alpha) &=&
\frac{1}{\pi}\frac{2}{1-s}\exp \left\{
  -\frac{1}{1-s} \left( \left|\alpha-{\rm e}^{{\rm
          i}\phi_k}\alpha_0\right|^2+ \left|\alpha-{\rm e}^{{\rm
          i}\phi_l}\alpha_0\right|^2 \right)\right.
\nonumber\\ && \quad\quad\quad\quad\quad\quad+ \left.
2\frac{1+s}{1-s}|\alpha_0|^2
\sin^2\left(\frac{\phi_k-\phi_l}{2} \right)\right\}
\nonumber\\ &&\times\;
  \cos\left\{ \gamma_k-\gamma_l-\frac{1+s}{1-s} |\alpha_0|^2
  \sin(\phi_k-\phi_l) \right.  \nonumber\\ &&+\, \left.  \frac{4}{1-s}
  |\alpha|\;|\alpha_0|
  \cos\left(\frac{\phi_k+\phi_l}{2}+\vartheta_0-\theta\right)
  \sin\left(\frac{\phi_k-\phi_l}{2}\right) \right\}.
\end{eqnarray}
%%%%%%%%%%%%%%%%%%%%%%%%%%%%%%%%%%%%%%%%%%%%%%%%%%%%%%%%%%%%%%%%%%%%%%%%%%
\begin{figure}  % Fig. 4
\vspace*{-2.1cm} \hspace*{5mm} \epsfxsize=14cm
\epsfbox{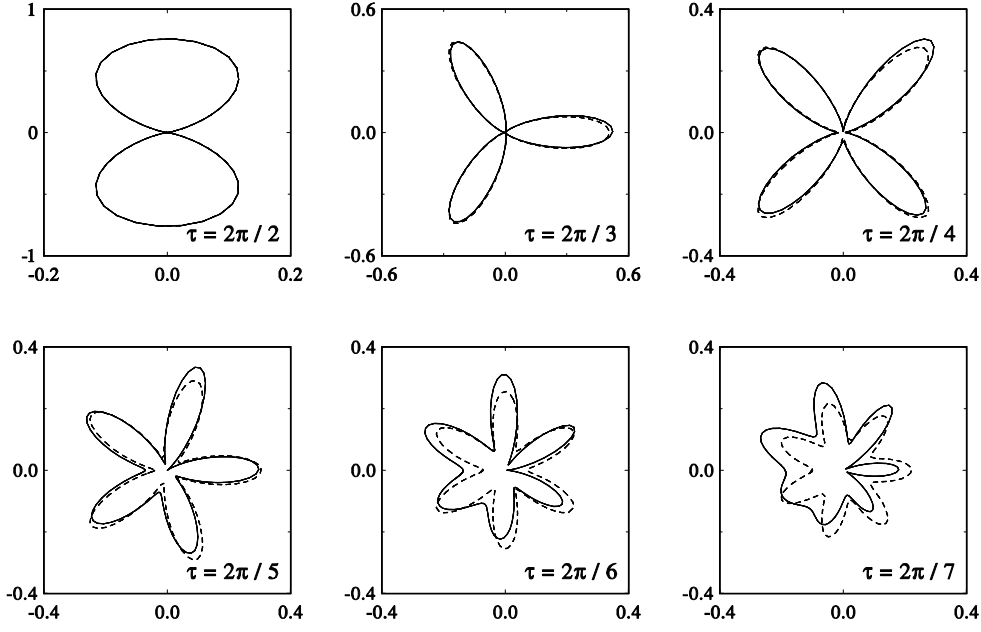} \vspace*{-3mm} \caption{ The Pegg-Barnett
phase distributions in polar coordinates for the discrete
superpositions of coherent states (3.57) with $N$=2--7
components in the anharmonic oscillator model for the initial
mean photon number $|\alpha_0|^2=4$; the exact phase
distributions $P(\theta)$, eq. (3.10) [solid lines], and the
functions $P_0(\theta)$, eq. (3.11) [dashed lines], without
interference contribution. }
\end{figure}
In eq.~\mref{N92} the phases are $\gamma_k = {\rm Arg}\,
c_k$, $\theta = {\rm Arg}\, \alpha$, $\vartheta_0 = {\rm
Arg}\, \alpha_0$, and $\phi_k$ appears in the
definition~\mref{N82}. On integration, we obtain the
following form of the $s$-parametrized phase distribution
$\spd$~\cite{Mir94}:
%----------------------------------------------------------------------
\begin{eqnarray}
  \spd &=& P_0^{(s)}(\theta) \,+\, P_{\rm int}^{(s)}(\theta),
\label{N93}
\end{eqnarray}
i.e., a simple sum
%----------------------------------------------------------------------
\begin{eqnarray}
  P_0^{(s)}(\theta) &=& \sum_{k=1}^{N} |c_k|^2\; P_k\us(\theta)
\label{N94}
\end{eqnarray}
of coherent terms
%----------------------------------------------------------------------
\begin{eqnarray}
  P_k\us(\theta) &=& \frac{1}{2\pi} \exp[-(X_{0}^2-X_k^2)] \left\{
    \exp(-X_k^2)+\sqrt{\pi} X_k\;[1+{\rm erf}(X_k)] \right\},
\label{N95}
\end{eqnarray}
where
%----------------------------------------------------------------------
\begin{eqnarray}
  X_k &=& X_{kk}^{(s)}(\theta) \,=\, \sqrt{\frac{2}{1-s}} |\alpha_0|
  \cos(\theta-\vartheta_0-\phi_k), \;\;\; X_{0} \;=\;
  X_{kk}^{(s)}(\vartheta_0+\phi_k)
\label{N96}
\end{eqnarray}
and the sum
%----------------------------------------------------------------------
\begin{eqnarray}
\label{N97} P_{\rm int}^{(s)}(\theta) &=& \sum_{k\neq l} c_k
c_l^* P_{kl}^{(s)}(\theta)
\end{eqnarray}
of the interference terms
%----------------------------------------------------------------------
\begin{eqnarray}
\label{N98} P_{kl}^{(s)}(\theta) &=&
\frac{1}{2\pi}\exp[-(X^2_{0} \kappa_{kl}-X_{kl}^2)]
\nonumber\\ &&\times\, \left\{
  \exp(-X_{kl}^2)+\sqrt{\pi} X_{kl}\;[1+{\rm erf}(X_{kl})] \right\},
\end{eqnarray}
with
%----------------------------------------------------------------------
\begin{eqnarray}
\label{N99} X_{kl} \;=\; X_{kl}^{(s)}(\theta) &=&
\frac{1}{2}\sqrt{\frac{2}{1-s}} |\alpha_0| \{ \exp[\I
(\phi_k+\vartheta_0-\theta)] \nonumber\\ &&+\; \exp[-\I
(\phi_l+\vartheta_0-\theta)]\},
\end{eqnarray}
%----------------------------------------------------------------------
\begin{eqnarray}
  \kappa_{kl} &=& \frac{1-s}{2}+\frac{1+s}{2} \exp\{\I
  (\phi_k-\phi_l)\},
\label{N100}
\end{eqnarray}
The Schr\"odinger cat of the form:
%----------------------------------------------------------------------
\begin{eqnarray}
\label{N101} |\psi\rangle &\equiv& |\alpha,\gamma\rangle
\;=\; {\cal N}_{\gamma} \left(|\alpha\rangle +
\exp(\I\gamma)|-\alpha\rangle\right),
\end{eqnarray}
with the normalization
%----------------------------------------------------------------------
\begin{eqnarray}
\label{N102} {\cal N}_{\gamma} &=&
\left[2\left(1+\cos\gamma\exp(-2|\alpha|^2)\right)
\right]^{-1/2},
\end{eqnarray}
is a special case of the superposition state~\mref{N82}.  The
cat~\mref{N101} consists of two coherent states
$|\alpha\rangle$ and $|-\alpha\rangle$, which are $\phi=\pi$
out of phase with respect to each other. The
state~\mref{N101} is not only of theoretical interest, since
several methods were proposed for generation and measurement
of this Schr\"odinger cat (e.g.,~\citeAY{BHRDZ92,GSMKK94}).
The state $|\alpha,\pi/2\rangle$ (i.e., for $\gamma=\pi/2$)
is called the Yurke-Stoler coherent state (\citeA{YS86}
\citeY{YS86,YS88}).  This state can be generated in the
anharmonic oscillator model (see \S 3.5). For other choices
of $\gamma$, the state~\mref{N101} goes over into the even
coherent state $|\alpha,0\rangle$ or the odd coherent state
$|\alpha,\pi\rangle$, which have the following Fock
representations~\cite{Per91}:
%----------------------------------------------------------------------
\begin{eqnarray}
\label{N103} |\alpha,0\rangle &=& \cosh^{-1/2}(|\alpha|^2)
\sum_{n=0}^{\infty} \frac{\alpha^{2n}}{\sqrt{(2n)!}}
|2n\rangle,
\end{eqnarray}
%----------------------------------------------------------------------
\begin{eqnarray}
\label{N104} |\alpha,\pi\rangle &=& \sinh^{-1/2}(|\alpha|^2)
\sum_{n=0}^{\infty} \frac{\alpha^{2n+1}}{\sqrt{(2n+1)!}}
|2n+1\rangle.
\end{eqnarray}
The dissimilar phase properties of the even/odd coherent
states were analyzed by~\citeAY{GK93} (see
also~\citeAY{BK95}). Their phase distributions $\pbd$ and
$\spd$ can be obtained readily from the general
expressions~\mref{N83} and~\mref{N93}, respectively.
Obviously, they consist of the normalized sum of the phase
distributions $P_{1,2}(\theta)$ (or $P_{1,2}^{(s)}(\theta)$)
for coherent states located at $\alpha$ and ($-\alpha$) in
the phase space plane, together with an additional
interference term $P_{12}(\theta)$ (or
$P_{12}^{(s)}(\theta)$).  As was shown by~\citeAY{GK93}, the
Wigner phase distribution $P^{(0)}(\theta)$ for the even
coherent state [eq.~\mref{N103}] can exhibit negative values,
in contrast to $P^{(0)}(\theta)$ for the odd coherent state
[eq.~\mref{N104}], which never does. The Fock expansion
[eq.~\mref{N103}] of the even coherent state contains only
even photon numbers similar to the superposition $|n_{\rm
even}\rangle+|n_{\rm even}+2\rangle$ discussed by us in \S
2.1 [see eq.~\mref{N68} and fig.~2]. Analogously, the odd
coherent state [eq.~\mref{N104}] and $|n_{\rm
odd}\rangle+|n_{\rm
  odd}+2\rangle$ contain only odd number states.  Hence, these
dissimilar features of the functions $P^{(0)}(\theta)$ for
$|\alpha,0\rangle$ and $|\alpha,\pi\rangle$ are well
understood for the same reasons as those given in \S 2.1 in
the analysis of the Wigner phase distribution for the
superposition of the two number states and the interpretation
of the oscillatory behavior of the coefficients
$G^{(0)}(m,n)$ [fig.~1a].

%%%%%%%%%%%%%%%%%%%%%%%%%%%%%%%%%%%%%%%%%%%%%%%%%%%%%%%%%%%%%%%%%%%%%%%%
\subsection{Squeezed states}
%%%%%%%%%%%%%%%%%%%%%%%%%%%%%%%%%%%%%%%%%%%%%%%%%%%%%%%%%%%%%%%%%%%%%%%%

Squeezed states have phase-sensitive noise properties, and it
is particularly interesting to study their phase properties.
\citeAY{SBK86,Yao87,LK87}, and~\citeAY{FZ88} used the
Susskind-Glogower formalism in a description of the phase
fluctuations of squeezed states.  \citeAY{Lyn87} applied the
measured-phase formalism of \citeAY{BP86}.  \citeAY{VP89}
and~\citeAY{VBP92} investigated phase properties of a
single-mode squeezed state from the point of view of the new
Pegg-Barnett phase formalism.~\citeAY{GCR89} made comparisons
of the phase properties of a single-mode squeezed state
obtained according to different phase formalisms, including
that of Pegg and Barnett.  \citeAY{BW92} introduced a
phase-space propensity description of quantum-phase
fluctuations and analyzed, in particular, squeezed vacuum.
The phase properties of the squeezed states have recently
been studied by \citeNP{CBM92}, and by
\citeA{Col93}~\citeY{Col93,Col93a}.  Various measures of
phase uncertainty and their dependence on the average number
of photons were studied by \citeNP{FS94}.

Squeezed states (ideal squeezed states, two-photon coherent
states) are defined by (see~\citeAY{LK87}):
%---------------------------------------------------------------------------
\begin{eqnarray}
  |\alpha_0,\zeta\rangle &=& \hat{D}(\alpha_0) \hat{S}(\zeta)
  |0\rangle,
\label{N105}
\end{eqnarray}
where $\hat{S}(\zeta)$ is the squeezing operator
%---------------------------------------------------------------------------
\begin{eqnarray}
  \hat{S}(\zeta) &=& \exp\left(
    \frac{1}{2}\zeta^*\hat{a}^2-\frac{1}{2}\zeta \hat{a}^{\dagger 2}
  \right),
\label{N106}
\end{eqnarray}
and $\zeta$ is the complex squeeze parameter
%---------------------------------------------------------------------------
\begin{eqnarray}
  \zeta &=& r {\rm e}^{2\I\eta},\ \ \ \ r\;=\;|\zeta|.
\label{N107}
\end{eqnarray}
The direct integrations lead to the $s$-parametrized
quasiprobability distribution (for $\eta=0$):
%---------------------------------------------------------------------------
\begin{eqnarray}
  \qpd &=& \frac{2}{\sqrt{(\mu-s)(\mu^{-1}-s)}} \nonumber \\
  &&\times\; \exp\left\{ -\frac{2}{\mu-s}[{\rm Im}(\alpha-\alpha_0)]^2
    -\frac{2}{\mu^{-1}-s}[{\rm Re}(\alpha-\alpha_0)]^2 \right\},
\label{N108}
\end{eqnarray}
where we have used the notation $\mu = {\rm e}^{2r}$.  After
integration over $|\alpha|$, assuming that $\alpha_0$ is
real, we arrive at the formula \cite{TMG93}:
%---------------------------------------------------------------------------
\begin{eqnarray}
  \spd &=& \frac{1}{2\pi} \frac{\sqrt{(\mu-s)(\mu^{-1}-s)}}
  {(\mu-s)\cos^2\theta+(\mu^{-1}-s)\sin^2\theta} \nonumber \\
  &&\times\; \exp[-(X_0^2-X^2)] \left\{\exp(-X^2)+\sqrt{\pi}X\,(1+{\rm
      erf}(X))\right\},
\label{N110}
\end{eqnarray}
where
%---------------------------------------------------------------------------
\begin{eqnarray}
  X &=& X\us(\theta) \:=\: \sqrt{\frac{2}{\mu^{-1}-s} }\;
  \frac{\alpha_0\sqrt{\mu-s}\cos\theta}{\sqrt{(\mu-s)\cos^2\theta
      +(\mu^{-1}-s)\sin^2\theta} }.
\label{N111}
\end{eqnarray}
Although the variable $X$ is slightly different, the main
structure of the phase distribution is preserved.
Formula~\mref{N110} is valid for both small and large
$\alpha_0$. For $\alpha_0=0$ we have the result for squeezed
vacuum. After appropriate approximations, one easily obtains
the formula derived by~\citeAY{SHV89a} for a highly squeezed
state.

The exact analytical formula for the $s$-parametrized phase
distribution for squeezed states, as given by
eqs.~\mref{N110} and \mref{N111}, for the squeezed vacuum
takes the form:
%---------------------------------------------------------------------------
\begin{eqnarray}
  \spd &=& \frac{1}{2\pi} \frac{\sqrt{(\mu-s)(\mu^{-1}-s)}}
  {(\mu-s)\cos^2\theta+(\mu^{-1}-s)\sin^2\theta},
\label{N112}
\end{eqnarray}
where $\mu=\exp(2r)$. This formula exhibits a two-peak
structure with peaks for $\theta=\pm\pi/2$ (for $r>0$). It is
easy to find that the peak heights are
%---------------------------------------------------------------------------
\begin{eqnarray}
  P\us(\pi/2) &=& \frac{1}{2\pi} \sqrt{ \frac{\mu-s}{\mu^{-1}-s} },
\label{N113}
\end{eqnarray}
meaning that for $s=0$, the peak height is proportional to
$\mu$.  One can easily check that the Pegg-Barnett result
lies between the $s=0$ and $s=-1$ results. Qualitatively, all
three distributions give the same two-peak phase
distributions, but they differ quantitatively: the sharpest
peaks are those of $P^{(0)}(\theta)$, and the broadest those
of $P^{(-1)}(\theta)$.

For squeezed states with non-zero displacement $\alpha_0$, an
additional factor of a form identical with that for coherent
states, except for the different meaning of $X(\theta)$,
appears in the phase distribution $P\us(\theta)$. Since this
extra factor shows a peak at $\theta=0$, a competition arises
between the two-peak structure of the squeezed vacuum and the
one peak structure of the coherent component. This
competition leads to the bifurcation in the phase
distribution discussed
by~\citeA{SHV89a}~\citeY{SHV89,SHV89a}. Figure~5 illustrates
such a bifurcation for $\alpha_0=1$, as exhibited by the
Wigner and Husimi phase distributions plotted on the same
scale to visualize the differences. Qualitatively, the
pictures are quite similar, and differ only in the widths of
the peaks. The Pegg-Barnett distribution in this case is very
close to the Wigner phase distribution, and for this reason
we omit it here. To calculate the Pegg-Barnett phase
distribution one can apply formula \mref{N36} with $c_n$
given by (see~\citeAY{LK87}):
%%%%%%%%%%%%%%%%%%%%%%%%%%%%%%%%%%%%%%%%%%%%%%%%%%%%%%%%%%%%%%%%%%%%%%%%%%%%
\begin{figure} % Fig. 5
\vspace*{-3.5cm} \hspace*{-3mm}
\epsfbox{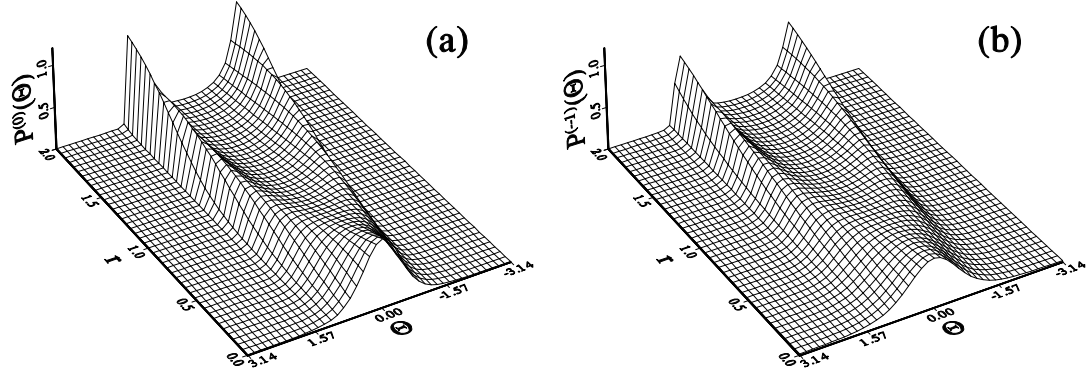} \vspace*{-4.5cm} \caption{ Pictures of
the phase bifurcation for the squeezed state with the mean
number of photons $|\alpha_0|^2=1$. The distributions are:
(a) $P^{(0)}(\theta)$, and (b) $P^{(-1)}(\theta)$. The
Pegg-Barnett distribution is very close to (a). }
\end{figure}
%---------------------------------------------------------------------------
\begin{eqnarray}
  c_n &=& \langle n|\alpha_0,\zeta\rangle \:=\: \frac{1}{\sqrt{n!\cosh
      r}} \left[\frac{1}{2}{\rm e}^{2\I\eta}\tanh r \right]^{n/2}
  H_n\left[ \frac{\alpha_0+\alpha_0^*{\rm e}^{2\I\eta}\tanh r}{\sqrt{2
        {\rm e}^{2\I\eta}\tanh r}}\right] \nonumber \\ &&\times\;
  \exp\left\{ -\frac{1}{2} [|\alpha_0|^2+\alpha_0^{*2}{\rm
      e}^{2\I\eta}\tanh r] \right\},
\label{N114}
\end{eqnarray}
assuming $\eta=0$ (results for $\eta=\pi/2$ can be obtained
on replacing $r$ by $-r$).

Approximate analytical formulas for the phase variance as
well as cosine and sine variances were obtained
by~\citeNP{VP89} for weakly squeezed vacuum. For large
squeezing the squeezed vacuum phase variance asymptotically
approaches $\pi^2/4$, which corresponds to the phase
distribution with two symmetrically-placed delta-functions
%----------------------------------------------------------------------
\begin{eqnarray}
\label{N114a} P(\theta) &=& {1\over 2}[\delta(\theta
-\pi/2)+\delta(\theta+\pi/2)].
\end{eqnarray}
Ideal squeezed vacuum is generated in the parametric
down-conversion process, in which the pump field is treated
as a constant classical quantity. Taking into account the
quantum character of the pump one finds that the signal field
is no longer the ideal squeezed vacuum and its phase
properties are different \cite{TG92a} [see \S 4.5].

%%%%%%%%%%%%%%%%%%%%%%%%%%%%%%%%%%%%%%%%%%%%%%%%%%%%%%%%%%%%%%%%%%%%%%%%
\subsection{Jaynes-Cummings model}
%%%%%%%%%%%%%%%%%%%%%%%%%%%%%%%%%%%%%%%%%%%%%%%%%%%%%%%%%%%%%%%%%%%%%%%%

%%%%%%%%%%%%%%%%%%%%%%%%%%%%%%%%%%%%%%%%%%%%%%%%%%%%%%%%%%%%%%%%%%%%%%%%%%%%
\begin{figure} % Fig. 6
\vspace*{-2.5cm} \hspace*{1cm} \epsfxsize=13cm
\epsfbox{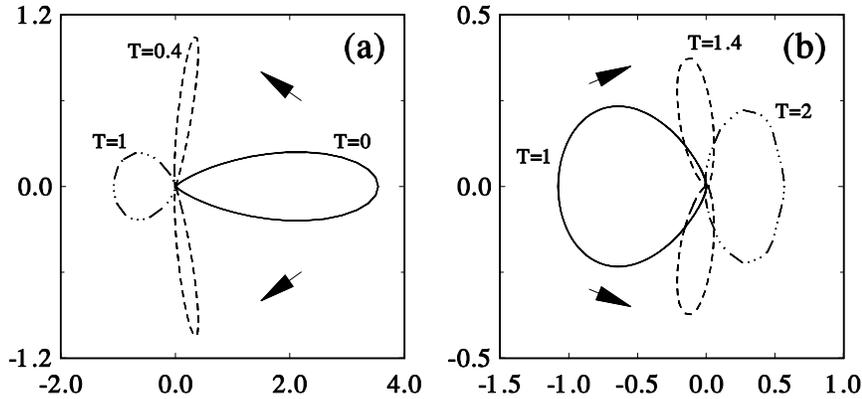} \vspace*{-1.5cm} \caption{ The
Pegg-Barnett phase distribution (3.44) of the Jaynes-Cummings
model as a function of scaled time $T=gt/(2\pi|\alpha_0|)$
for the initial mean photon number $|\alpha_0|^2=20$. }
\end{figure}

The Jaynes-Cummings model~\cite{JC63} (see reviews
by~\citeAY{YE85} and~\citeAY{SK93}) is the most popular model
used to describe the resonant interaction of single two-level
atom with single mode of the electromagnetic field. One of
the most remarkable effects predicted
theoretically~\cite{ENS80,NSE81} and then observed
experimentally~\cite{RWK87} in the Jaynes-Cummings model are
collapses and revivals of the atomic
inversion.~\citeAY{ER89}, using the $Q$-function, have shown
that the collapses and revivals can be understood in terms of
interferences in phase space.~\citeAY{PK90} mentioned the
splitting of the phase probability distribution into two
counter-rotating satellite distributions in a model
consisting of two degenerate atomic levels, coupled through a
virtual level by a Raman-type transition.~\citeAY{DTS90}
discussed the collapses and revivals in this model from the
point of view of the field-mode phase properties studied in
the framework of the Pegg-Barnett formalism.

The model is described by the Hamiltonian (at exact
resonance):
%----------------------------------------------------------------------
\begin{eqnarray}
\label{N117} \hat{H} &=& \hbar\omega (\hat{a}\plus\hat{a} +
\hat{R}^z) + \hbar g
(\hat{R}\plus\hat{a}+\hat{R}\hat{a}\plus),
\end{eqnarray}
where $\hat{a}\plus$ and $\hat{a}$ are the creation and
annihilation operators for the field mode; the two-level atom
is described by the raising, $\hat{R}\plus$, and lowering,
$\hat{R}$, operators and the inversion operator $\hat{R}^z$,
and $g$ is the coupling constant.

To study the phase properties of the field mode, we must know
the state evolution of the system. After dropping the free
evolution terms, which change the phase in a trivial way, and
assuming that the atom is initially in its ground state and
that the field is in a coherent state $|\alpha_0\rangle$, the
state of the system is found to be
%----------------------------------------------------------------------
\begin{eqnarray}
\label{N118} |\psi(t)\rangle &=& \sum_{n=0}^{\infty} b_n
\exp(\I n\vartheta_0) \left[ \cos (\sqrt{n}\,gt) |n,g\rangle
-\I
  \sin(\sqrt{n}\,gt)|n-1,e\rangle \right],
\end{eqnarray}
where $|g\rangle$ and $|e\rangle$ denote the ground and
excited states of the atom, the coefficients $b_n$ are given
by eq.~\mref{N81} and $\vartheta_0$ is the coherent state
phase.

According to the Pegg-Barnett formalism, one obtains the
phase distribution $\pbd$ in the form~\cite{DTS90}:
%----------------------------------------------------------------------
\begin{eqnarray}
\label{N119} \pbd &=& \frac{1}{2\pi} \left\{ 1+ 2\sum_{n>k}
b_n b_k
  \cos[(n-k)\theta]\cos[(\sqrt{n}-\sqrt{k})gt] \right\}.
\end{eqnarray}
This formula can be rewritten into the form:
%----------------------------------------------------------------------
\begin{eqnarray}
\label{N120} \pbd &=& \frac{1}{2} [P_+(\theta)+P_-(\theta)],
\end{eqnarray}
%%%%%%%%%%%%%%%%%%%%%%%%%%%%%%%%%%%%%%%%%%%%%%%%%%%%%%%%%%%%%%%%%%%%%%%%%%%%
\begin{figure} % Fig. 7
\vspace*{-0.5cm} \hspace*{1cm} \epsfxsize=12cm
\epsfbox{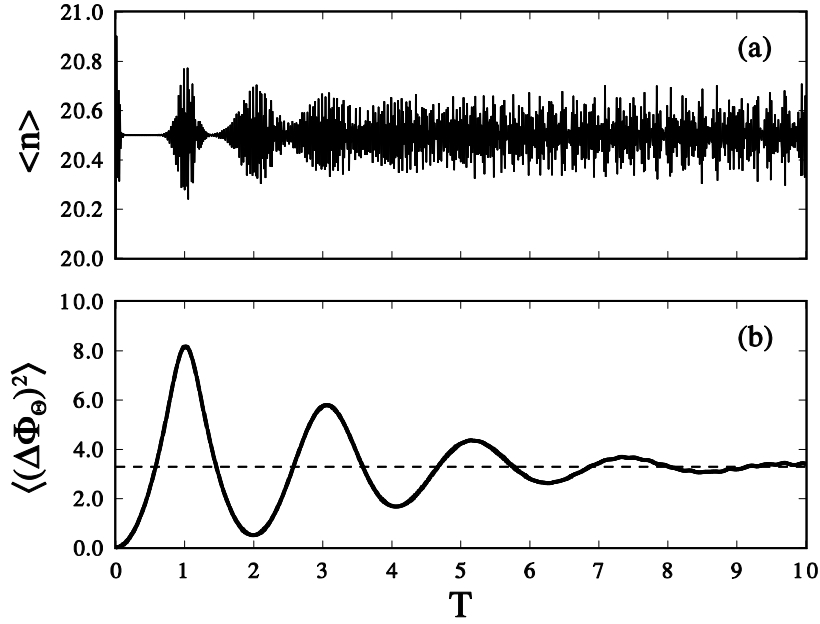} \vspace*{-3mm} \caption{ Evolution of (a)
the mean photon number $\langle\hat{n}\rangle$ and (b) the
variance $\langle(\Delta{\hat{\Phi}_{\theta}})^2\rangle$ of
the Pegg-Barnett phase operator for the Jaynes-Cummings model
as a function of scaled time $T=gt/(2\pi|\alpha_0|)$ for
$|\alpha_0|^2=20$. }
\end{figure}
where
%----------------------------------------------------------------------
\begin{eqnarray}
\label{N121} P_{\pm}(\theta) &=& \frac{1}{2\pi} \left\{ 1 + 2
\sum_{n>k} b_n b_k
  \cos\left[(n-k)\theta \mp (\sqrt{n}-\sqrt{k})gt\right] \right\},
\end{eqnarray}
which shows explicitly that as time elapses, the phase
distribution $\pbd$ splits into two distinct, right and left
rotating, distributions in the polar coordinate system. Polar
plots of the phase distribution are shown in fig.~6 (the time
$T=gt/(2\pi|\alpha_0|)$ is scaled in the revival times). So,
after a certain interval of time, the two counter-rotating
distributions ``collide'', and at that time the components of
the field oscillate in phase and one can expect the revival
of the atomic inversion.  The numerical calculations
corroborate this statement~\cite{DTS90}. This behavior of the
phase distribution resembles the behavior of the $Q$-function
studied by \citeNP{ER91}. The time behavior of the phase
variance together with the phase probability density
distribution carries certain information about the collapses
and revivals.  To show this, we first give the explicit
expression for the variance.  Using eqs.~\mref{N36}
and~\mref{N119}, one obtains
%----------------------------------------------------------------------
\begin{eqnarray}
\label{N122} \var{\phaseop} &=& \frac{\pi^2}{3} +4 \sum_{n>k}
\frac{(-1)^{n-k}}{(n-k)^2} b_n b_k
\cos[(\sqrt{n}-\sqrt{k})gt].
\end{eqnarray}
Variance~\mref{N122} is illustrated graphically in fig.~7 for
$|\alpha_{0}|^2=20$. The variance goes up initially and
reaches a maximum at the scaled time $T=1$, which is the
first revival time. The revival times correspond to the
extrema of the phase variance. In this way, the well known
phenomenon of collapses and revivals has obtained clear
interpretation in terms of the cavity-mode phase. More
details can be found in the paper by~\citeNP{DTS90}. The
dynamical properties of the field phase in the
Jaynes-Cummings model were studied by~\citeNP{DTS91a}, and
the effects of cavity damping by \citeNP{DS92}. Some
generalizations of this simple model were also considered
from the point of view of their phase properties
(\citeAY{DTS91b,MC91,MSW91,DHS92,MCZ92,PL92,PLZ92},
\citeA{WBSW92}~\citeY{WBSW92,WBSW93},
\citeAY{Fan93,JMK93,DGJ94,FW94,MGX94}).

%%%%%%%%%%%%%%%%%%%%%%%%%%%%%%%%%%%%%%%%%%%%%%%%%%%%%%%%%%%%%%%%%%%%%%%%
\subsection{Anharmonic oscillator model}
%%%%%%%%%%%%%%%%%%%%%%%%%%%%%%%%%%%%%%%%%%%%%%%%%%%%%%%%%%%%%%%%%%%%%%%%

The anharmonic oscillator model is described by the
Hamiltonian
%---------------------------------------------------------------------------
\begin{eqnarray}
  \hat{H} &=& \hbar\omega \hat{a}^{\dagger} \hat{a} +
  {1\over2}\hbar\kappa \hat{a}^{\dagger 2} \hat{a}^2,
\label{N123}
\end{eqnarray}
where $\hat{a}$ and $\hat{a}^{\dagger}$ are the annihilation
and creation operators of the field mode, and $\kappa$ is the
coupling constant, which is real and can be related to the
nonlinear susceptibility $\chi^{(3)}$ of the medium if the
anharmonic oscillator is used to describe propagation of
laser light (with right or left circular polarization) in a
nonlinear Kerr medium. If the state of the incoming beam is a
coherent state $|\alpha_0\rangle$, the resulting state of the
outgoing beam is given by
%---------------------------------------------------------------------------
\begin{eqnarray}
  |\psi(\tau)\rangle &=& \hat{U}(\tau)|\alpha_0\rangle \;=\;
  \exp(-|\alpha_0|^2/2)\sum_{n=0}^{\infty}
  {\alpha_{0}^{n}\over{\sqrt{n!}}}\exp{\left [ \I{\tau\over 2}n(n-1)
    \right]}|n\rangle,
\label{N124}
\end{eqnarray}
where $\tau=-\kappa t$. In the problem of light propagating
in a Kerr medium, one can make the replacement $t=-z/v$ to
describe the spatial evolution of the field instead of the
time evolution.  On introducing the notation
$\alpha_0=|\alpha_0|\exp(\I \vartheta_0)$ the
state~\mref{N124} can be written as
%---------------------------------------------------------------------------
\begin{eqnarray}
  |\psi(\tau)\rangle=\sum_{n=0}^{\infty}b_n\exp{\left\{\I{\left[n\vartheta_0
          +{\tau\over 2}n(n-1)\right]}\right\}}|n\rangle,
\label{N125}
\end{eqnarray}
where $b_n$ is given by eq.~\mref{N81}.

The appearance of the nonlinear phase factor in the
state~\mref{N125} modifies essentially the properties of the
field represented by such a state with respect to the initial
coherent state $|\alpha_0\rangle$. It was shown
by~\citeAY{Tan84} that a high degree of squeezing can be
obtained in the anharmonic oscillator model.  Squeezing in
the same process was later considered by~\citeAY{KY86}, who
used the name {\em
  crescent squeezing} because of the crescent shape of the
quasiprobability distribution contours obtained in the
process. The evolution of the quasiprobability distribution
$Q(\alpha,\alpha^*)$ in the anharmonic oscillator model was
considered by~\citeAY{Mil86},~\citeAY{MH86},~\citeA{PL88}
\citeY{PL88,PL90},~\citeAY{DM89}, and~\citeAY{PLK90}.  The
states that differ from coherent states by extra phase
factors, as in eq.~\mref{N125}, are the generalized coherent
states introduced by~\citeAY{TG66} and discussed
by~\citeAY{Sto71}.~\citeAY{Bia68} has shown that, under
appropriate periodic conditions, such states become discrete
superpositions of coherent states. \citeAY{YS86}, and
\citeAY{TM87} discussed the possibility of generating
quantum-mechanical superpositions of macroscopically
distinguishable states in the course of evolution for the
anharmonic oscillator.~\citeAY{MTK90} have shown that
superpositions with not only even but also odd numbers of
components can be obtained.

The Pegg-Barnett Hermitian phase formalism has been applied
to the study of the phase properties of the
states~\mref{N125} by~\citeAY{Ger90}, who discussed the
limiting cases of very low and very high light intensities,
and by~\citeAY{GT91f}, who gave a more systematic discussion
of the exact results. Phase fluctuations in the anharmonic
oscillator model were also analyzed within former phase
formalisms (\citeAY{Ger87},~\citeAY{Lyn87}).

The continuous Pegg-Barnett phase probability
distribution~\mref{N36} for the field state~\mref{N125} takes
the following form:
%---------------------------------------------------------------------------
\begin{eqnarray}
  \pbd &=& {1\over{2\pi}} {\left \{ 1+2 \sum_{n>k} b_n b_k \cos {\left
          [(n-k) \theta - {\tau\over 2}[n(n-1)-k(k-1)]\right ]} \right
    \}}.
\label{N126}
\end{eqnarray}
and the $s$-parametrized quasiprobability distribution
function is now given by ~\cite{Mir94}:
%----------------------------------------------------------------------
\begin{eqnarray}
\label{N127} {\cal W}^{(s)}(\alpha,\tau) &=& \frac{2}{1-s}
\exp\left\{
  -\frac{2}{1-s}(|\alpha_0|^2+|\alpha|^2) \right\} \mbox{\huge\{}
2\exp\left(\frac{1+s}{1-s}|\alpha_0|^2\right) \nonumber\\
&&\times\; \sum_{m>n}
\frac{|\alpha|^{m-n}|\alpha_0|^{n+m}}{m!} \left(\frac{2}{1-s}
\right)^{m-n} \left(\frac{s+1}{s-1} \right)^{n}
L_n^{m-n}\left(\frac{4|\alpha|^2}{1-s^2}\right) \nonumber\\
&&\times\; \cos\left\{
(m-n)(\vartheta_0-\theta)+\frac{\tau}{2}
  [\;m(m-1)-n(n-1)\;] \right\} \nonumber \\ &&+\; J_0(\I
\frac{4}{1-s}|\alpha|\;|\alpha_0|) \mbox{\huge\}},
\end{eqnarray}
where $J_0(x)$ is the Bessel function.  For $\tau=0$, $\qpd$,
given by eq.~\mref{N127}, is the coherent-state distribution
[eq.~\mref{N75}]. In the special case, for $Q$-function
($s=-1$), eq.~\mref{N127} reduces to
%----------------------------------------------------------------------
\begin{eqnarray}
  Q(\alpha,\tau) &=& \exp(-|\alpha|^2-|\alpha_0|^2) \left|
    \sum_{n=0}^{\infty} \frac{(\alpha^*\alpha_0)^n}{n!} \exp\left(\I
      \frac{\tau}{2}n(n-1)\right) \right|^2.
\label{N128}
\end{eqnarray}
The $s$-parametrized phase distribution, resulting from
eq.~\mref{N127} is
%----------------------------------------------------------------------
\begin{eqnarray}
\label{N129} \spd &=& \frac{1}{2\pi} \left\{ 1+2 \sum_{m>n}
b_m b_n\, G^{(s)}(m,n) \right.  \nonumber\\ &&\times\; \left.
\cos \left[
    (m-n)(\theta-\vartheta_0) -\frac{\tau}{2} \mbox{\Large (}
    m(m-1)-n(n-1) \mbox{\Large )} \right] \right\},
\end{eqnarray}
where the coefficients $G^{(s)}(m,n)$ are given by
eq.~\mref{N60}. Symmetrization of the phase window with
respect to the phase $\vartheta_0$ as done for the
Pegg-Barnett phase distribution [eq.\mref{N126}] is
equivalent to introduction of the relative phase variable
$\theta-\vartheta_0$, and the two formulas differ only by the
coefficients $G^{(s)}(m,n)$, as in eq.~\mref{N59}.  For
$\tau=0$, eqs.~\mref{N126} and~\mref{N129} describe the phase
probability distributions for the initial coherent state
$|\alpha_0\rangle$.  When the nonlinear evolution is on
($\tau\neq 0$), the distributions $\pbd$ and $\spd$ acquire
some new and very interesting features. A systematic
discussion of the properties as well as the plots of $\pbd$
and $P^{(-1)}(\theta)$ are given by~\cite{TGMK91} and
by~\cite{GT91f}.

The phase distribution $\pbd$ can be used to calculate the
mean and the variance of the phase operator, defined by
eqs.~\mref{N24} and~\mref{N25}. The results are~\cite{GT91f}:
%---------------------------------------------------------------------------
\begin{eqnarray}
  \lefteqn{\langle\psi (\tau)|\phaseop|\psi(\tau)\rangle=\vartheta_0 +
    \int\limits_{-\pi}^{\pi}\theta \pbd \,{\rm d}\theta} \nonumber\\
  &&=\vartheta_0 - 2\sum_{n>k}b_n b_k{{(-1)^{n-k}}\over{n-k}}\sin
  {\left\{{\tau\over2}[n(n-1)-k(k-1)]\right\}},
\label{N130}
\end{eqnarray}
%---------------------------------------------------------------------------
\begin{eqnarray}
  \lefteqn{\var{\phaseop}=\int\limits_{-\pi}^{\pi}\theta^2 \pbd \,{\rm
      d}\theta - {\left[\int\limits_{-\pi}^{\pi}\theta \pbd \,{\rm
          d}\theta\right]}^2} \nonumber\\ &=&{{\pi^2}\over
    3}+4\sum_{n>k}b_nb_k{{(-1)^{n-k}}\over{(n-k)^2}}\cos
  {\left\{{\tau\over2}[n(n-1)-k(k-1)]\right\}} \nonumber\\
  &&-\;{\left\{ 2\sum_{n>k}b_nb_k{{(-1)^{n-k}}\over{n-k}}\sin
      {\left\{{\tau\over2}[n(n-1)-k(k-1)]\right\}}\right\}}^2.
\label{N131}
\end{eqnarray}
For $\tau=0$, we recover the results for a coherent state
with the phase $\vartheta_0$ [eqs.~\mref{N31}
and~\mref{N39}]. The nonlinear evolution of the system leads
to a nonlinear shift of the mean phase and essentially
modifies the variance. An example is illustrated in fig.~8,
where the evolution of the mean phase (fig.~8a) and its
variance (fig.~8b) are plotted against $\tau$ for various
values of $|\alpha_{0}|^2$.  We have assumed $\vartheta_0=0$,
and the window of the phase values is taken between $-\pi$
and $\pi$. The evolution is periodic with the period $2\pi$,
so the initial values are restored for $\tau=2\pi$. Figure~8a
shows the intensity-dependent phase shift. The amplitude of
the mean phase oscillation becomes larger with increasing
mean number of photons. The line $\pi^2/3$ in fig.~8b marks
the variance for the state with random distribution of phase.
It is seen clearly from fig.~8b that the stronger the initial
field, the higher the phase variance.  For $|\alpha_{0}|^2=4$
the phase variance increases rapidly and most of the period
oscillates around $\pi^2/3$ -- the value for uniform phase
distribution. This means that the phase is randomized during
the evolution, although it periodically reproduces its
initial values. This tendency is even more pronounced when
the mean number of photons increases. The periodicity of the
evolution is destroyed by damping \cite{GT91b}. The sine and
cosine functions of the phase were also calculated and the
results compared with other approaches~\cite{GT91f}.
%%%%%%%%%%%%%%%%%%%%%%%%%%%%%%%%%%%%%%%%%%%%%%%%%%%%%%%%%%%%%%%%%%%%%%%%%%%%
\begin{figure} % Fig. 8
\vspace*{-12.5cm} \hspace*{0cm}
\epsfbox{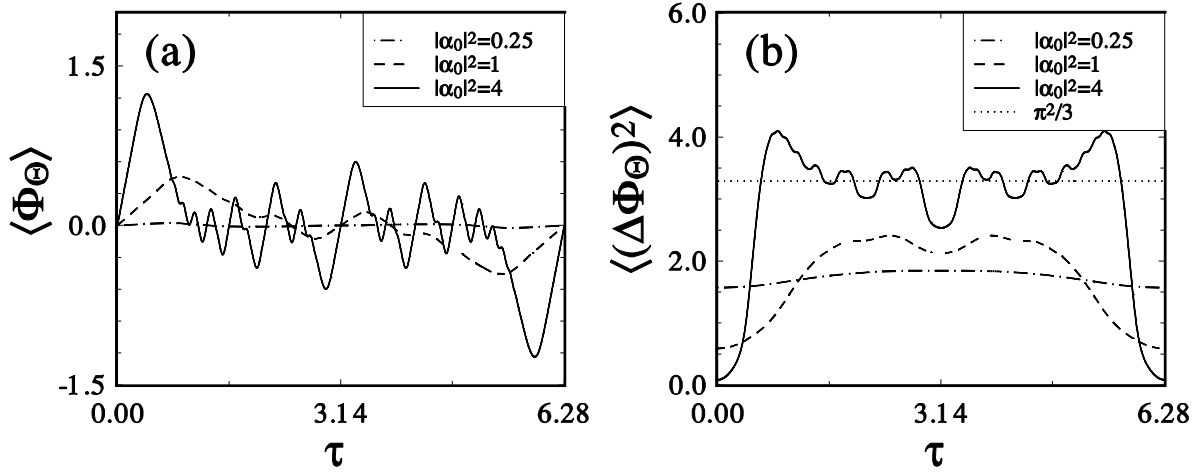} \vspace*{-4.7cm} \caption{ Evolution of
(a) the mean phase (3.55) and (b) the phase variance (3.56)
for the anharmonic oscillator model. }
\end{figure}

The local minima in $\var{\phaseop}$ apparent in fig.~8
indicate the points in the evolution in which superpositions
of coherent states are formed, and the phase variance
decreases at these points.  This occurs for $\tau=2\pi M/N$
($N$=2, 3, 4,\dots, and $M$, $N$ are mutually prime numbers),
for which the $\pbd$ and $\spd$ plotted in polar coordinates
show $N$-fold symmetry, confirming the generation of discrete
superpositions of coherent states with 2, 3, 4, \ldots
components:
%----------------------------------------------------------------------
\begin{eqnarray}
  \left|\psi\left(\tau=2\pi \frac{M}{N}\right)\right\rangle &=&
  \sum_{k=1}^{2N} c_k |\exp(\I \phi_k)\alpha_0\rangle,
\label{N132}
\end{eqnarray}
where the phases $\phi_k$ are simply
%----------------------------------------------------------------------
\begin{eqnarray}
  \phi_k &=& \frac{\pi}{N}k, \hspace{2cm} k=1,\ldots,2N,
\label{N133}
\end{eqnarray}
and the superposition coefficients $c_k$, representing the
so-called fractional revivals, are given by
(\citeAY{AP89},~\citeAY{TGMK91}):
%----------------------------------------------------------------------
\begin{eqnarray}
\label{N134} c_k &=& \frac{1}{2N}\sum_{n=1}^{2N} \exp\left(
-\I
  \frac{\pi}{N}[nk-Mn(n-1)] \right).
\end{eqnarray}
Such superpositions, created during the evolution of the
anharmonic oscillator model, have very specific phase
properties discussed in \S 3.2. Plots the phase
distributions~\mref{N83} and~\mref{N84} (where $N$ should be
replaced by $2N$) for the superpositions~\mref{N133} with
several components are presented in fig.~4.  The phase
distribution indicates the superpositions in a very
spectacular way, as shown by \citeNP{TGMK91}, \citeNP{GT91f}
and~\citeAY{San92} for the anharmonic oscillator model, and
by \citeNP{PT92} for the model with higher nonlinearities.

%%%%%%%%%%%%%%%%%%%%%%%%%%%%%%%%%%%%%%%%%%%%%%%%%%%%%%%%%%%%%%%%%%%%%%%%
\subsection{Displaced number states}
%%%%%%%%%%%%%%%%%%%%%%%%%%%%%%%%%%%%%%%%%%%%%%%%%%%%%%%%%%%%%%%%%%%%%%%%

Other states which are interesting from the point of view of
their phase properties are the displaced number states
$|\alpha_0,n_0\rangle$ generated by the action of the
displacement operator $\hat{D}(\alpha_0)$ on a Fock state
$|n_0\rangle$ (see~\citeNP{DKKB90});
%----------------------------------------------------------------------
\begin{eqnarray}
  |\alpha_0,n_0\rangle &=& \hat{D} (\alpha_0) |n_0\rangle.
\label{N135}
\end{eqnarray}
In a special case, when $n_0=0$, the states~\mref{N135}
become a coherent state $|\alpha_0\rangle$. The
$s$-parametrized quasiprobability distribution for the
state~\mref{N135} is
%---------------------------------------------------------------------------
\begin{eqnarray}
  \qpd &=& \frac{1}{\pi} \frac{2}{1-s} (-1)^n
  \left(\frac{1+s}{1-s}\right)^{n} \nonumber \\ &&\times\; \exp\left\{
    -\frac{2}{1-s} |\alpha-\alpha_0|^2 \right\}
  L_n\left(\frac{4|\alpha-\alpha_0|^2}{1-s^2}\right),
\label{N136}
\end{eqnarray}
whereas the $s$-parametrized phase distribution
becomes~\cite{TMG93}:
%---------------------------------------------------------------------------
\begin{eqnarray}
  \spd &=& \left( \frac{2}{1-s}\right) ^{n} \sum_{k=0}^{n}
  \frac{(-1)^{n-k}}{k!} \left( \frac{1+s}{2}\right)^{n-k} \left( n
    \atop k \right) \nonumber \\ &&\times\; \sum_{l=0}^{k} \left( k
    \atop l \right) \frac{N_{k-l}(2k-2l)!}{2^{2k-2l}(k-l)!}
  (X_0^2-X^2)^l P_{k-l}(X),
\label{N137}
\end{eqnarray}
here
%---------------------------------------------------------------------------
\begin{eqnarray}
  P_n(X) &=& \frac{N_n^{-1}}{2\pi} \exp[-(X_0^2-X^2)]
  \left\{\exp(-X^2)\, Q_n(X) +\sqrt{\pi}X\,(1+{\rm erf}(X))\right\},
\label{N138}
\end{eqnarray}
%---------------------------------------------------------------------------
\begin{eqnarray}
  Q_n(X) &=& \frac{2^{2n}(n!)^2}{(2n)!} \sum_{k=0}^{n}
  \frac{1}{k!}X^{2k} -\sum_{k=1}^{n} \frac{2^{2k}k!}{(2k)!}X^{2k},
\label{N139}
\end{eqnarray}
%%%%%%%%%%%%%%%%%%%%%%%%%%%%%%%%%%%%%%%%%%%%%%%%%%%%%%%%%%%%%%%%%%%%%%%%%%%%
\begin{figure} % Fig. 9
\vspace*{-1.8cm} \hspace*{1.5cm} \epsfxsize=12cm
\epsfbox{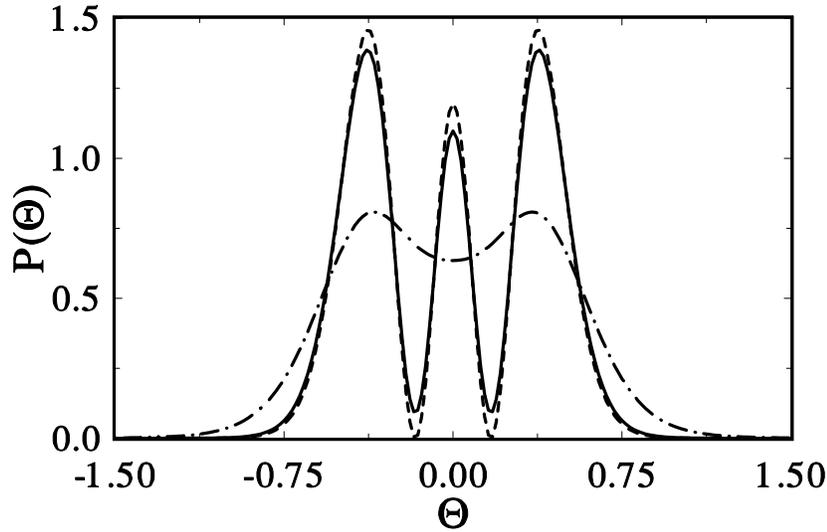} \vspace*{-2mm} \caption{ Phase
distributions for the displaced number state with $n=2$ and
$\alpha_0=3$. Meaning of the lines is the same as in fig. 3.
}
\end{figure}
and the normalization constant is equal to
%---------------------------------------------------------------------------
\begin{eqnarray}
  N_m &=& 1+ \exp(-X_0^2)\, \frac{1}{2\pi}
  \int\limits_{-\pi}^{\pi}[Q_n(X)-1]\,{\rm d}\theta \nonumber \\ &=&
  1+\exp(-X_0^2)\left\{ -1+\frac{2^{2n}(n!)^2}{(2n)!}
    \sum_{k=0}^{n}\frac{(2k)!}{2^{2k}(k!)^3} X_0^{2k}
    -\sum_{k=1}^{n}\frac{1}{k!} X_0^{2k} \right\}.
\label{N140}
\end{eqnarray}
The $X$ variable in this case is
%---------------------------------------------------------------------------
\begin{eqnarray}
  X \;=\; X\us(\theta)\;=\; \sqrt{\frac{2}{1-s}} \alpha_0 \cos\theta,
\label{N141}
\end{eqnarray}
and we have assumed $\alpha_0$ as real. Despite its more
complex structure, formula~\mref{N137} contains phase
distributions $P_n(X)$ that exhibit the main features of the
previous phase distributions $\spd$, i.e., eq.~\mref{N76} for
a coherent state and eq.~\mref{N110} for a squeezed state.

Displaced number states have the following Fock expansion:
%----------------------------------------------------------------------
\begin{eqnarray}
  |\alpha_0,n_0\rangle &=& \sum_{n} b_n {\rm e}^{\I\phi_n} |n\rangle,
\label{N142}
\end{eqnarray}
where the amplitudes $b_n$ and phase-factors $\phi_n$ are
%----------------------------------------------------------------------
\begin{eqnarray}
\label{N143} b_n &=& \langle n|\hat{D}(|\alpha_0|) |
n_0\rangle \nonumber\\ &=&
\exp\left(-\frac{1}{2}|\alpha_0|^2\right)
\left(\frac{n_-!}{n_+!} \right)^{1/2} \,(-1)^{n_+-n}
|\alpha_0|^{n_+ -n_-} L_{n_-}^{n_+-n_-} (|\alpha_0|^2),
\end{eqnarray}
%----------------------------------------------------------------------
\begin{eqnarray}
\begin{array}{rll}
  n_- &=& \min\{n,n_0\}, \\ n_+ &=& n+n_0-n_-=\max\{n,n_0\},
\end{array}
\label{N144}
\end{eqnarray}
%----------------------------------------------------------------------
\begin{eqnarray}
  \phi_n &=\: (n-n_0)\: {\rm Arg} \alpha_0 &=\: (n-n_0) \vartheta_0,
\label{N145}
\end{eqnarray}
which on insertion into eq.~\mref{N36} give explicitly the
Pegg-Barnett distribution $\pbd$.

Both for coherent states and squeezed states, there was no
qualitative difference between various phase distributions.
Thus, one could say that at least qualitatively, all the
phase distributions carried the same phase information. Here,
we give an example of states for which the above statement is
no longer true. These are displaced number states.  The phase
properties of such states were discussed earlier by
\citeAY{TMGC92}. It was shown that there is a qualitative
difference between the Husimi phase distribution on one side,
and the Pegg-Barnett and Wigner phase distributions on the
other. There is an essential loss of information in the case
of the Husimi phase distribution. The differences can be
interpreted easily when the concept of the area of overlap in
phase space introduced by~\citeAY{SW87} is invoked.
Formula~\mref{N137} provides the possibility of deeper
insight into the structure of the $s$-parametrized phase
distributions. The phase distribution $P\us(\theta)$ is a
result of competition between the functions $P_n(X)$, which
are peaked at $\theta=0$, and the functions $(X_0^2-X^2)^l$,
which have peaks at $\theta=\pm\pi/2$. For $s=-1$ only the
term with $n-k=0$ survives, and there is no modulation due to
the $(-1)^{n-k}$ factor. This is the reason why the Husimi
phase distribution can have at most two peaks, no matter how
large is $n$. Both for the Pegg-Barnett phase distribution
and $P^{(0)}(\theta)$ there are $n+1$ peaks. It is also worth
noting that despite the fact that the Wigner function ${\cal
W}^{(0)}$, [eq.~\mref{N136}] oscillates between positive and
negative values, the Wigner phase distribution
$P^{(0)}(\theta)$ [eq.~\mref{N137}] is positive definite. An
illustration of the differences between the phase
distributions for the displaced number states with $n=2$ and
$|\alpha_0|^2=9$ is shown in fig.~9. It is seen that the
Pegg-Barnett phase distribution is very close to
$P^{(0)}(\theta)$, and that they carry basically the same
phase information, while there is an essential loss of phase
information carried by $P^{(-1)}(\theta)$. The Pegg-Barnett
and $P^{(0)}(\theta)$ distributions are very similar for
given $n$, while $P^{(-1)}(\theta)$ has at most two peaks
that become broader as $n$ increases. This example shows the
difference between a ``pure'' canonical phase distribution
such as the Pegg-Barnett distribution, which could be
associated with a ``pure'' phase measurement, and a
``measured'' phase distribution such as $P^{(-1)}(\theta)$,
which can be associated with the noisy measurement of the
phase. The noise introduced by the measurement process
reduces the phase information that can be inferred from the
measured data.

%%%%%%%%%%%%%%%%%%%%%%%%%%%%%%%%%%%%%%%%%%%%%%%%%%%%%%%%%%%%%%%%%%%%%%%
%         Phase properties of two-mode optical fields                 %
%%%%%%%%%%%%%%%%%%%%%%%%%%%%%%%%%%%%%%%%%%%%%%%%%%%%%%%%%%%%%%%%%%%%%%%
\section{Phase properties of two-mode optical fields}
\setcounter{equation}{0}

The single-mode version of the Pegg-Barnett phase formalism
can be extended easily into the two-mode fields \cite{BP90}
that are often a subject of consideration in quantum optics.
This leads to the joint phase probability distribution for
the phases of the two modes, and allows the study of not only
the individual mode phase characteristics discussed above but
also essentially two-mode phase characteristics such as
correlation between the phases of the two modes.  The phase
properties of a two-mode field are simply constructed from
the single-mode phases (see \S 2.1). The two-mode joint phase
distribution is given by
%----------------------------------------------------------------------
\begin{eqnarray}
\label{N146} P(\theta_{a},\theta_{b}) \:=\: \lim_{\sigma
\rightarrow \infty} \left(\frac{\sigma+1}{2\pi}\right)^2
|\langle\theta_{m_a},\theta_{m_b}|f\rangle|^{2}.
\end{eqnarray}
This phase distribution can be applied, similar to the
single-mode case, to calculations of the mean values of the
phase-dependent quantities, such as individual phases, their
variances, etc.  We are often interested not in the
individual phases corresponding to either mode, but rather in
the operators or distributions representing the sum and
difference of the single-mode phases, which can also be
calculated using the joint phase distribution
[eq.~\mref{N146}]. However, the phase sum and difference
values will cover the $4\pi$ range, and the integrations over
the phase sum and difference variable should be performed
over the whole range. This approach, although fully
justified, is not compatible with the idea that the
individual phase should be $2\pi$-periodic, and there should
be a way to cast the phase sum and difference into the $2\pi$
range. Such a casting procedure was proposed by
\citeNP{BP90}. The two approaches, however, give different
values for the phase sum and difference variances, for
example, and one should be aware of the differences.
Sometimes the original calculations based on the joint phase
distribution~\mref{N146} have a more transparent
interpretation, especially when one considers the intermode
phase correlations. We shall adduce here examples of both
approaches (the quantities obtained with the use of the
casting procedure will be distinguished by the subscript
$2\pi$). The casting procedure is described briefly below.

The possible eigenvalues of the phase-sum operator are
%----------------------------------------------------------------------
\begin{eqnarray}
\label{N147} \theta_{m_+} \:=\: \theta_{0a} + \theta_{0b} +
\frac{2\pi}{\sigma+1} m_{+},
\end{eqnarray}
where $m_+=0,1,\ldots,2\sigma$, and the eigenvalues of the
phase-difference operator are
%----------------------------------------------------------------------
\begin{eqnarray}
\label{N148} \theta_{m_-} \:=\: \theta_{0a} - \theta_{0b} +
\frac{2\pi}{\sigma+1} m_{-},
\end{eqnarray}
where $m_-=-\sigma,-\sigma+1,\ldots,\sigma$. It is seen that
the eigenvalue spectra~\mref{N147}--\mref{N148} of the phase
sum and difference operators have widths of $4\pi$. Since
phases differing by $2\pi$ are physically indistinguishable,
the phase sum and difference operators and distributions
should be cast into a $2\pi$-range~\cite{BP90}.  The casting
procedure can be applied to the joint continuous-phase
distribution $P_{4\pi}(\theta_{+},\theta_{-})$, defined as
%----------------------------------------------------------------------
\begin{eqnarray}
\label{N149} P_{4\pi}(\theta_{+},\theta_{-}) &=&
\frac{1}{2}\lim_{\sigma
  \rightarrow \infty} \left(\frac{\sigma+1}{2\pi}\right)^2
|\langle\theta_{m_+},\theta_{m_-}|f\rangle|^{2},
\end{eqnarray}
where
%----------------------------------------------------------------------
\begin{eqnarray}
\label{N150} \theta_{\pm} &=& \theta_{a} \pm \theta_{b}.
\end{eqnarray}
As was stressed by~\citeAY{BP90}, there are many ways to
apply the casting procedure. However, if the distribution
$P_{4\pi}(\theta_{+},\theta_{-})$ is sharply peaked, we must
avoid splitting the original single peak into two parts, one
at each end of the $2\pi$-interval. Such a poor choice of the
$2\pi$-range leads to the same interpretation problems
encountered for a poor choice of $\theta_0$ in the
single-mode case~\cite{BP89}. The casting procedure can be
applied as follows:
%----------------------------------------------------------------------
\begin{eqnarray}
\label{N151} P_{2\pi}(\theta_+,\theta_-) &=&
P_{4\pi}(\theta_+,\theta_-)+P_{4\pi}(\theta_++\delta_1,\theta_-+\delta_2),
\end{eqnarray}
where the shifts $\delta_1$ and $\delta_2$ are dependent on
the values of $\theta_-$ and $\theta_+$:
%----------------------------------------------------------------------
\begin{eqnarray}
\label{N152}
\begin{array}{lll}
  {\rm I.} & \delta_1 =2\pi,\; \delta_2=0 & \\ \;\;\;{\rm for} &
  \theta_+\in\langle\theta_{0_+}+\pi,\theta_{0_+}+3\pi\rangle, &
  \theta_-\in\langle |\theta_+-\theta_{0_+}-2\pi|+\theta_{0_-}-2\pi,
  \theta_{0_-}-\pi\rangle \\ {\rm II.} & \delta_1 =0,\; \delta_2=2\pi
  & \\ \;\;\;{\rm for} &
  \theta_-\in\langle\theta_{0_-}-\pi,\theta_{0_-}+\pi\rangle, &
  \theta_+\in\langle |\theta_--\theta_{0_-}|+\theta_{0_+},
  \theta_{0_+}+\pi\rangle \\ {\rm III.} & \delta_1 =-2\pi,\;
  \delta_2=0 & \\ \;\;\;{\rm for} &
  \theta_+\in\langle\theta_{0_+}+\pi,\theta_{0_+}+3\pi\rangle, &
  \theta_-\in\langle\theta_{0_-}+\pi, 2\pi-
  |\theta_+-\theta_{0_+}-2\pi|+\theta_{0_-}\rangle \\ {\rm IV.} &
  \delta_1 =0,\; \delta_2=-2\pi & \\ \;\;\;{\rm for} &
  \theta_-\in\langle\theta_{0_-}-\pi,\theta_{0_-}+\pi\rangle, &
  \theta_+\in\langle\theta_{0_+}+3\pi,
  4\pi-|\theta_--\theta_{0_-}|+\theta_{0_+}\rangle.
\end{array}
\end{eqnarray}
This analysis of four regions in the
($\theta_+,\theta_-$)-plane to be cut and shifted is close to
the original idea of~\citeAY{BP90}, and can be easily
understood in a geometrical representation of the variable
transformation. Moreover, as a further consequence of the
$2\pi$-periodicity of eq.~\mref{N151}, one can keep the same
shifts $\delta_1$ and $\delta_2$ in the whole
($\theta_+,\theta_-$)-plane without distinguishing any
regions. Let us only mention some of the possible simplified
castings:
%----------------------------------------------------------------------
\begin{eqnarray}
\label{N153} P_{2\pi}(\theta_+,\theta_-) &=&
P_{4\pi}(\theta_+,\theta_-)+P_{4\pi}(\theta_+,\theta_-\pm
2\pi) \nonumber\\ &=&
P_{4\pi}(\theta_+,\theta_-)+P_{4\pi}(\theta_+\pm
2\pi,\theta_-) \nonumber\\ &=&
P_{4\pi}(\theta_+,\theta_-)+P_{4\pi}(\theta_+\pm\pi,\theta_-\pm\pi) \nonumber\\
&=&\ldots,
\end{eqnarray}
and combinations of the shifts satisfying the condition
$|\delta_1|+|\delta_2|=2\pi$ or
$||\delta_1|-|\delta_2||=2\pi$.  The resulting joint
distribution $P_{2\pi}(\theta_+,\theta_-)$ is $2\pi$-periodic
in $\theta_+$ and $\theta_-$. Alternatively, one can apply
the casting procedure to phase distribution~\mref{N146}:
%----------------------------------------------------------------------
\begin{eqnarray}
\label{N154} P_{2\pi}(\theta_+,\theta_-) \:=\: \frac{1}{2}
\left[
  P_{4\pi}\left(\frac{\theta_++\theta_-}{2},\frac{\theta_+-\theta_-}{2}\right)
  +P_{4\pi}\left(\frac{\theta_++\theta_-}{2}+\delta_1,\frac{\theta_+-\theta_-}{2}
    +\delta_2\right) \right],
\end{eqnarray}
The factor $1/2$ occurring in eqs.~\mref{N149}
and~\mref{N154} comes from the Jacobian of the
transformation~\mref{N150} for the variables. The marginal
mod($2\pi$) phase-sum, $P_{2\pi}(\theta_+)$, and phase-
difference, $P_{2\pi}(\theta_-)$, distributions are given by
%----------------------------------------------------------------------
\begin{eqnarray}
\label{N155} P_{2\pi}(\theta_{\mp}) &=&
\int\limits_{\theta'_{0\pm}}^{\theta'_{0\pm}+2\pi}
P_{2\pi}(\theta_{+},\theta_-) {\rm d} \theta_{\pm},
\end{eqnarray}
where
%----------------------------------------------------------------------
\begin{eqnarray}
\label{N156} \theta'_{0\pm} &=& \theta_{0\pm} \pm \pi.
\end{eqnarray}
In the above approach, the casting was prior to the
integration.  There is another equivalent manner of obtaining
mod($2\pi$) marginal phase sum and difference distributions
in which the casting is applied after integration. In this
approach~\cite{BP90}, one starts from eq.~\mref{N149} to
calculate the $4\pi$-periodic marginal distributions
$P_{4\pi}(\theta_{\pm})$:
%----------------------------------------------------------------------
\begin{eqnarray}
\label{N157} P_{4\pi}(\theta_+) &=&
\int\limits_{|\theta_+-\theta_{0_+}-2\pi|+\theta_{0_-}-2\pi}^{2\pi-
  |\theta_+-\theta_{0_+}-2\pi|+\theta_{0_-}}
P_{4\pi}(\theta_+,\theta_-) {\rm d} \theta_-,
\end{eqnarray}
%----------------------------------------------------------------------
\begin{eqnarray}
\label{N158} P_{4\pi}(\theta_-) &=&
\int\limits_{|\theta_--\theta_{0_-}|+\theta_{0_+}}^{4\pi-
  |\theta_--\theta_{0_-}|+\theta_{0_+}} P_{4\pi}(\theta_+,\theta_-)
{\rm d} \theta_+.
\end{eqnarray}
Contrary to the former approach, the casting procedure is now
applied to the single-mode distributions
$P_{4\pi}(\theta_{\pm})$~\cite{BP90}:
%----------------------------------------------------------------------
\begin{eqnarray}
\label{N159} P_{2\pi}(\theta_+) &=& \left\{
\begin{array}{ll}
  P_{4\pi}(\theta_+) + P_{4\pi}(\theta_+ + 2\pi)
  & {\rm if}\;\;\; \theta_{0_+}+\pi
  \leq \theta_+ \leq \theta_{0_+} +2\pi \\
  P_{4\pi}(\theta_+) + P_{4\pi}(\theta_+ -
  2\pi) & {\rm if}\;\;\; \theta_{0_+}+2\pi \leq \theta_+ \leq
  \theta_{0_+} +3\pi
\end{array}
\right.
\end{eqnarray}
and
%----------------------------------------------------------------------
\begin{eqnarray}
\label{N160} P_{2\pi}(\theta_-) &=& \left\{
\begin{array}{ll}
  P_{4\pi}(\theta_-) + P_{4\pi}(\theta_- + 2\pi) & {\rm if}\;\;\;
  \theta_{0_-}-\pi \leq \theta_- \leq \theta_{0_-} \\
  P_{4\pi}(\theta_-) + P_{4\pi}(\theta_- - 2\pi) & {\rm if}\;\;\;
  \theta_{0_-} \leq \theta_- \leq \theta_{0_-} +\pi\,.
\end{array}
\right.
\end{eqnarray}
Again, due to the $2\pi$-periodicity of
$P_{2\pi}(\theta_{\pm})$ in $\theta_{\pm}$, one can simplify
the recipes~\mref{N159} and \mref{N160} to one of the forms:
%----------------------------------------------------------------------
\begin{eqnarray}
\label{N161} P_{2\pi}(\theta_{\pm}) &=&
P_{4\pi}(\theta_{\pm}) + P_{4\pi}(\theta_{\pm} + 2\pi)
\nonumber\\ &=& P_{4\pi}(\theta_{\pm}) +
P_{4\pi}(\theta_{\pm} - 2\pi)
\end{eqnarray}
in the whole intervals $\theta_{0\pm} \leq \theta_{\pm} <
\theta_{0\pm} +2\pi$.

One can analyze analogously the two-mode $s$-parametrized
phase distributions. Here we give only one expression for the
mod($2\pi$) $s$-parametrized phase-difference distribution
for arbitrary density matrix $\hat{\rho}$ and any $s$:
%----------------------------------------------------------------------
\begin{eqnarray}
\label{N162} P^{(s)}_{2\pi}(\theta_{-}) &=& \frac{1}{2\pi}
\sum_{n=0}^{\infty}\sum_{k=0}^{n}\sum_{l=0}^{n} G^{(s)}(k,l)
G^{(s)}(n-k,n-l) \nonumber\\ &&\times\; \exp[\I
(k-l)\theta_{-}] \langle l,n-l|\hat{\rho}|k,n-k\rangle
\end{eqnarray}
with the coefficients $G^{(s)}(k,l)$ given by eq.~\mref{N60}.
Also, by putting $G^{(s)}(k,l)\rightarrow 1$, the mod($2\pi$)
Pegg-Barnett phase-difference distribution is obtained as
derived by~\citeAY{LST95}.

%%%%%%%%%%%%%%%%%%%%%%%%%%%%%%%%%%%%%%%%%%%%%%%%%%%%%%%%%%%%%%%%%%%%%%%%
\subsection{Two-mode squeezed vacuum}
%%%%%%%%%%%%%%%%%%%%%%%%%%%%%%%%%%%%%%%%%%%%%%%%%%%%%%%%%%%%%%%%%%%%%%%%

Single-mode squeezed states, discussed in \S 3.3, differ
essentially from the two-mode squeezed states discussed
extensively by~\citeAY{CS85} and~\citeAY{SC85}.  The
Pegg-Barnett phase formalism was applied by~\citeAY{BP90},
and by \citeAY{GT91g} to study the phase properties of the
two-mode squeezed vacuum, and some of the results are adduced
here.

The two-mode squeezed vacuum state is defined by applying the
two-mode squeeze operator $\hat{S}(r,\varphi)$ on the
two-mode vacuum, and is given by~\cite{SC85}:
%---------------------------------------------------------------------------
\begin{eqnarray}
  |0,0\rangle_{(r,\varphi)}&=&\hat{S}(r,\varphi)|0,0\rangle
  \nonumber\\ &=&(\cosh r)^{-1}\exp\left(\mbox{e}^{2\I\varphi} \tanh
    r\:\hat{a}_{1}^{\dagger} \hat{a}_{2}^{\dagger} \right)|0,0\rangle
  \nonumber\\ &=&(\cosh
  r)^{-1}\sum_{n=0}^{\infty}\left(\mbox{e}^{2\I\varphi} \tanh
    r\right)^n|n,n\rangle,
\label{N163}
\end{eqnarray}
where $\hat{a}_{1}^{\dagger} $ and $\hat{a}_{2}^{\dagger} $
are the creation operators for the two modes, $r$ ($0\leq
r<\infty$) is the strength of squeezing, and $\varphi$
($-\pi/2\leq\varphi<\pi/2$) is the phase (note the shift in
phase by $\pi/2$ with respect to the usual choice of
$\varphi$).

The state~\mref{N163}, when the procedure described earlier
is applied to it, leads to the joint probability distribution
for the phases $\theta_{1}$ and $\theta_{2}$ of the two modes
in the form~\cite{BP90}:
%---------------------------------------------------------------------------
\begin{eqnarray}
  P(\theta_{1},\theta_{2})&=&(4\pi^2\cosh^2 r)^{-1}
  \left(1+\tanh^2r-2\tanh r\cos(\theta_{1}+\theta_{2})\right)^{-1}.
\label{N164}
\end{eqnarray}
One important property of the two-mode squeezed vacuum, which
is apparent from eq.~\mref{N164}, is that
$P(\theta_{1},\theta_{2})$ depends on the sum of the two
phases only. Integrating $P(\theta_{1},\theta_{2})$ over one
of the phases gives the marginal phase distribution
$P(\theta_{1})$ or $P(\theta_{2})$ for the phase $\theta_{1}$
or $\theta_{2}$:
%---------------------------------------------------------------------------
\begin{eqnarray}
  P(\theta_{1}) &=& \int\limits_{-\pi}^{\pi}
  P(\theta_{1},\theta_{2})\,{\rm d}\theta_{2}\;=\;
  P(\theta_{2})\;=\;\frac{1}{2\pi},
\label{N165}
\end{eqnarray}
meaning that the phases $\theta_{1}$ and $\theta_{2}$ of the
individual modes are distributed uniformly. This gives
%---------------------------------------------------------------------------
\begin{eqnarray}
  \langle\hat{\Phi}_{\theta_{1}}\rangle &=&
  \varphi+\int\limits_{-\pi}^{\pi} \theta_{1}P(\theta_{1})\,{\rm
    d}\theta_{1} \;=\; \langle\hat{\Phi}_{\theta_{2}}\rangle \;=\;
  \varphi,
\label{N166}
\end{eqnarray}
and
%---------------------------------------------------------------------------
\begin{eqnarray}
  \langle\hat{\Phi}_{\theta_{1}}+\hat{\Phi}_{\theta_{2}}\rangle=2\varphi,
  \hspace{1cm}
  \langle\hat{\Phi}_{\theta_{1}}-\hat{\Phi}_{\theta_{2}}\rangle=0.
\label{N167}
\end{eqnarray}
Thus, the phase-sum operator is related to the phase
2$\varphi$ defining the two-mode squeezed vacuum
state~\mref{N163}.

%%%%%%%%%%%%%%%%%%%%%%%%%%%%%%%%%%%%%%%%%%%%%%%%%%%%%%%%%%%%%%%%%%%%%%%%%%%
\begin{figure} % Fig. 10
\vspace*{-8.3cm} \hspace*{2cm} \epsfxsize=12cm
\epsfbox{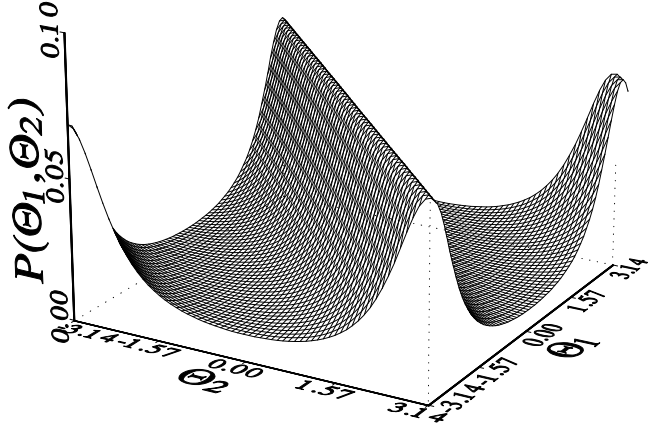} \vspace*{-.5cm} \caption{ The joint
probability distribution $P(\theta_1,\theta_2)$, eq. (4.19),
for the two-mode squeezed vacuum with $r=0.5$. }
\end{figure}
The two-mode squeezed vacuum has very specific phase
properties: the individual phases as well as the phase
difference are random, and the only non-random phase is the
phase sum.

Figure~10 shows an example of the joint phase probability
distribution $P(\theta_{1},\theta_{2})$. The ridge, which is
parallel to the diagonal of the phase window square reflects
the dependence of $P(\theta_{1},\theta_{2})$ on
$\theta_{1}+\theta_{2}$ only.

The phase distribution $P(\theta_1,\theta_2)$,
[eq.~\mref{N164}] is an explicit function of the phase sum,
but not of the phase difference. This suggests expression of
eq.~\mref{N164} in new variables $(\theta_+,\theta_-)$. After
applying the casting procedure (see introduction to \S 4) the
joint mod($2\pi$) phase distribution is~\cite{BP90}:
%---------------------------------------------------------------------------
\begin{eqnarray}
  P_{2\pi}(\theta_{+},\theta_{-})&=&(4\pi^2\cosh^2 r)^{-1}
  \left(1+\tanh^2r-2\tanh r\cos\theta_{+}\right)^{-1},
\label{N168}
\end{eqnarray}
whereas the marginal phase distributions are
%---------------------------------------------------------------------------
\begin{eqnarray}
  P_{2\pi}(\theta_{+})&=&(2\pi\cosh^2 r)^{-1} \left(1+\tanh^2r-2\tanh
    r\cos\theta_{+}\right)^{-1},
\label{N169}
\end{eqnarray}
%---------------------------------------------------------------------------
\begin{eqnarray}
  P_{2\pi}(\theta_{-}) &=& \frac{1}{2\pi}.
\label{N170}
\end{eqnarray}
The uniform shape of function~\mref{N170} signifies
randomness of the phase difference in the field
[eq.~\mref{N163}]. If the casting procedure is not applied,
the marginal distributions $P(\theta_{\pm})\equiv
P_{4\pi}(\theta_{\pm})$ have more complicated
structures~\cite{BP90}. In particular, $P_{4\pi}(\theta_{-})$
is not uniform because of the integration limits in
eq.~\mref{N158}.  In general, the mod($4\pi$) distribution
has no unique shape signifying randomness of the phase sum or
difference. There are many distributions in the $4\pi$-range
leading to a flat mod($2\pi$) function.

%%%%%%%%%%%%%%%%%%%%%%%%%%%%%%%%%%%%%%%%%%%%%%%%%%%%%%%%%%%%%%%%%%%%%%%%%%%%
\begin{figure} % Fig. 11
\vspace*{-1.5cm} \hspace*{2cm} \epsfxsize=11cm
\epsfbox{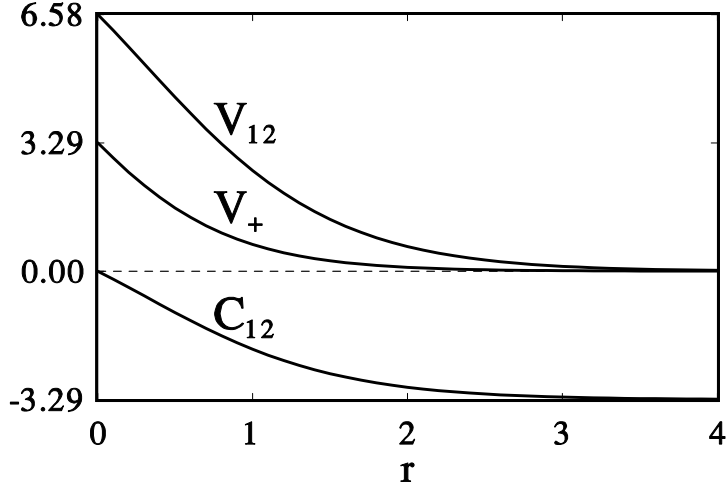} \vspace*{0cm} \caption{ The phase
variances $V_{12}\equiv \langle[\Delta(\hat{\Phi}
_{\theta_{1}}+ \hat{\Phi}_{\theta_{2}})]^2\rangle$, eq.
(4.29), and $V_{+}\equiv
\langle[\Delta(\hat{\Phi}_{\theta_{1}}+ \hat{\Phi}
_{\theta_{2}})]^2\rangle_{2\pi}$, eq. (4.30), and the phase
correlation function $C_{12}$, eq. (4.28), against the
squeeze parameter $r$ for the two-mode squeezed vacuum. }
\end{figure}
The two-mode variance of the phase-sum operator can be
calculated according to the general formula:
%---------------------------------------------------------------------------
\begin{eqnarray}
  \langle[\Delta(\hat{\Phi}_{\theta_{1}}+\hat{\Phi}_{\theta_{2}})]^2\rangle
  &=& \var{\hat{\Phi}_{\theta_{1}}}+
  \var{\hat{\Phi}_{\theta_{2}}}+2C_{12}
\label{N171}
\end{eqnarray}
in terms of the individual phase variances
$\var{\hat{\Phi}_{\theta_{1,2}}}$ and the phase correlation
function (correlation coefficient)
%---------------------------------------------------------------------------
\begin{eqnarray}
  C_{12} &\equiv&
  \langle\hat{\Phi}_{\theta_{1}}\hat{\Phi}_{\theta_{2}}\rangle
  -\langle\hat{\Phi}_{\theta_{1}}\rangle
  \langle\hat{\Phi}_{\theta_{2}}\rangle \nonumber\\ &=&
  \int\limits_{-\pi}^{\pi}\int\limits_{-\pi}^{\pi}\theta_{1}\theta_{2}
  P(\theta_{1},\theta_{2})\,{\rm d}\theta_{1}\,{\rm d}\theta_{2} -
  \bigg(\int\limits_{-\pi}^{\pi}\theta_{1} P(\theta_{1})\,{\rm
    d}\theta_{1}\bigg) \: \bigg(\int\limits_{-\pi}^{\pi}\theta_{2}
  P(\theta_{2})\,{\rm d}\theta_{2} \bigg).
\label{N172}
\end{eqnarray}
The variances $\var{\hat{\Phi}_{\theta_{1,2}}}$ are simply
$\pi^2/3$ [because of eq.~\mref{N165}], and the phase
correlation function $C_{12}$ is equal to:
%---------------------------------------------------------------------------
\begin{eqnarray}
  C_{12} &=& -2
  (\cosh\,r)^{-2}\sum_{n>k}\frac{(\tanh\,r)^{n+k}}{(n-k)^2} \;=\; - 2
  \,{\rm dilog}(1-\tanh\,r).
\label{N173}
\end{eqnarray}
This correlation function describes the correlation between
the phases of the two modes of the two-mode squeezed vacuum.
In fig.~11 the correlation coefficient as well as the phase
variances are plotted against the squeeze parameter $r$.  The
correlation is negative and, as $r$ tends to infinity,
approaches $-\pi^2/3$ asymptotically. Finally, phase
variance~\mref{N171} has the form:
%---------------------------------------------------------------------------
\begin{eqnarray}
  \langle[\Delta(\hat{\Phi}_{\theta_{1}}+\hat{\Phi}_{\theta_{2}})]^2\rangle
  &=& 2\,{\pi^2\over 3}-4\,{\rm dilog}(1-\tanh\,r).
\label{N171a}
\end{eqnarray}
The strong negative correlation between the two phases lowers
the variance~\mref{N171a} of the phase-sum operator. For
$r\rightarrow\infty$, this variance tends asymptotically to
zero, which means that for very high squeezing the sum of the
two phases becomes well defined (phase locking effect).

The (``single-mode'') mod($2\pi$) phase-sum variance
is~\cite{BP90}:
%---------------------------------------------------------------------------
\begin{eqnarray}
  \langle[\Delta(\hat{\Phi}_{\theta_{1}}+
  \hat{\Phi}_{\theta_{2}})]^2\rangle_{2\pi} &=&
  \int\limits_{-\pi}^{\pi} \theta_+^2 P_{2\pi}(\theta_{+}) {\rm d}
  \theta_+ \nonumber\\ &=& \frac{\pi^2}{3} + 4 \,{\rm
    dilog}(1+\tanh\,r).
\label{N174}
\end{eqnarray}
As the squeezing parameter $r$ varies from 0 to $\infty$, the
mod($2\pi$) variance [eq.~\mref{N174}] decreases from
$\pi^2/3$ to zero, whereas the two-mode phase-sum variance~
[eq.~\mref{N171a}] changes from $2\pi^2/3$ to zero with
increasing $r$.  Hence, both variances~\mref{N171a}
and~\mref{N174}, reveal the fact that the phase sum becomes
perfectly locked in the limit of large squeezing
($r\rightarrow\infty$).  The value $\pi^2/3$ of the
variance~\mref{N174} describes random phase sum for zero
squeezing. In this case of $r=0$, the two-mode
variance~\mref{N171a} is twice as much as the mod($2\pi$)
phase-sum variance~\mref{N174}, since it shows randomization
of the two phases, $\hat{\Phi}_{\theta_1}$ and
$\hat{\Phi}_{\theta_2}$, separately.

As was stressed in \S 4, both the original distributions,
given by eqs.~\mref{N164} and~\mref{N165}, and the
mod($2\pi$) distributions, given by
eqs.~\mref{N168}--\mref{N170}, are valid and useful. However,
some care is required when interpreting the results obtained
in both ways. The phase-sum variance has generally different
values, as seen from fig.~11, in the two approaches.  The
original distributions are better for understanding the
intermode phase correlation, which can be calculated
explicitly from eq.~\mref{N172}, while for the mod($2\pi$)
distribution the correlation is concealed in the value of the
phase variance~\mref{N174} and is not seen explicitly. On the
other hand, the mod($2\pi$) results have a clear
interpretation for the sum and difference of the individual
phases treated as single phase variables.

Generalizing formula~\mref{N15} and taking into account the
fact that the two-mode squeezed vacuum is a ``physical
state'', we can calculate the expectation values of the phase
exponential operators in the following way~\cite{GT91g}:
%----------------------------------------------------------------------
\begin{eqnarray}
\label{N175} \lefteqn{ \langle \exp(\I m_{1}
\hat{\Phi}_{\theta_{1}}) \exp(\I m_{2}
  \hat{\Phi}_{\theta_{2}}) \rangle \;=\; \langle \widehat{\exp}(\I
  m_{1} \Phi_{{1}{\rm SG}})\, \widehat{\exp}(\I m_{2} \Phi_{{2}{\rm
      SG}}) \rangle \null} \nonumber\\ &=& \null (\cosh r)^{-2}
\sum_{n,k=0}^{\infty}\sum_{m,l=0}^{\infty} \left({\rm
    e}^{2\I\varphi}\tanh r\right)^{l+m} \langle l,l|n,k\rangle \langle
n+m_{1},k+m_{2}|m,m\rangle \nonumber\\ &=& \null \left({\rm
    e}^{2\I\varphi}\tanh r\right)^{m_{1}} \delta _{m_{1},m_{2}}\, ,
\end{eqnarray}
where for brevity we denote $\langle(...)\rangle \:\equiv\:
_{(r,\varphi)}\langle 0,0|(...)  |0,0\rangle_{(r,\varphi)}$.
The operators
%----------------------------------------------------------------------
\begin{eqnarray}
\label{N176} \widehat{\exp}(\I m_{{1},{2}}\Phi_{{1},{2}{\rm
SG}}) &=& \sum_{n=0}^{\infty} |n\rangle\langle n+m_{{1},{2}}|
\end{eqnarray}
are the Susskind-Glogower phase operators for the two modes.
Formula~\mref{N175} is strikingly simple, and shows that only
exponentials of the phase sum have nonzero expectation
values.

Using eq.~\mref{N175}, the following results for the cosine
and sine of the phase-sum operator are obtained~\cite{GT91g}:
%----------------------------------------------------------------------
\begin{eqnarray}
\label{N177} \langle
\cos(\hat{\Phi}_{\theta_{1}}+\hat{\Phi}_{\theta_{2}}) \rangle
&=& \tanh r \cos 2\varphi, \nonumber\\ \langle \sin(
\hat{\Phi}_{\theta_{1}} +\hat{\Phi}_{\theta_{2}}) \rangle &=&
\tanh r \sin 2\varphi,
\end{eqnarray}
%----------------------------------------------------------------------
\begin{eqnarray}
\label{N178} \langle \cos^2( \hat{\Phi}_{\theta_{1}}
+\hat{\Phi}_{\theta_{2}}) \rangle &=&
\frac{1}{2}+\frac{1}{2}(\tanh r)^2 \cos 4\varphi, \nonumber\\
\langle \sin^2( \hat{\Phi}_{\theta_{1}}
+\hat{\Phi}_{\theta_{2}}) \rangle &=&
\frac{1}{2}-\frac{1}{2}(\tanh r)^2 \cos 4\varphi,
\end{eqnarray}
%----------------------------------------------------------------------
\begin{eqnarray}
\label{N179} \langle [\Delta \cos( \hat{\Phi}_{\theta_{1}}
+\hat{\Phi}_{\theta_{2}})]^2\rangle &=& \langle [\Delta \sin(
\hat{\Phi}_{\theta_{1}} +\hat{\Phi}_{\theta_{2}})]^2\rangle
\;=\;{1\over 2} (\cosh r)^{-2}.
\end{eqnarray}
For very large squeezing ($r\rightarrow\infty$, $\tanh
r\rightarrow 1$, $\cosh r\rightarrow\infty$), the expectation
values~\mref{N177} and~\mref{N178} of the functions of the
phase-sum operator become asymptotically corresponding
functions of the phase $2\varphi$, confirming the relation
between the phase sum and $2\varphi $ that is already
apparent from eq.~\mref{N167}.  It is interesting that the
expectation value of the phase-sum operator is equal to
$2\varphi$ irrespective of the value of $r$, whereas for the
sine and cosine functions correspondence is obtained only
asymptotically.  The variances~\mref{N179} then become zero
and the sine and cosine of the phase sum are well defined.

It should, however, be emphasized that the expectation values
calculated according to the Pegg-Barnett formalism depend on
the choice of the particular window of the phase eigenvalues.
If a choice different from that made above were made, the
clear picture of the phase properties of the two-mode
squeezed vacuum would be disturbed. For example, the value of
the correlation coefficient~\mref{N173} would be different,
and the phase-sum variance~\mref{N171} would not tend
asymptotically to zero. However,
formulas~\mref{N175}--\mref{N179}, because of the way they
have been calculated do not, in fact, depend on the choice of
the phase window. This gives us the opportunity to make a
choice which introduces consistency in the behavior of the
phase itself and its sine and cosine functions.  Another way
of making the choice is to minimize the variance~\mref{N171}
of the phase-sum operator.

%%%%%%%%%%%%%%%%%%%%%%%%%%%%%%%%%%%%%%%%%%%%%%%%%%%%%%%%%%%%%%%%%%%%%%%%
\subsection{Pair coherent states}
%%%%%%%%%%%%%%%%%%%%%%%%%%%%%%%%%%%%%%%%%%%%%%%%%%%%%%%%%%%%%%%%%%%%%%%%

Pair coherent states introduced
by~\citeA{Aga86}~\citeY{Aga86,Aga88} are quantum states of
the two-mode electromagnetic field, which are simultaneous
eigenstates of the pair annihilation operator and the
difference in the number operators of the two modes of the
field. \citeAY{Aga88} has discussed the nonclassical
properties of such states, showing that they exhibit
remarkable quantum features such as sub-Poissonian
statistics, correlations in the number fluctuations,
squeezing, and violations of the Cauchy-Schwarz inequalities.
He has also presented results for fluctuations in the phase
of the field using the Susskind-Glogower phase formalism. The
phase properties of such states on the basis of the
Pegg-Barnett formalism were studied by~\citeAY{GT91e}, and by
\citeAY{Gou93}. Phase distributions for squeezed pair
coherent states were analyzed by~\citeAY{Ger95}.

The pair coherent states are defined by \citeAY{Aga88} as
eigenstates of the pair-annihilation operator
%---------------------------------------------------------------------------
\begin{eqnarray}
  \hat{a}\hat{b}|\zeta,q\rangle &=& \zeta|\zeta,q\rangle,
\label{N180}
\end{eqnarray}
where $\zeta$ is a complex eigenvalue and $q$ is the
degeneracy parameter, which can be fixed by the requirement
that $|\zeta,q\rangle$ is an eigenstate of the difference of
the number operators for the two modes:
%---------------------------------------------------------------------------
\begin{eqnarray}
  (\hat{a}^{\dagger} \hat{a}-\hat{b}^{\dagger} \hat{b})|\zeta,q\rangle
  &=& q|\zeta,q\rangle.
\label{N181}
\end{eqnarray}

The solution to the above eigenvalue problem, assuming $q$ to
be positive, is given by~\cite{Aga88}:
%---------------------------------------------------------------------------
\begin{eqnarray}
  |\zeta,q\rangle=N_q\sum_{n=0}^{\infty}\frac{\zeta^n}{[n!(n+q)!]^{1/2}}
  |n+q,n\rangle,
\label{N182}
\end{eqnarray}
where $N_q$ is the normalization constant
%---------------------------------------------------------------------------
\begin{eqnarray}
  N_q=\left(\sum_{n=0}^{\infty}\frac{|\zeta|^{2n}}{n!(n+q)!}\right)^{-1/2}
  =[(\I |\zeta|)^{-q}J_q(2\I|\zeta|)]^{-1/2}.
\label{N183}
\end{eqnarray}
The state $|n+q,n\rangle$ is a Fock state with $n+q$ photons
in mode $a$ and $n$ photons in mode $b$. If the complex
number $\zeta$ is written in the form:
%---------------------------------------------------------------------------
\begin{eqnarray}
  \zeta=|\zeta|\exp(\I \varphi),
\label{N184}
\end{eqnarray}
the state~\mref{N182} can be written as
%---------------------------------------------------------------------------
\begin{eqnarray}
  |\zeta,q\rangle=\sum_{n=0}^{\infty}b_n\mbox{e}^{\I
    n\varphi}|n+q,n\rangle,
\label{N185}
\end{eqnarray} where
%---------------------------------------------------------------------------
\begin{eqnarray}
  b_n=N_q\frac{|\zeta|^n}{[n!(n+q)!]^{1/2}}\geq 0.
\label{N186}
\end{eqnarray}
Now, the phase properties of the state~\mref{N185} can be
studied easily using the Pegg-Barnett formalism in a standard
way as described above.  The resulting joint probability
distribution for the phases $\theta_a$ and $\theta_b$ of the
two modes is given by~\cite{GT91e}:
%---------------------------------------------------------------------------
\begin{eqnarray}
  P(\theta_a,\theta_b)&=&\frac{1}{(2\pi)^2}\biggl\{1+2
  \sum_{n>k}b_nb_k\cos[(n-k)(\theta_a+\theta_b)]\biggl\},
\label{N187}
\end{eqnarray}
where $b_n$ is given by eq.~\mref{N186}.  For $q=0$,
formula~\mref{N187} can be written in the following simple
form:
%---------------------------------------------------------------------------
\begin{eqnarray}
  P(\theta_a,\theta_b)&=&\frac{N_{0}^{2}}{(2\pi)^2}
  \exp[2|\zeta|\cos(\theta_a+\theta_b)].
\label{N188}
\end{eqnarray}

As in the case of the two-mode squeezed vacuum, the joint
phase probability distribution depends on the sum of the two
phases only, which means strong correlation between the two
phases. Again, the only non-uniformly distributed phase
quantity is the phase sum $\theta_a+\theta_b$.  This suggests
 re-expression of the phase distribution~\mref{N187} in new
variables of the phase sum, $\theta_+=\theta_a+\theta_b$, and
phase difference, $\theta_-=\theta_a-\theta_b$. After
applying the casting procedure, the mod($2\pi$) Pegg-Barnett
distribution $P_{2\pi}(\theta_{+},\theta_{-})$ takes the
form:
%----------------------------------------------------------------------
\begin{eqnarray}
\label{N189} P_{2\pi}(\theta_{+},\theta_{-}) &=&
\frac{1}{4\pi^2} \left\{1+
  2\sum_{n>k}^{\infty} b_n b_k \cos[(n-k)\theta_+] \right\},
\end{eqnarray}
and the marginal distributions are:
%----------------------------------------------------------------------
\begin{eqnarray}
\label{N190} P_{2\pi}(\theta_{+}) &=& \frac{1}{2\pi}
\left\{1+2\sum_{n>k}^{\infty}
  b_n b_k \cos[(n-k)\theta_+] \right\},
\end{eqnarray}
and
%----------------------------------------------------------------------
\begin{eqnarray}
\label{N191} P_{2\pi}(\theta_{-}) &=& \frac{1}{2\pi}.
\end{eqnarray}

For completeness of our discussion and by analogy with our
presentation of the single-mode models, we now give
expressions for various $s$-parametrized phase distributions.
Thus, the mod($2\pi$) two-mode $s$-parametrized phase
distribution is equal to
%----------------------------------------------------------------------
\begin{eqnarray}
\label{N192} P^{(s)}_{2\pi}(\theta_{+},\theta_-) \:=\:
\frac{1}{4\pi^2} \left\{1+
  2\sum_{n>k}^{\infty} b_n b_k G^{(s)}(n,k) G^{(s)}(n+q,k+q)
  \cos[(n-k)\theta_+] \right\},
\end{eqnarray}
where the coefficients $G^{(s)}(n,k)$ are given by
eqs.~\mref{N60}--\mref{N62}.  The mod($2\pi$) marginal
$s$-parametrized phase-sum distribution is
%----------------------------------------------------------------------
\begin{eqnarray}
\label{N193} P^{(s)}_{2\pi}(\theta_{+}) &=& \frac{1}{2\pi}
\left\{1+
  2\sum_{n>k}^{\infty} b_n b_k G^{(s)}(n,k) G^{(s)}(n+q,k+q)
  \cos[(n-k)\theta_+] \right\}.
\end{eqnarray}
The mod($2\pi$) $s$-parametrized phase-difference
distribution $P^{(s)}_{2\pi}(\theta_{-})$ and the single-mode
ones, $P^{(s)}(\theta_{a})$ and $P^{(s)}(\theta_{b})$, are
uniform:
%----------------------------------------------------------------------
\begin{eqnarray}
\label{N194} P^{(s)}_{2\pi}(\theta_{-}) &=&
P^{(s)}(\theta_{a}) \;=\; P^{(s)}(\theta_{b}) \;=\;
\frac{1}{2\pi}.
\end{eqnarray}
The distributions~\mref{N192}--\mref{N194}, similar to the
distributions~\mref{N189}--\mref{N191}, reveal the
fundamental phase properties of pair coherent states.

The correlation coefficient $C_{ab}$, eq.~\mref{N172},
[subscripts $1,2$ should be replaced by $a$ and $b$,
respectively] is given in this case by the formula:
%---------------------------------------------------------------------------
\begin{eqnarray}
  C_{ab} &=&-\,2\sum_{n>k}\frac{b_nb_k}{(n-k)^2},
\label{N195}
\end{eqnarray}
where $b_n$ is given by eq.~\mref{N186}. This correlation is
negative and lowers the variance of the phase-sum operator.
For $|\zeta|\rightarrow\infty$, this coefficient approaches
$-\pi^2/3$, the phase-sum variance becomes zero, and we have
the classical situation of perfectly defined phase sum (the
phase locking effect). This phase correlation coefficient can
be contrasted with the photon number correlation coefficient,
considered by~\citeAY{Aga88}, which increases as $|\zeta|$
increases. The sine and cosine functions of the phase-sum
operator were also obtained by~\citeAY{GT91e} and compared to
their counterparts obtained by~\citeAY{Aga88}, who used the
Susskind-Glogower approach.

%%%%%%%%%%%%%%%%%%%%%%%%%%%%%%%%%%%%%%%%%%%%%%%%%%%%%%%%%%%%%%%%%%%%%%%%
\subsection{Elliptically polarized light propagating in a nonlinear Kerr medium}
%%%%%%%%%%%%%%%%%%%%%%%%%%%%%%%%%%%%%%%%%%%%%%%%%%%%%%%%%%%%%%%%%%%%%%%%

To describe propagation of elliptically polarized light in a
nonlinear Kerr medium, a two-mode description of the field is
needed. The quantum nature of the field results in the
appearance of such quantum effects as photon
antibunching~\cite{RB79,TK79,Rit80} and squeezing
(\citeA{TK83} \citeY{TK83,TK84}).~\citeA{TK83} have shown
that as much as 98 percent of squeezing can be obtained when
intense light propagates in a nonlinear Kerr medium. They
referred to this effect as self-squeezing. \citeAY{AP89a}
re-examined the problem of propagation of elliptically
polarized light through a Kerr medium, considering not only
the Heisenberg equations of motion for the field operators,
but also the evolution of the states themselves. Quantum
fluctuations in the Stokes parameters of light propagating in
a Kerr medium were discussed by~\citeAY{TK90}, and by
\citeAY{TG92b}.

The following effective interaction Hamiltonian can be used
to describe the propagation of elliptically polarized light
in a Kerr medium (\citeA{TK83}~\citeY{TK83,TK84}):
%---------------------------------------------------------------------------
\begin{eqnarray}
  \hat{H}_{\rm I} &=& {1\over 2}\hbar\kappa\left(\hat{a}_{1}^{\dagger
      2} \hat{a}_{1}^{2}+\hat{a}_{2}^{\dagger 2}
    \hat{a}_{2}^{2}+4d\hat{a}_{1}^{\dagger}\hat{a}_{2}^{\dagger}
    \hat{a}_{1}\hat{a}_{2}\right),
\label{N196}
\end{eqnarray}
where $\hat{a}_{1}$ and $\hat{a}_{2} $ are the annihilation
operators for the circularly right- (``1'') and left- (``2'')
polarized modes of the field, $\kappa$ is the coupling
constant, which is real and related to the nonlinear
susceptibility tensor $\chi^{(3)}$ of the medium, and $d$ is
the asymmetry parameter describing the coupling between the
two modes.  For a fully symmetrical susceptibility tensor,
$d=1$. Otherwise, $d\neq 0$ and describes the asymmetry of
the nonlinear properties of the medium (\citeAY{Rit80},
\citeA{TK83}~\citeY{TK83,TK84}).

Using the Hamiltonian~\mref{N196}, one can obtain the
evolution operator $\hat{U}(\tau)$, and assuming that the
initial state of the field is a coherent state of the
elliptically polarized light, one obtains for the resulting
state of the field~\cite{AP89a}:
%---------------------------------------------------------------------------
\begin{eqnarray}
  |\psi(\tau)\rangle &=&\hat{U}(\tau)|\alpha_{1},\alpha_{2}\rangle
  \nonumber\\ &=&\sum_{n_{1},n_{2}}b_{n_{1}}b_{n_{2}} \exp\biggl\{\I
  (n_{1}\varphi_{1}+n_{2}\varphi_{2}) \nonumber\\ &&+\;\I {\tau\over
    2}\left[n_{1}(n_{1}-1)
    +n_{2}(n_{2}-1)+4dn_{1}n_{2}\right]\biggr\}|n_{1},n_{2}\rangle ,
\label{N197}
\end{eqnarray}
where $\tau=n(\omega)kz/c$ ($n(\omega)$ with the refractive
index), and the coefficients $b_{n_{1,2}}$ are given by
eq.~\mref{N81} with $|\alpha_{1}|^2$ and $|\alpha_{2}|^2$ as
the mean numbers of photons for the circularly right- and
left-polarized modes, respectively, whereas $\varphi_{1,2}$
are the phases of the coherent states of the two modes.

The state~\mref{N197} is the two-mode state of the field, and
the two-mode generalization of the Pegg-Barnett formalism
used by~\citeAY{GT91c} leads to the following joint
probability distribution for the continuous-phase variables,
$\theta_{1}$ and $\theta_{2}$, of the two modes:
%---------------------------------------------------------------------------
\begin{eqnarray}
  P(\theta_{1},\theta_{2})&=&\frac{1}{(2\pi)^2}\Biggl
  |\sum_{n_{1}=0}^{\infty} \sum_{n_{2}=0}^{\infty}b_{n_{1}}b_{n_{2}}
  \exp\biggl\{-\I n_{1}\theta_{1}-\I n_{2}\theta_{2} \nonumber\\
  &&+\;\I
  \frac{\tau}{2}[n_{1}(n_{1}-1)+n_{2}(n_{2}-1)+4dn_{1}n_{2}]\biggr\}
  \Biggl|^2.
\label{N198}
\end{eqnarray}
The phase distribution function $P(\theta_{1},\theta_{2})$
describes the phase properties of elliptically polarized
light propagating through a Kerr medium, which were discussed
in detail by~\citeAY{GT91c}. Figure~12 shows an example of
the evolution of $P(\theta_{1},\theta_{2})$.  It is seen that
the peak is shifted and broadened during the evolution. Since
the numbers of photons in the two modes are different, one
can see that the shift of the peak and its broadening is
asymmetric. The intensity-dependent phase shift is bigger for
the mode with higher number of photons. This corresponds to
the classical effect of self-phase modulation in a nonlinear
Kerr medium. The quantum description shows not only the shift
but also the broadening of the phase distribution (phase
diffusion).

Integration of the distribution function
$P(\theta_{1},\theta_{2})$ over one of the phases
$\theta_{1}$ or $\theta_{2}$ leads to the marginal
distribution $P(\theta_{2})$ or $P(\theta_{1})$ for the
individual phases.  All single-mode phase characteristics of
the field can be calculated using these distributions, and
the corresponding formulas were given by~\citeAY{GT91c}.

%%%%%%%%%%%%%%%%%%%%%%%%%%%%%%%%%%%%%%%%%%%%%%%%%%%%%%%%%%%%%%%%%%%%%%%%%%%
\begin{figure} % Fig. 12
\vspace*{-5.5cm} \hspace*{1.8cm} \epsfxsize=14cm
\epsfbox{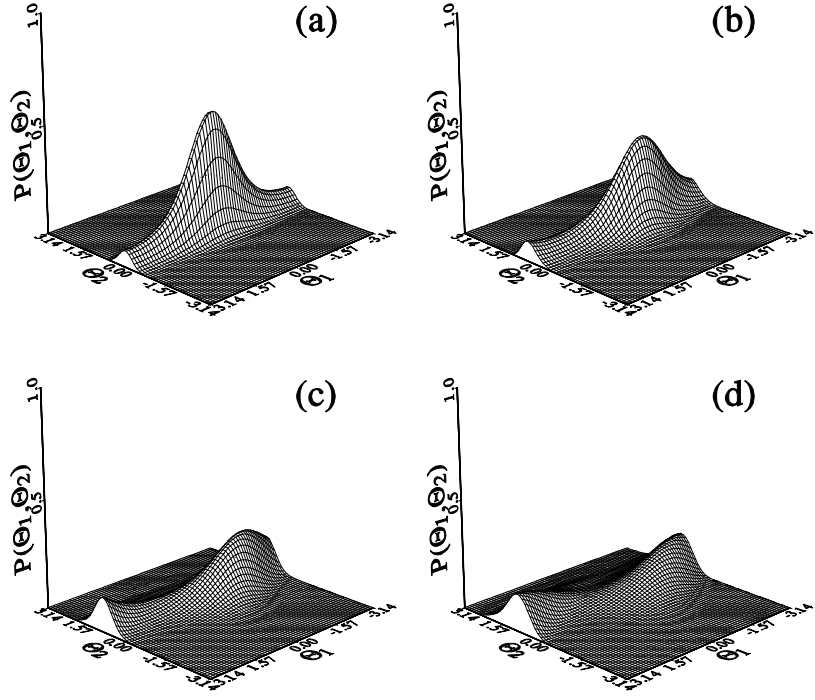} \vspace*{-1.5cm} \caption{ Evolution of
the joint probability distribution $P(\theta_1,\theta_2)$,
eq. (4.54), of light propagating in a Kerr medium:
$|\alpha_1|^2=0.25$, $|\alpha_2|^2=4$ and $d=1$; (a)
$\tau=0$, (b) $\tau=0.1$, (c) $\tau=0.2$, (d) $\tau=0.3$. }
\end{figure}
In addition to the phase properties of the individual modes,
it is interesting, in the two-mode case, to study the
behavior of the phase difference between the two modes. In
the Pegg-Barnett formalism, the phase-difference operator is
simply the difference of the phase operators for the two
modes, so the mean value of the phase-difference operator is
the difference of the mean values of the single-mode phase
operators. To calculate the variance of the phase-difference
operator, we can use the relation:
%---------------------------------------------------------------------------
\begin{eqnarray}
  \langle[\Delta(\hat{\Phi}_{\theta_{1}}-\hat{\Phi}_{\theta_{2}})]^2\rangle
  &=& \var{\hat{\Phi}_{\theta_{1}}}+
  \var{\hat{\Phi}_{\theta_{2}}}-2C_{12},
\label{N199}
\end{eqnarray}
where the last term is the correlation coefficient between
the phases of the two modes and can be calculated by
integration of $P(\theta_{1},\theta_{2})$ according to
eq.~\mref{N172}.  Thus, the resulting formula
is~\cite{GT91c}:
%----------------------------------------------------------------------
\begin{eqnarray}
\label{N200} C_{12}(\tau) \:=\: \sum_{n_1>n_1'}
\sum_{n_2>n_2'} f_{12} f_{21} - \left( \sum_{n_1= n_1'}
\sum_{n_2>n_2'} f_{21}
  b_{n_1}^2 \right) \left( \sum_{n_2=n_2'}
  \sum_{n_1>n_1'} f_{12} b_{n_2}^2 \right), \end{eqnarray}
%%%%%%%%%%%%%%%%%%%%%%%%%%%%%%%%%%%%%%%%%%%%%%%%%%%%%%%%%%%%%%%%%%%%%%%%%%%%
\begin{figure} % Fig. 13
\vspace*{-12.5cm} \hspace*{-1cm}
\epsfbox{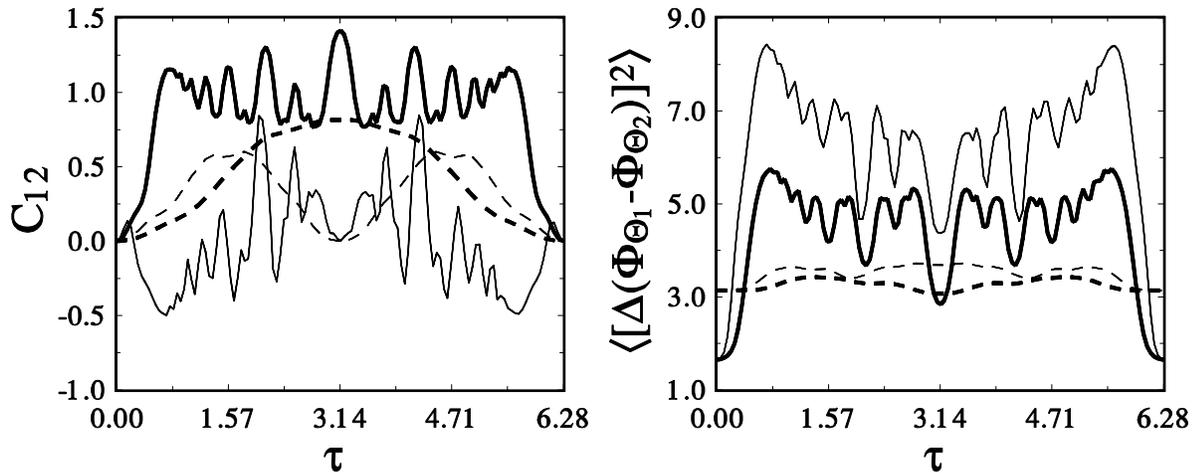} \vspace*{-4.6cm} \caption{ Evolution of
the intermode phase correlation function $C_{12}(\tau)$, eq.
(4.56), and the phase-difference variance
$\langle[\Delta(\hat{\Phi}_{\theta_{1}}-\hat{\Phi}_{\theta_{2}})]^2\rangle$,
eq. (4.55), of light propagating in a Kerr medium. Thin solid
line:
--- $|\alpha_1|^2=0.25$, $|\alpha_2|^2=4$ and $d=1$; bold solid
line --- $|\alpha_1|^2=0.25$, $|\alpha_2|^2=4$ and $d=1/2$;
thin dashed line --- $|\alpha_1|^2=0.25$, $|\alpha_2|^2=0.25$
and $d=1$; bold dashed line --- $|\alpha_1|^2=0.25$,
$|\alpha_2|^2=0.25$ and $d=1/2$. }
\end{figure}
where
%----------------------------------------------------------------------
\begin{eqnarray}
\label{N201} f_{ij} &=& 2 b_{n_i} b_{n'_i}
\frac{(-1)^{n_i-n'_i}}{n_i-n'_i} \sin\left\{ \frac{\tau}{2}
(n_i-n'_i) \left[n_i+n'_i
    -1+2d(n_{j}+n'_{j})\right] \right\}.
\end{eqnarray}
A graphic illustration of the correlation
function~\mref{N200} is shown in the left-hand panel of
fig.~13. The strength of the correlation depends crucially on
the value of the asymmetry parameter $d$.  The highest values
of the correlation are obtained for $d=1/2$.  This means that
the minimum of the phase-difference variance, in view of
eq.~\mref{N199}, is obtained for $d=1/2$. The
phase-difference variance is shown in the right-hand panel of
fig.~13. It was shown \cite{TG91} that, similar to the
single-mode case, dissipation destroys the periodicity of the
evolution and broadens the phase distribution.

Recently, the phase properties of light propagating in a Kerr
medium have been reconsidered \cite{LST95} from the point of
view of the Hermitian phase-difference operator introduced by
\citeA{LS94}~\citeY{LS93b,LS94}, which is based on the polar
decomposition of the Stokes operators. This example shows
clearly the difference between the Pegg-Barnett and
Luis--S\'anchez-Soto phase-difference formalism, which is
most visible for weak fields. The Luis--S\'anchez-Soto
phase-difference operator differs from the Pegg-Barnett
phase-difference operator, which is simply the difference of
the phase operators of the two modes.  For strong fields both
formalisms give the same results. The nonlinear Kerr medium
appears to be a good testing ground for different phase
approaches.

As shown by~\citeAY{GT91a}, superpositions with any number of
components can be obtained in the process of light
propagation in the Kerr medium (similar to the anharmonic
oscillator model described in \S 3.5) if the evolution time
$\tau$ is taken as a fraction $M/N$ of the period, where $M$
and $N$ are mutually prime integers.  Exact analytical
formulas for finding the superposition coefficients were
given for any $N$. The joint phase probability distribution
$P(\theta_{1},\theta_{2})$ splits into separate peaks if the
state of the field becomes a discrete superposition of
coherent states, and this is a very spectacular way of
presenting such superpositions. Some examples are shown in
fig.~14.
%%%%%%%%%%%%%%%%%%%%%%%%%%%%%%%%%%%%%%%%%%%%%%%%%%%%%%%%%%%%%%%%%%%%%%%%%%%%
\begin{figure} % Fig. 14
\vspace*{-13.2cm} \hspace*{-1cm}
\epsfbox{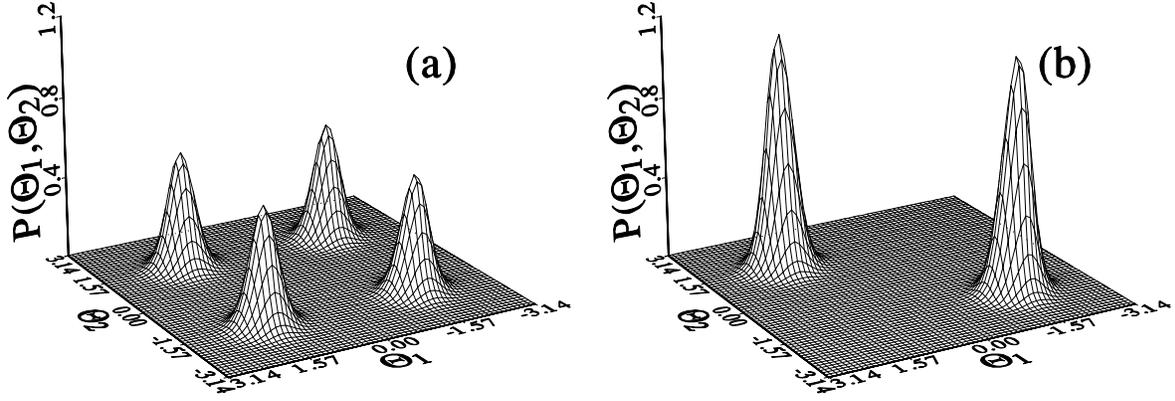} \vspace*{-4.4cm} \caption{ The joint
probability distribution $P(\theta_1,\theta_2)$,
eq. (4.54), of light propagating in a Kerr medium. $|\alpha_1|^2=|%
\alpha_2|^2=4$, $\tau=2\pi/2$, and (a) $d=0$; (b) $d=1/2$. }
\end{figure}

%%%%%%%%%%%%%%%%%%%%%%%%%%%%%%%%%%%%%%%%%%%%%%%%%%%%%%%%%%%%%%%%%%%%%%%%
\subsection{Second-harmonic generation}
%%%%%%%%%%%%%%%%%%%%%%%%%%%%%%%%%%%%%%%%%%%%%%%%%%%%%%%%%%%%%%%%%%%%%%%%
%%%%%%%%%%%%%%%%%%%%%%%%%%%%%%%%%%%%%%%%%%%%%%%%%%%%%%%%%%%%%%%%%%%%%%%%%%%%
\begin{figure} % Fig. 15
\vspace*{-2cm} \hspace*{0cm} \epsfxsize=17.5cm
\epsfbox{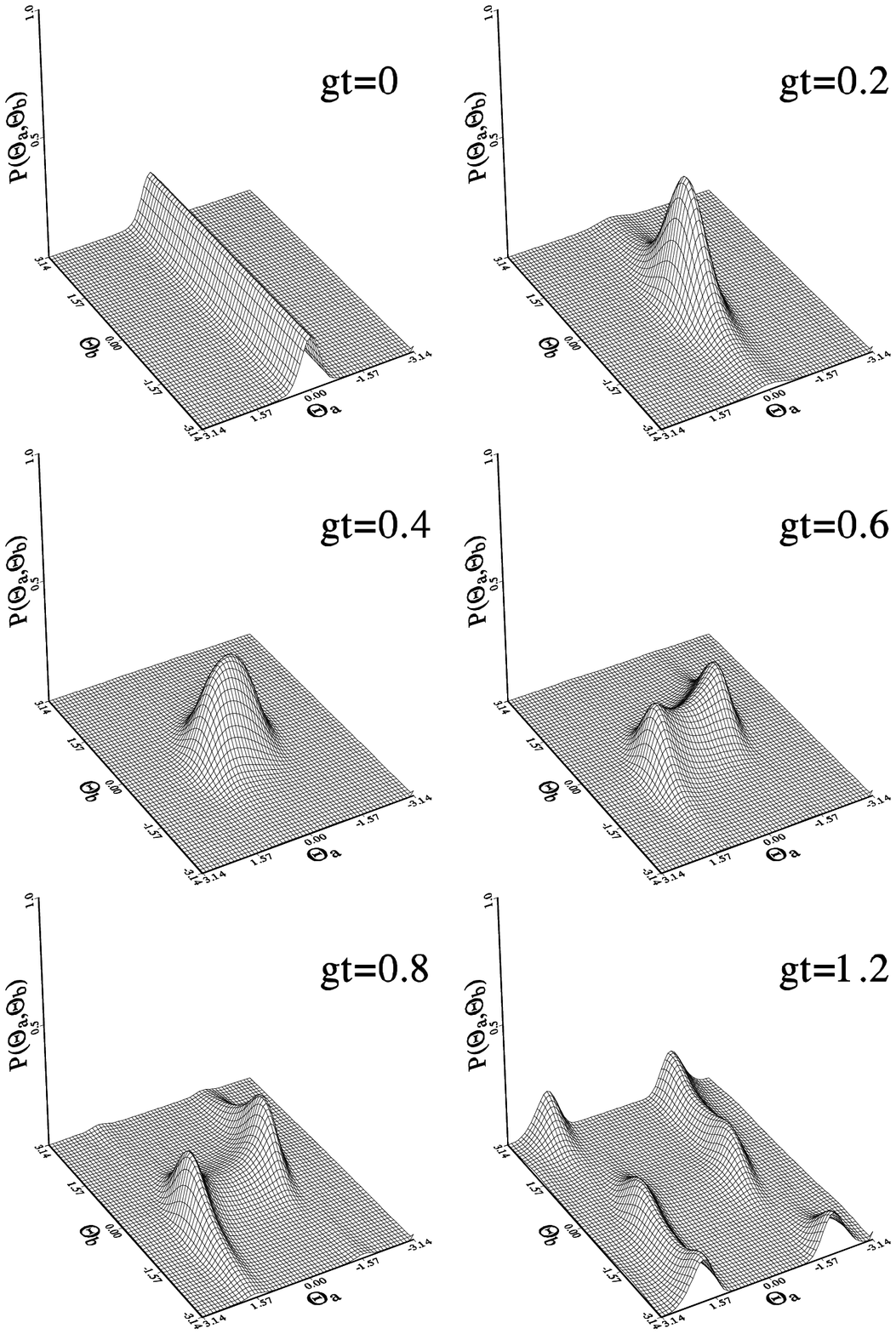} \vspace*{0cm} \caption{ Evolution of the
joint phase probability distribution $P(\theta_a,\theta_b)$,
eq. (4.71), in the second-harmonic generation. The initial
mean number of photons of the fundamental beam is
$|\alpha_0|^2=4$, and $gt$ is the dimensionless scaled time.
}
\end{figure}

Second-harmonic generation is probably the best known
nonlinear optical process. In the quantum picture we deal
with a nonlinear process in which two photons are annihilated
and single photon with doubled frequency is created. The
quantum states of the field generated in the process exhibit
a number of unique quantum features such as photon
antibunching~\cite{KT77} and squeezing~\cite{Man82,WKH86} for
both the fundamental and second-harmonic modes (for a review
and literature see~\citeAY{KKT85}).~\citeAY{NM91} discussed
the properties of the quantum state of the fundamental mode,
calculating numerically the quasiprobability distribution
function $Q(\alpha,\alpha^*)$ for it. They suggested that the
quantum state of the fundamental mode evolves, in the course
of the second-harmonic generation, into a superposition of
two macroscopically distinguishable states, similar to the
superpositions obtained for the anharmonic oscillator
model~\cite{YS86,TM87,MTK90,GT91f}, or a Kerr
medium~\cite{AP89a,GT91a}.~\citeAY{GTZ91b} discussed the
phase properties of the field produced in the second-harmonic
generation process.

To describe second-harmonic generation, the following model
Hamiltonian is used:
%---------------------------------------------------------------------------
\begin{eqnarray}
  \hat{H}\;=\; \hat{H}_0 + \hat{H}_{\rm I} &=& \hbar \omega
  \hat{a}^{\dagger}\hat{a} + 2\hbar \omega \hat{b}^{\dagger}\hat{b} +
  \hbar g(\hat{b}^{\dagger}\hat{a}^{2} + \hat{b}\hat{a}^{\dagger 2}),
\label{N202}
\end{eqnarray}
where $\hat{a}$ ($\hat{a}^{\dagger}$) and $\hat{b}$
($\hat{b}^{\dagger}$) are the annihilation (creation)
operators of the fundamental mode of frequency $\omega$ and
the second-harmonic mode at frequency $2\omega$,
respectively. The coupling constant $g$, which is real,
describes the coupling between the two modes. Since
$\hat{H}_0$ and $\hat{H}_{\rm I}$ commute, there are two
constants of motion: $\hat{H}_0$ and $\hat{H}_{\rm I}$,
$\hat{H}_0$ determines the total energy stored in both modes,
which is conserved by the interaction $\hat{H}_{\rm I}$.  The
free evolution can be thus factored out, and the resulting
state of the field can be written as
%---------------------------------------------------------------------------
\begin{eqnarray}
  |\Psi(t)\rangle &=& \exp(-\I \hat{H}_{\rm I}t/ \hbar ) | \Psi (0)
  \rangle,
\label{N203}
\end{eqnarray}
where $|\Psi(0) \rangle$ is the initial state of the field.
Since the interaction Hamiltonian $\hat{H}_{\rm I}$ is not
diagonal in the number-state basis, the numerical method of
diagonalization of $\hat{H}_{\rm I}$ may be applied to find
the state evolution \cite{WB70}.

Let us assume that initially there are $n$ photons in the
fundamental mode and no photons in the second-harmonic mode;
i.e., the initial state of the field is
$|n,0\rangle=|n\rangle |0\rangle$.  Since $\hat{H}_0$ is a
constant of motion, we have the relation:
%---------------------------------------------------------------------------
\begin{eqnarray}
  \langle \hat{a}^{\dagger}\hat{a}\rangle + 2\langle
  \hat{b}^{\dagger}\hat{b}\rangle &=& {\rm constant}\;=\;n,
\label{N204}
\end{eqnarray}
which implies that the creation of $k$ photons of the
second-harmonic mode requires annihilation of $2k$ photons of
the fundamental mode. Thus, for given $n$, we can introduce
the states
%---------------------------------------------------------------------------
\begin{eqnarray}
  |\psi_{k}^{(n)}\rangle &=& |n-2k,k\rangle, \ \ \ \ \ \ \ \
  k=0,1,\dots,[n/2],
\label{N205}
\end{eqnarray}
where $[n/2]$ denotes the integer part of $n/2$, which form a
complete basis of states of the field for given $n$. We have
%---------------------------------------------------------------------------
\begin{eqnarray}
  \langle\psi_{k'}^{(n')}|\psi_{k}^{(n)}\rangle &=&
  \delta_{nn'}\delta_{kk'},
\label{N206}
\end{eqnarray}
meaning that the constant of motion $\hat{H}_0$ splits the
field space into orthogonal subspaces, which for given $n$
have the number of components equal to $[n/2]+1$. In such a
basis, the interaction Hamiltonian has the following nonzero
matrix elements:
%---------------------------------------------------------------------------
\begin{eqnarray}
  \langle\psi_{k+1}^{(n)}|\hat{H}_{\rm I}|\psi_{k}^{(n)}\rangle
  \;\equiv\; \left(\hat{H}_{\rm I}\right)_{k+1,k}^{(n)} &=&
  \langle\psi_{k}^{(n)}|\hat{H}_{\rm I}|\psi_{k+1}^{(n)}\rangle
  \;\equiv\; \left(\hat{H}_{\rm I}\right)_{k,k+1}^{(n)} \nonumber\\
  &=&\hbar g \sqrt{(k+1)(n-2k)(n-2k-1)},
\label{N207}
\end{eqnarray}
which form a symmetric matrix of the dimension
$([n/2]+1)\times ([n/2]+1)$ with real nonzero elements (we
have assumed $g$ real) located on the two diagonals
immediately above and below the principal diagonal. Such a
matrix can be easily diagonalized numerically~\cite{WB70}.

To find the state evolution, we need the matrix elements of
the evolution operator:
%---------------------------------------------------------------------------
\begin{eqnarray}
  d_{n,k}(t)&\equiv&\langle\psi_{k}^{(n)}|\exp(-\I \hat{H}_{\rm I}
  t/\hbar)|\psi_{0}^{(n)}\rangle.
\label{N208}
\end{eqnarray}
If the matrix $\hat{U}$ is the unitary matrix that
diagonalizes the interaction Hamiltonian matrix given by
eq.~\mref{N207}, i.e.:
%---------------------------------------------------------------------------
\begin{eqnarray}
  \hat{U}^{-1}\hat{H}_{\rm I}^{(n)}\hat{U} &=& \hbar g\times {\rm
    diag}\left(\lambda_0,\lambda_1,\ldots,\lambda_{[n/2]}\right),
\label{N209}
\end{eqnarray}
then the coefficients $d_{n,k}(t)$ can be written as
%---------------------------------------------------------------------------
\begin{eqnarray}
  d_{n,k}(t) &=& \sum_{i=0}^{[n/2]}e^{-\I
    gt\lambda_i}U_{ki}U_{0i}^{*},
\label{N210}
\end{eqnarray}
where $\lambda_i$ are the eigenvalues of the interaction
Hamiltonian in units of $\hbar g$. Of course, the matrix
$\hat{U}$ as well as the eigenvalues $\lambda_i$ are defined
for given $n$ and should have the additional index $n$, which
we have omitted to shorten the notation. Moreover, for real
$g$ the interaction Hamiltonian matrix is real, and the
transformation matrix $\hat{U}$ is a real orthogonal matrix,
so the asterisk can also be dropped.

The numerical diagonalization procedure gives the eigenvalues
$\lambda_i$ as well as the elements of the matrix $\hat{U}$,
and thus the coefficients $d_{n,k}(t)$ can be calculated
according to eq.~\mref{N210}. It is worthwhile to note,
however, that due to the symmetry of the Hamiltonian the
eigenvalues $\lambda_i$ are distributed symmetrically with
respect to zero, with single eigenvalue equal to zero if
there is an odd number of them.  When the eigenvalues are
numbered from the lowest to the highest value, there is an
additional symmetry relation:
%---------------------------------------------------------------------------
\begin{eqnarray}
  U_{ki}U_{0i} &=& (-1)^kU_{k,[n/2]-i}U_{0,[n/2]-i}\,,
\label{N211}
\end{eqnarray}
which makes the coefficients $d_{n,k}(t)$ either real ($k$
even) or imaginary ($k$ odd). This property of the
coefficients $d_{n,k}(t)$ is very important and in some cases
allows exact analytical results to be obtained.

With the coefficients $d_{n,k}(t)$ available, the resulting
state of the field~\mref{N203} can be written, for the
initial state $|n,0\rangle$, as
%---------------------------------------------------------------------------
\begin{eqnarray}
  |\psi^{(n)}(t)\rangle &=&
  \sum_{k=0}^{[n/2]}d_{n,k}(t)|\psi_{k}^{(n)}\rangle.
\label{N212}
\end{eqnarray}
The typical initial conditions for the second-harmonic
generation are: a coherent state of the fundamental mode and
the vacuum of the second-harmonic mode. The initial state of
the field can thus be written as
%---------------------------------------------------------------------------
\begin{eqnarray}
  |\psi (0) \rangle &=& \sum_{n=0}^{\infty} c_n |n,0 \rangle,
\label{N213}
\end{eqnarray}
where $c_n=b_n {\rm e}^{\I n\varphi_a}$ is the Poissonian
weighting factor~\mref{N81} of the coherent state
$|\alpha_0\rangle$ with the phase $\varphi_a={\rm
Arg}\,\alpha_0$.  With these initial conditions, the
resulting state~\mref{N203} is given by
%---------------------------------------------------------------------------
\begin{eqnarray}
  |\psi(t)\rangle&=&\sum_{n=0}^{\infty}c_n|\psi^{(n)}(t)\rangle
  \;=\;\sum_{n=0}^{\infty}c_n\sum_{k=0}^{[n/2]}d_{n,k}(t)|n-2k,k\rangle.
\label{N214}
\end{eqnarray}
Equation~\mref{N214}, describing the evolution of the system
is the starting point for a further discussion of
second-harmonic generation. If the initial state of the
fundamental mode is not a coherent state, but has a
decomposition into number states of the form~\mref{N213} with
different amplitudes $c_n$, eq.~\mref{N214} is still valid if
appropriate $c_n$'s are taken. This is true, for example, for
an initially squeezed state of the fundamental mode. The
coefficients $d_{n,k}(t) $ have been calculated numerically
to find the evolution of the field state~\mref{N203}, and
consequently, its phase properties
(\citeA{TGZ91a}~\citeY{TGZ91a,TGZ91b}, ~\citeAY{GTZ91b}).

Repeating the standard procedure of the Pegg-Barnett
formalism with the field state~\mref{N203}, the joint phase
probability distribution is obtained in the form:
%---------------------------------------------------------------------------
\begin{eqnarray}
  P(\theta_a,\theta_b) \:=\:
  {1\over{(2\pi)^2}}\Biggl|\sum_{n=0}^{\infty}
  b_n\sum_{k=0}^{[n/2]}d_{n,k}(t) \exp\{-\I [(n-2k)\theta_a +
  k\theta_b - k(2\varphi_a - \varphi_b)]\}\Biggr|^2,
\label{N215}
\end{eqnarray}
where $\theta_a$ and $\theta_b$ are continuous-phase
variables for the fundamental and second-harmonic modes, and
the phases $\varphi_a$ and $\varphi_b$ are the initial phases
with respect to which the distribution is symmetrized. It is
interesting, that formula~\mref{N215} depends, in fact, on
the difference $2\varphi_a-\varphi_b$, which reproduces the
classical phase relation for second-harmonic generation.
Classically, for the initial conditions chosen here, this
phase difference takes the value $\pi/2$, which turns out to
be a good choice to fix the phase windows in the quantum
description as well.

The evolution of the joint probability distribution
$P(\theta_a,\theta_b)$, given by eq.~\mref{N215}, is
illustrated graphically in fig.~15. At the initial stage of
the evolution the phase distribution in the $\theta_a$
direction (fundamental mode) is broadened, while a peak of
the second-harmonic mode phase starts to grow. The emergence
of the peak at $\theta_b=0$ confirms the classical relation
$2\varphi_a-\varphi_b=\pi/2$, which has been applied to fix
the phase window.  The phase distribution in the $\theta_b$
direction narrows at the beginning of the evolution, meaning
less uncertainty in the phase of the second harmonic.
However, for later times the distribution
$P(\theta_a,\theta_b) $ splits into two peaks, which
resembles the splitting of the $Q(\alpha,\alpha^*)$ function
found by~\citeAY{NM91}.  For still later times, more and more
peaks appear in the distribution $P(\theta_a,\theta_b) $, and
this distribution becomes more and more uniform, which means
randomization of the phase. The route to the random phase
distribution, however, goes through a sequence of increasing
numbers of peaks. The splitting of the joint phase
distribution can be understood if one realizes that the mean
number of photons of the second harmonic oscillates and after
reaching the maximum the second-harmonic generation becomes,
as a matter of fact, the down-conversion process which
exhibits a two-peak structure of the phase distribution in
the direction of the fundamental mode (see \S 4.5). The
appearance of new peaks may thus be interpreted as a
transition of the process from the second-harmonic to the
down-conversion regime, and vice versa. The phase variances
for both modes tend asymptotically to the value $\pi^2/3 $ of
the randomly distributed phase~\cite{GTZ91b}; however, it has
turned out that partial revivals of the phase structure can
be observed during the evolution \cite{DJ92}.

It is also interesting to study the phase distribution of the
field produced by second-harmonic generation with other than
coherent initial states of the fundamental mode. Such studies
were performed by \citeAY{TGZ91b}, showing for example that
even for a second harmonic generated by an initial number
state the joint phase probability distribution has a
modulation structure owing to the intermode correlation that
develops in the process of the evolution.

%%%%%%%%%%%%%%%%%%%%%%%%%%%%%%%%%%%%%%%%%%%%%%%%%%%%%%%%%%%%%%%%%%%%%%%%
\subsection{Parametric down conversion with quantum pump}
%%%%%%%%%%%%%%%%%%%%%%%%%%%%%%%%%%%%%%%%%%%%%%%%%%%%%%%%%%%%%%%%%%%%%%%%
%%%%%%%%%%%%%%%%%%%%%%%%%%%%%%%%%%%%%%%%%%%%%%%%%%%%%%%%%%%%%%%%%%%%%%%%%%%%
\begin{figure} % Fig. 16
\vspace*{-2cm} \hspace*{0cm} \epsfxsize=17.5cm
\epsfbox{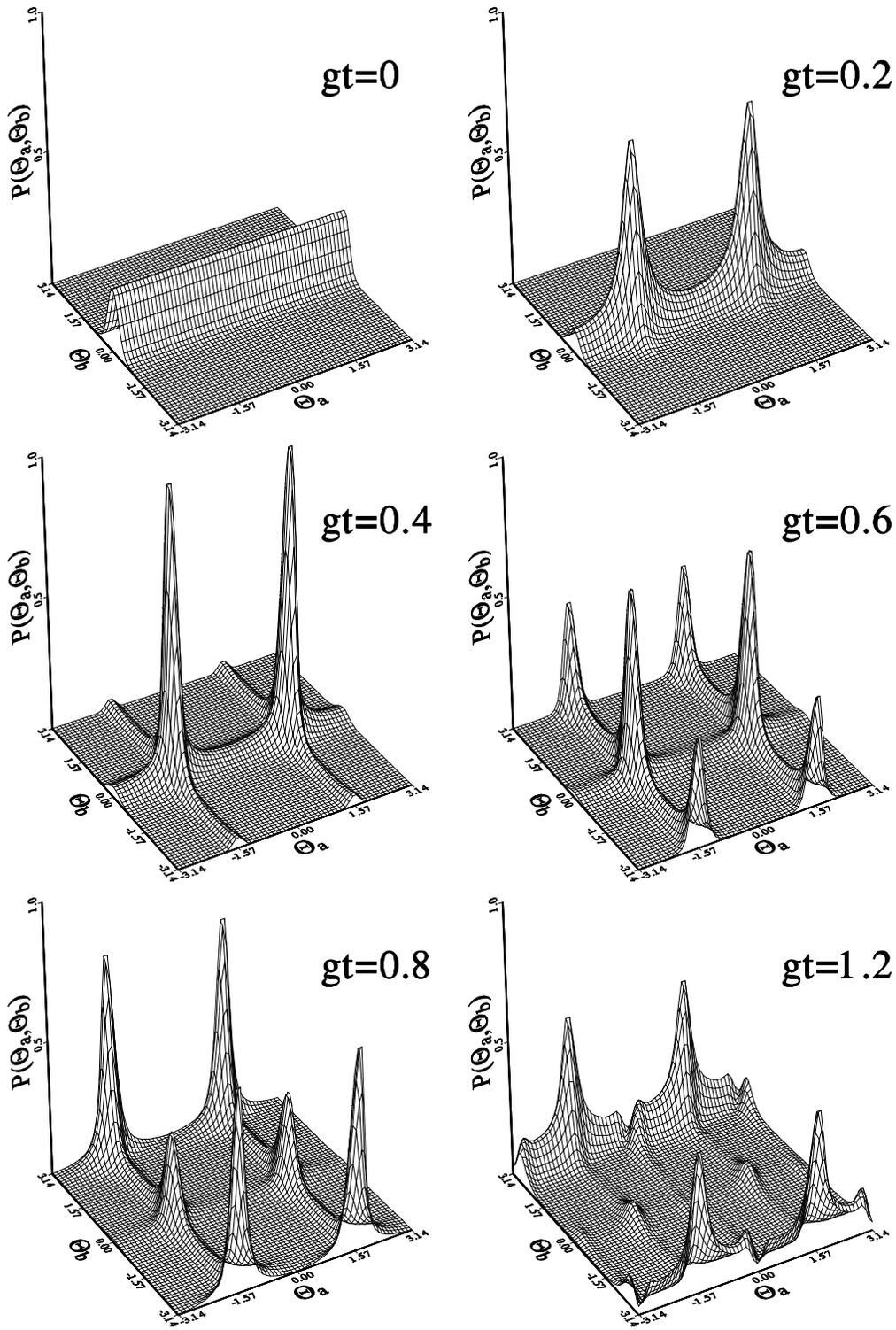} \vspace*{0cm} \caption{ Evolution of the
joint phase probability distribution $P(\theta_a,\theta_b)$,
eq. (4.76), in the process of parametric down conversion with
quantum pump. The initial mean number of photons in the mode
$b$ is $|\beta_0|^2=4$. }
\end{figure}
%%%%%%%%%%%%%%%%%%%%%%%%%%%%%%%%%%%%%%%%%%%%%%%%%%%%%%%%%%%%%%%%%%%%%%%%%%%%
\begin{figure} % Fig. 17
\vspace*{-0.5cm} \hspace*{1.5cm} \epsfxsize=11cm
\epsfbox{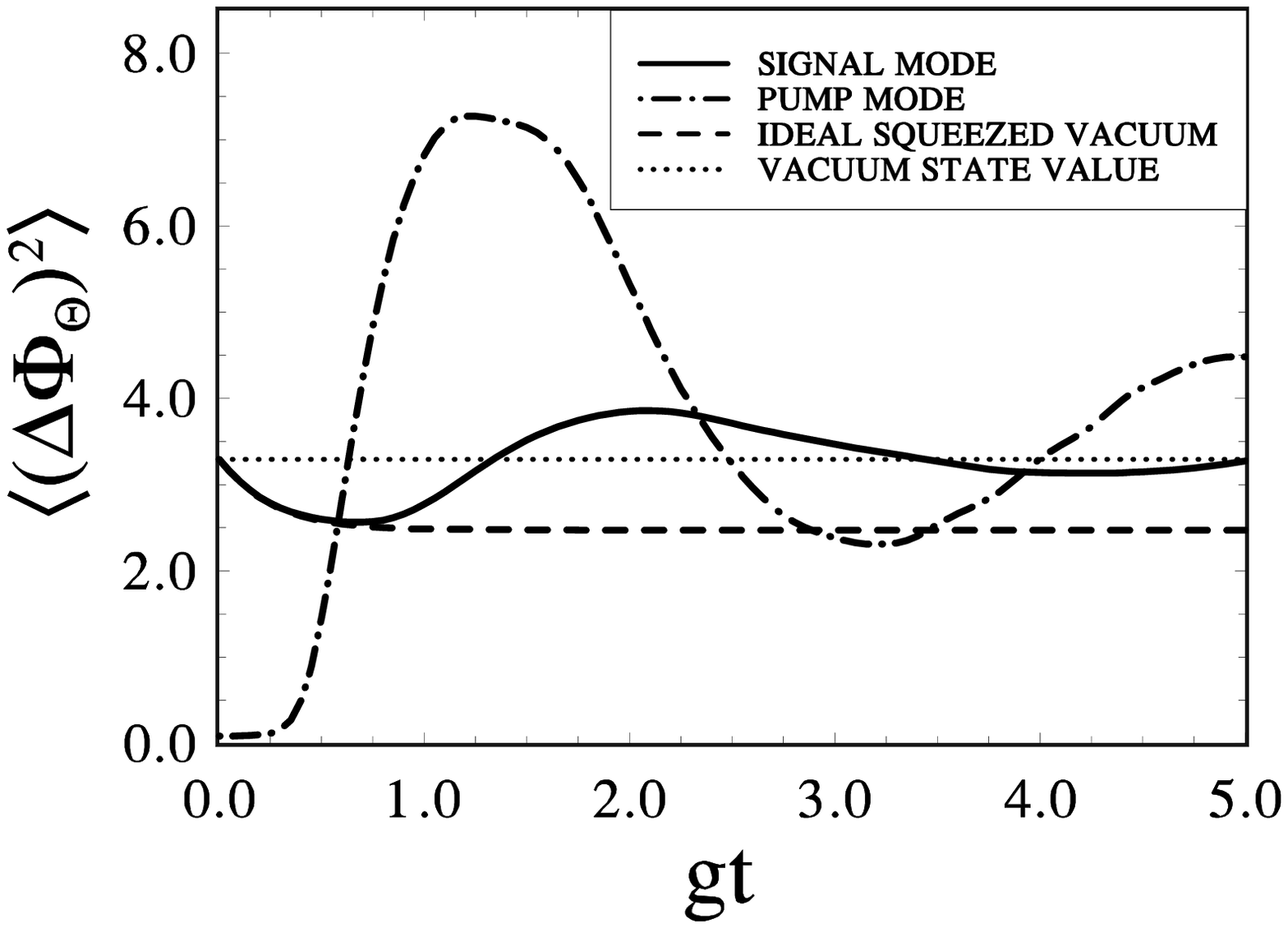} \vspace*{-5mm} \caption{ Evolution of the
phase variances
$\langle(\Delta{\hat{\Phi}_{\theta_{a}}})^2\rangle$, eq.
(4.77), and
$\langle(\Delta{\hat{\Phi}_{\theta_{b}}})^2\rangle$, eq.
(4.78), in the parametric down conversion. The initial mean
number of photons in the mode $b$ is $|\beta _0|^2=4$. }
\end{figure}

The parametric down-conversion process with quantum pump,
which is a subharmonic generation process, can be described
by the same model Hamiltonian, [eq.~\mref{N202}] as the
second-harmonic generation. The initial conditions
distinguishing the two processes are the following: For the
subharmonic generation process mode $b$ is initially
populated while mode $a$ is in the vacuum state. The
distinction between the two processes is far from trivial,
and the states generated in the two processes have quite
different properties
(\citeNP{GTZ91a,JDM92},\citeA{TG92a}~\citeY{TG92a,TG92b},
\citeNP{GTZ93}).

Let us assume, in analogy to our analysis of second-harmonic
generation, that initially there are $n$ photons in the pump
mode $(b)$ and no photons in the signal mode $(a)$; i.e., the
initial state of the field is $|0,n\rangle=|0\rangle_a
|n\rangle_b$.  Since $\hat{H}_0$ is a constant of motion, we
have the relation
%---------------------------------------------------------------------------
\begin{eqnarray}
  \langle \hat{a}^{\dagger}\hat{a}\rangle + 2\langle
  \hat{b}^{\dagger}\hat{b}\rangle &=&{\rm constant}\;=\;2n,
\label{N216}
\end{eqnarray}
which implies that the annihilation of $k$ photons of the
pump mode requires creation of $2k$ photons of the signal
mode. Thus, for given $n$, we can introduce the states
%---------------------------------------------------------------------------
\begin{eqnarray}
  |\psi_{n-k}^{(2n)}\rangle&=&|2k,n-k\rangle, \ \ \ \ \ \ \ \
  k=0,1,\ldots,n \ ,
\label{N217}
\end{eqnarray}
which should be compared to the corresponding
expression~\mref{N205} for the second-harmonic generation.

Proceeding along the same lines as in second-harmonic
generation, the resulting state of the field can be written
as
%---------------------------------------------------------------------------
\begin{eqnarray}
  |\psi(t)\rangle &=&\sum_{n=0}^{\infty}b_n\mbox{e}^{\I n\varphi_b}
  \sum_{k=0}^{n}d_{2n,k}(t)|2k,n-k\rangle,
\label{N218}
\end{eqnarray}
where the coefficients $d_{2n,k}(t)$ are given by
%---------------------------------------------------------------------------
\begin{eqnarray}
  d_{2n,k}(t)&\equiv&\langle 2k,n-k|\exp(-\I \hat{H}_{\rm I}
  t)\,|0,n\rangle,
\label{N219}
\end{eqnarray}
whereas now the $c_n=b_n {\rm e}^{\I n\varphi_b}$ are the
Poissonian weighting factors~\mref{N81} for the initially
coherent state $|\beta_0=|\beta_0|\exp{(\I\varphi_b)}\rangle$
of the mode $b$. Again, the method of numerical
diagonalization is used to calculate the coefficients
$d_{2n,k}(t) $ and, in effect, the phase properties of the
state~\mref{N218}.
%%%%%%%%%%%%%%%%%%%%%%%%%%%%%%%%%%%%%%%%%%%%%%%%%%%%%%%%%%%%%%%%%%%%%%%%%%%%
\begin{figure} % Fig. 18
\vspace*{-5mm} \hspace*{1.5cm} \epsfxsize=11cm
\epsfbox{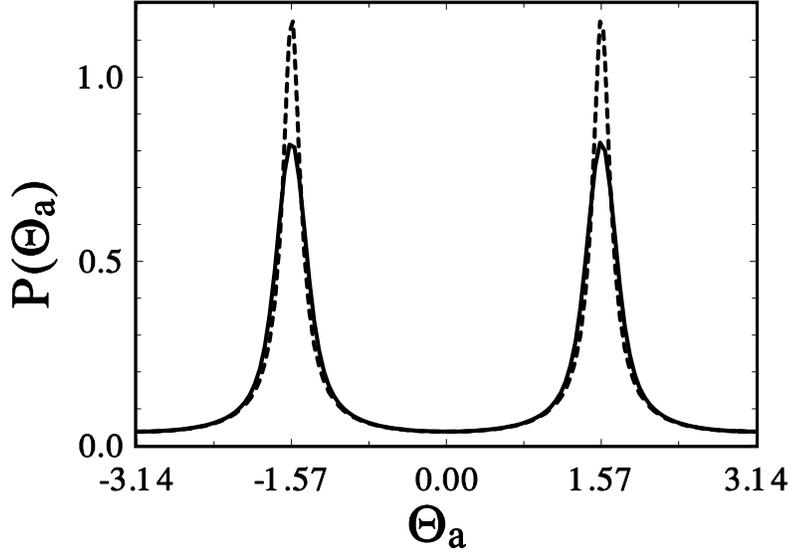} \vspace*{-5mm} \caption{ Phase
distribution $P(\theta_a)$, eq. (4.79), for $gt=0.3$, in the
parametric down conversion. }
\end{figure}

The joint phase probability distribution in this case is
given by
%---------------------------------------------------------------------------
\begin{eqnarray}
  P(\theta_a,\theta_b) \:=\:
  {1\over{(2\pi)^2}}\Biggl|\sum_{n=0}^{\infty}
  b_n\sum_{k=0}^{n}d_{2n,k}(t) \exp\{-\I [2k\theta_a +(n-k)\theta_b +
  k(2\varphi_a - \varphi_b)]\}\Biggr|^2.
\label{N220}
\end{eqnarray}
As for the second-harmonic generation, we similarly take
$2\varphi_a - \varphi_b=\pi/2$ to fix the phase windows. The
evolution of the joint probability distribution
$P(\theta_a,\theta_b) $ for parametric down conversion with
the mean number of photons $|\beta_0|^2=4$ is shown in
fig.~16. Comparison of figs.~15 and~16 shows immediately a
striking difference between the phase properties of the
fields obtained in the two processes.  The state produced in
the down-conversion process acquires from the very beginning
the two-peak structure in the $\theta_a$ direction, which
suggests the appearance of a superposition of two states in
the resulting field. The two peaks which appear at the
beginning of the evolution correspond, in fact, to the
two-peak phase distribution of the squeezed states (see \S
3.3). At later stages of the evolution randomization of the
two phases takes place, similarly as for the second-harmonic.
However, the symmetry with respect to $\theta_a$ is
preserved. The two-peak structure of the phase distribution
has already appeared, although not in its pure form, in the
phase distribution for second harmonic generation (fig.~15).
Its appearance can be ascribed to the down-conversion process
that has overcome second-harmonic generation at this stage of
the evolution. The transition from the one-peak phase
distribution to the two-peak distribution makes a qualitative
difference between the two field states, and is a sort of
``phase transition''.

Once the joint phase distribution $P(\theta_{a},\theta_{b})$
is known, all quantum-mechanical phase expectation values can
be calculated.  In particular, the phase variance for the
signal mode can be calculated according to the formula:
%---------------------------------------------------------------------------
\begin{eqnarray}
  \var{\hat{\Phi}_{\theta_{a}}}&=& \int\limits_{-\pi}^{\pi}{\rm
    d}\theta_{a}\;\theta_{a}^{2} \int\limits_{-\pi}^{\pi}\, {\rm
    d}\theta_{b}\; P(\theta_{\rm a},\theta_{b}) \nonumber\\
  &=&\frac{\pi^2}{3}+\mbox{Re}\sum_{n>n'}b_nb_{n'} \frac{\exp[-\I
    (n-n')(2\varphi_{\rm a}-\varphi_{b})]}{(n-n')^2} \nonumber\\
  &&\times\;\sum_{k=0}^{n'}d_{2n,k+n-n'}(t)d_{2n',k}^{*}(t),
\label{N221}
\end{eqnarray}
and for the pump mode we have
%---------------------------------------------------------------------------
\begin{eqnarray}
  \var{\hat{\Phi}_{\theta_{b}}}&=& \int\limits_{-\pi}^{\pi}{\rm
    d}\theta_{b}\;\theta_{b}^{2} \int\limits_{-\pi}^{\pi}\, {\rm
    d}\theta_{a}\; P(\theta_{a},\theta_{b}) \nonumber\\
  &=&\frac{\pi^2}{3}+4\mbox{Re}\sum_{n>n'}b_nb_{n'}
  \frac{(-1)^{n-n'}}{(n-n')^2}
  \sum_{k=0}^{n'}d_{2n,k}(t)d_{2n',k}^{*}(t),
\label{N222}
\end{eqnarray}
where we have used eq.~\mref{N220}, and we take $2\varphi_a -
\varphi_b=\pi/2$.  The time evolution of the phase variances
can be calculated numerically using these expressions for
given initial field states.  The dynamical behavior of the
phase variances calculated from eqs.~\mref{N221}
and~\mref{N222} is illustrated graphically in fig.~17 for
$|\beta_0|^2=4$.  The dashed line $\pi^2/3$ marks the
variance for the state with random distribution of phase.  It
is apparent that the phase variance of the signal mode starts
from the value $\pi^2/3$, dips into the minimum, and after a
few oscillations again becomes close to $\pi^2/3$. For
comparison, the phase variance for the ideal squeezed state
is also shown.  The two variances are initially
indistinguishable, but the phase variance for the squeezed
state approaches monotonically its asymptotical value
$\pi^2/4$, while for the quantum-pump case the phase variance
of the signal mode begins to oscillate at later times. This
confirms the statement that there a limit is imposed by the
quantum fluctuations of the pump on the applicability of the
ideal down-converter model. The phase variance of the pump
mode increases rapidly from its initial value for the
coherent state, and also shows oscillatory behavior
approaching the value $\pi^2/3$ at the long-time limit. Thus,
the long-time effect of the quantum fluctuations of the pump
mode is the randomization of the phase distribution for both
signal and pump modes. This randomization process is not
monotonous, and it turned out that at least partial revivals
of the phase structure are possible during the evolution
\cite{Gan92,GTZ93}.

Integrating $P(\theta_a,\theta_b)$ over one of the phases
leads to the marginal phase distributions $P(\theta_a)$ and
$P(\theta_b)$ for the phases $\theta_a$ and $\theta_b$ of the
individual modes. We have
%---------------------------------------------------------------------------
\begin{eqnarray}
  P(\theta_a) &=&{1\over{2\pi}}\Biggl\{1+2\mbox{Re}\sum_{n>n'}b_n
  b_{n'} \sum_{k=0}^{n}\sum_{k'=0}^{n'}d_{2n,k}(t)d_{2n',k'}^{*}(t)
  \nonumber\\ &&\times\;\exp\left[-\I
    (k-k')(2\theta_a+2\varphi_a-\varphi_b)\right]
  \delta_{n-n',k-k'}\Biggr\},
\label{N223}
\end{eqnarray}

%---------------------------------------------------------------------------
\begin{eqnarray}
  P(\theta_b)&=&{1\over{2\pi}}\Biggl\{1+2\mbox{Re}\sum_{n>n'} b_n
  b_{n'} \sum_{k=0}^{n'}d_{2n,k}(t)d_{2n',k}^{*}(t) \nonumber\\
  &&\times\;\exp\left[-\I (n-n')\theta_b\right] \Biggr\}.
\label{N224}
\end{eqnarray}
The phase distribution $P(\theta_a)$ for the signal mode is
shown in fig.~18 for $gt=0.3$; i.e., for the time at which
the squeezing in the signal mode has its maximum value. For
comparison we show the phase distribution for the squeezed
vacuum for $r=2|\beta_0|gt=1.2$. The effect of quantum
fluctuations of the pump is seen as the broadening of the
phase distribution with respect to that for the ideal
squeezed state.

%%%%%%%%%%%%%%%%%%%%%%%%%%%%%%%%%%%%%%%%%%%%%%%%%%%%%%%%%%%%%%%%%%%%%%%%
%                        Conclusion                                    %
%%%%%%%%%%%%%%%%%%%%%%%%%%%%%%%%%%%%%%%%%%%%%%%%%%%%%%%%%%%%%%%%%%%%%%%%
\section{Conclusion}
In this article we have reviewed some recent results
concerning the quantum-phase description of optical fields.
We have focused our attention on the real fields that can be
generated in practice in various nonlinear optical processes.
So, we rather avoided discussions of the phase formalisms as
such and tried to exploit their practical applicability in
the description of optical fields. In the description of the
phase properties we used two different, though related
formalisms: the Pegg-Barnett Hermitian phase formalism and
the formalism based on $s$-parametrized phase distributions.
The Pegg-Barnett Hermitian phase formalism is a good example
of the concept of the phase as a physical property of a
single field mode represented by a Hermitian phase operator
canonically conjugate one to the number operator. It allows
one to obtain the phase distributions for the fields, mean
values and variances of the phase, and other phase
characteristics of the field in a reasonably simple way, both
from the conceptual as well as the calculational point of
view. The phase distributions obtained from this formalism
are $2\pi$-periodic, positive definite and normalized. They
can be treated as a good representation of the quantum state
of the field and can be referred to as canonical phase
distributions.

Another description of the optical phase used by us is that
based on the $s$-parametrized quasiprobability distributions,
which can give phase distributions that can be both narrower
and broader (depending on $s$) than the Pegg-Barnett phase
distribution, but these distributions with $s\leq -1$ can be
associated with some noisy, real measurements of the phase
probability distribution and can be referred to as measured
phase distributions. Using the examples of real-field states
presented here, we tried to show the similarities and
differences that one encounters when various phase
distributions are applied to describe a particular field
state. Our choice of the field states is, of course, a bit
arbitrary, and we relied to a large extent on our own
results. We believe, however, that our review covers a number
of field states important for quantum optics, and that the
results presented here may prove interesting. We have also
attempted to give a more or less complete review of the
literature on the subject, but the subject of quantum phase
is still a ``hot'' one and the literature is growing rapidly.

%%%%%%%%%%%%%%%%%%%%%%%%%%%%%%%%%%%%%%%%%%%%%%%%%%%%%%%%%%%%%%%%%%%%%%%%
\section*{Acknowledgments}
%%%%%%%%%%%%%%%%%%%%%%%%%%%%%%%%%%%%%%%%%%%%%%%%%%%%%%%%%%%%%%%%%%%%%%%%
This work was partially supported by the Polish Committee for
Scientific Research (KBN) under the grants No. 2 P03B 128 8
and 2 P03B 188 8.  A.M. is particularly indebted to Professor
Stephen M. Barnett for his hospitality and scientific
guidance at Oxford University, and he is grateful to the
Foundation for Polish Science for the Fellowship. Ts.G. would
like to thank Professor Herbert Walther for his hospitality
at the Max-Planck-Institut f\"ur Quantenoptik, and the
Alexander von Humboldt Foundation for the Research
Fellowship.

%%%%%%%%%%%%%%%%%%%%%%%%%%%%%%%%%%%%%%%%%%%%%%%%%%%%%%%%%%%%%%%%%%%%%
%                           Appendices                              %
%%%%%%%%%%%%%%%%%%%%%%%%%%%%%%%%%%%%%%%%%%%%%%%%%%%%%%%%%%%%%%%%%%%%%
%\section*{Appendices}
%\appendix

%%%%%%%%%%%%%%%%%%%%%%%%%%%%%%%%%%%%%%%%%%%%%%%%%%%%%%%%%%%%%%%%%%%%%%%%
\section*{Appendix A. Garrison-Wong phase formalism}
%%%%%%%%%%%%%%%%%%%%%%%%%%%%%%%%%%%%%%%%%%%%%%%%%%%%%%%%%%%%%%%%%%%%%%%%
\def\theequation{A.\arabic{equation}} \setcounter{equation}{0}

\citeAY{GW70} constructed the phase operator $\hat{\Phi}_{\rm
GW}$ using the relation:
%---------------------------------------------------------------------------
\begin{eqnarray}
  \langle g | \hat{\Phi}_{\rm GW} | f \rangle &=&
  \int\limits_{\theta_{0}}^{\theta_{0} + 2\pi}\, {\rm d}\theta \,
  g^{*}({\rm e}^{-\I\theta}) \, \theta\,f({\rm e}^{-\I\theta})
\label{N239}
\end{eqnarray}
for any $g,f \in {\cal H}^{2}$, where ${\cal H}^{2}$ is the
Hilbert space in the unit disk of the complex plane, and
$\theta_{0}$ is arbitrary.  Here, we have changed the sign
with respect to the original Garrison-Wong paper and
introduced arbitrary $\theta_{0}$. The inner product in
${\cal H}^{2}$ is defined by
%---------------------------------------------------------------------------
\begin{eqnarray}
  \langle g | f \rangle &=& \int\limits_{\theta_{0}}^{\theta_{0} +
    2\pi}\, {\rm d}\theta \, g^{*}({\rm e}^{-\I \theta}) \, f({\rm
    e}^{-\I \theta}).
\label{N240}
\end{eqnarray}
The boundary value of $f$ is given by a convergent Fourier
series,
%---------------------------------------------------------------------------
\begin{eqnarray}
  f({\rm e}^{-\I \theta}) &=& \frac{1}{\sqrt{2\pi}}\sum_{n=0}^{\infty}
  c_{n} \, {\rm e}^{-\I n\theta},
\label{N241}
\end{eqnarray}
which does not contain coefficients $c_{n}$ with negative
$n$.

Subsequently,~\citeAY{PY73} established the connection of
this operator to the~\citeAY{SG64} exponential phase
operators $\hat{E}_{\pm}$ of the form:
%---------------------------------------------------------------------------
\begin{eqnarray}
  \hat{\Phi}_{\rm GW} &=& \theta_{0} \:+\: \pi \:+\: \I
  \,\left[\ln(1-{\rm e}^{\I \theta_{0}} \, \hat{E}_{+}) \:-\:
    \ln(1-{\rm e}^{-\I \theta_{0}} \, \hat{E}_{-})\right].
\label{N242}
\end{eqnarray}
The operators $\hat{E}_{-}$ and
$\hat{E}_{+}=(\hat{E}_{-})^{\dagger}$ are defined by the
annihilation and creation operators $\hat{a}$ and
$\hat{a}^{\dagger}$ of the mode:
%---------------------------------------------------------------------------
\begin{eqnarray}
  \hat{E}_{-} = (\hat{a}^{\dagger} \hat{a}+1)^{-1/2} \, \hat{a}
  ,\qquad \hat{E}_{+} = \hat{a}^{\dagger} \, (\hat{a}^{\dagger}
  \hat{a}+1)^{-1/2} ,\qquad [\hat{E}_{-},\hat{E}_{+}] = |0 \rangle
  \langle 0|,
\label{N243}
\end{eqnarray}
where $|0\rangle$ is the vacuum state [$\hat{E}_{-}$ is
another notation for the Susskind-Glogower exponential
operator~\mref{N05}].

Let us consider the ``phase states''
%---------------------------------------------------------------------------
\begin{eqnarray}
  |\theta\rangle &=& \frac{1}{\sqrt{2\pi}} \: \sum_{n=0}^{\infty}
  \exp(\I n\theta) \, |n\rangle,
\label{N244}
\end{eqnarray}
which are the right and left eigenstates of the operators
$\hat{E}_{-}$ and $\hat{E}_{+}$:
%---------------------------------------------------------------------------
\begin{eqnarray}
  \hat{E}_{-} \, |\theta\rangle &=& \exp(\I \theta) \, |\theta\rangle
  \:, \nonumber\\ \langle\theta| \, \hat{E}_{+} &=& \exp(-\I\theta)\,
  \langle\theta|.
\label{N245}
\end{eqnarray}
The states~\mref{N244} are not orthogonal but allow for the
resolution of the identity operator:
%---------------------------------------------------------------------------
\begin{eqnarray}
  \int\limits_{\theta_{0}}^{\theta_{0} + 2\pi}\, {\rm d}\theta \:
  |\theta\rangle\langle\theta| &=& \hat{1}.
\label{N246}
\end{eqnarray}
With the aid of eqs.~\mref{N245} and~\mref{N246} to the
operator, eq.~\mref{N242} can be rewritten in the
form~\cite{BE91}:
%---------------------------------------------------------------------------
\begin{eqnarray}
  \hat{\Phi}_{\rm GW} &=& \theta_{0} \:+\: \pi \:+\:
  \I\int\limits_{\theta_{0}}^{\theta_{0} + 2\pi}\, {\rm d}\theta \:
  |\theta\rangle\langle\theta| \, \left\{\ln[1-{\rm e}^{-\I
      (\theta-\theta_{0})}] \:-\: \ln[1-{\rm e}^{\I
      (\theta-\theta_{0})}]\right\} \nonumber\\ &=&\;
  \int\limits_{\theta_{0}}^{\theta_{0} + 2\pi}\, {\rm d}\theta \:
  |\theta\rangle\theta\langle\theta|.
\label{N247}
\end{eqnarray}
Since the states (A.6) are not orthogonal, they are not
eigenstates of the Garrison-Wong phase operator.

From eq.~\mref{N247}, we have
%---------------------------------------------------------------------------
\begin{eqnarray}
  \langle g | \hat{\Phi}_{\rm GW} | f \rangle &=&
  \int\limits_{\theta_{0}}^{\theta_{0} + 2\pi}\, {\rm d}\theta \:
  \langle g|\theta\rangle\theta\langle\theta|f\rangle.
\label{N248}
\end{eqnarray}
Taking the field states $|f\rangle$ in the form
\begin{eqnarray}
  |f\rangle &=& \sum_{n=0}^{\infty} c_{n} \, |n\rangle,
\label{N249}
\end{eqnarray}
we then have
%---------------------------------------------------------------------------
\begin{eqnarray}
  \langle\theta | f\rangle &=\: f({\rm e}^{-\I \theta}) &=\:
  \frac{1}{\sqrt{2\pi}}\sum_{n=0}^{\infty} c_{n} {\rm e}^{-\I
    n\theta},
\label{N250}
\end{eqnarray}
which has the same form as eq.~\mref{N241}, and we can
consider the phase operators~\mref{N239} and~\mref{N242} as
equivalent. However, we should keep in mind that the
Garrison-Wong phase operator is defined on a dense set of
state vectors, which for mathematical consistency and the
requirement that the number-phase commutator should be $-i$,
imply $f(-1)=0$. Unfortunately, when approximating even
simple physical states on this dense set, one finds rather
undesirable properties \cite{BE91}.

Since the states~\mref{N244} are not orthogonal, we have
%---------------------------------------------------------------------------
\begin{eqnarray}
  \hat{\Phi}_{\rm GW}^{k}
  &\neq\:\int\limits_{\theta_{0}}^{\theta_{0}+2\pi}\, {\rm d}\theta
  \:\theta^{k} |\theta\rangle\langle\theta| & (k>1),
\label{N251}
\end{eqnarray}
and for the expectation values
%---------------------------------------------------------------------------
\begin{eqnarray}
  \langle f| \hat{\Phi}_{\rm GW}^{k} |f\rangle &\neq\:
  \int\limits_{\theta_{0}}^{\theta_{0} + 2\pi}\, {\rm
    d}\theta\:\theta^{k} |\langle\theta|f\rangle|^{2} &(k>1).
\label{N252}
\end{eqnarray}
This means that the quantity $|\langle\theta|f\rangle|^{2}$
cannot be interpreted as a phase distribution function. To
find the Garrison-Wong phase distribution function, we must
calculate the quantity:
%---------------------------------------------------------------------------
\begin{eqnarray}
  P_{\rm GW}(\theta) &=\:\left|_{\rm
      GW}\!\langle\theta|f\rangle\right|^{2} & =\:\left|
    \sum_{n=0}^{\infty} c_{n} \:_{\rm GW}\!\langle\theta| n\rangle\
  \right|^{2},
\label{N253}
\end{eqnarray}
where the vector $|\theta\rangle_{\rm GW}$ is the eigenvector
of the Garrison-Wong phase operator. The function $_{\rm
  GW}\langle\theta|n\rangle$ has a quite complex structure
(\citeAY{GW70},~\citeA{PY73}~\citeY{PY73,PY92}), but it can
be found from the recursive formulas given by~\citeAY{GW70},
which are:
%---------------------------------------------------------------------------
\begin{eqnarray}
  _{\rm GW}\!\langle\theta|n\rangle &=&
  \left[\frac{1}{\pi}\,\sin\left(\frac{\theta - \theta_{0}}{2}\right)
  \right]^{1/2} \: \Phi_{n}(\theta),
\label{N254}
\end{eqnarray}
where, for $n\geq 1$:
%---------------------------------------------------------------------------
\begin{eqnarray}
  \Phi_{n}(\theta) &=& -\,\sum_{m=0}^{n-1} \left(1 -\frac{m}{n}\right)
  \:\gamma_{n-m}(\theta) \:\Phi_{m}(\theta),
\label{N255}
\end{eqnarray}
%---------------------------------------------------------------------------
\begin{eqnarray}
  \gamma_{n}(\theta) &=& \frac{1}{2\pi} \,
  \int\limits_{\theta_{0}}^{\theta_{0} + 2\pi}\, {\rm d}\theta' \:
  \ln|\theta' - \theta| \, {\rm e}^{\I n\theta'} \:-\:
  \frac{1}{2n}\left[{\rm e}^{\I n\theta_{0}} \:+\: {\rm e}^{\I
      n\theta}\right],
\label{N256}
\end{eqnarray}
%---------------------------------------------------------------------------
\begin{eqnarray}
  \Phi_{0}(\theta) &=& {\rm e}^{-\gamma_{0}(\theta)},
\label{N257}
\end{eqnarray}
%---------------------------------------------------------------------------
\begin{eqnarray}
  \gamma_{0}(\theta) &=& -\frac{1}{2} \:+\: \frac{1}{4\pi}
  \left[(2\pi+\theta_{0}-\theta) \, \ln(2\pi+\theta_{0} -\theta) \:+\:
    (\theta-\theta_{0}) \, \ln(\theta - \theta_{0})\right].
\label{N258}
\end{eqnarray}
The formulas~\mref{N254}--\mref{N258} were used
by~\citeAY{GMT92} to calculate the Garrison-Wong phase
distribution for some real states of the field showing that
their symmetry is incompatible with the symmetry of the phase
distributions obtained from the Pegg-Barnett as well as the
$s$-parametrized phase approaches. In the Garrison-Wong
approach even vacuum has a preferred phase, which is hardly
acceptable on physical grounds. The recursive
relation~\mref{N255} has the following solution~\cite{Mir94}:
%----------------------------------------------------------------------
\begin{eqnarray}
  \Phi_n(\theta) &=& {\rm e}^{-\gamma_0(\theta)} \sum_{\{n_i,m_i\}}
  \prod_{i=1}^{k} \frac{(-1)^{m_i}}{m_i!} \gamma_{n_i}^{m_i}(\theta),
\label{N259}
\end{eqnarray}
where the sum over $\{n_i,m_i\}$ is taken under the condition
$\sum_{i=1}^{k} n_i m_i=n$ and after integration the
functions $\gamma_n(\theta)$ [eq.~\mref{N256}] take the form:
%----------------------------------------------------------------------
\begin{eqnarray}
  \gamma_n(\theta) &=& \frac{1}{2\pi{\rm i}n} \mbox{\huge \{} {\rm
    e}^{{\rm i}n\theta_0} \left[ \ln
    \left(\frac{2\pi}{\theta-\theta_0}-1 \right) -{\rm i}\pi\right]
  \nonumber\\ &&+\;{\rm e}^{{\rm i}n\theta} \mbox{\Large (} {\rm
    Ei}[{\rm i}\,n(2\pi+\theta_0-\theta)] -{\rm Ei} [-{\rm
    i}\,n(\theta-\theta_0)] \mbox{\Large )} \mbox{\huge \}}
\label{N260}
\end{eqnarray}
in terms of the exponential integral Ei($x$).
Equations~\mref{N259}--\mref{N260} are more convenient for
numerical calculations than eqs.~\mref{N255}--\mref{N256}.

Substituting eq.~\mref{N244} into eq.~\mref{N247} and
performing the integration over $\theta$ yields the following
number-states expansion for the Garrison-Wong phase operator
[compare to eq.~\mref{N14}]:
%---------------------------------------------------------------------------
\begin{eqnarray}
  \hat{\Phi}_{\rm GW} &=& \theta_{0} \:+\: \pi \:+\: \sum_{n\neq
    n'}^{} \frac{\exp[\I(n-n')\theta_{0}]\, |n\rangle \langle n'|}{\I
    \,(n-n')},
\label{N261}
\end{eqnarray}
leading to the number phase commutator
%---------------------------------------------------------------------------
\begin{eqnarray}
  [\hat{\Phi}_{\rm GW},\hat{a}\plus \hat{a}] &=& -\I\, (1 -
  2\pi\,|\theta_{0}\rangle\langle\theta_{0}|),
\label{N262}
\end{eqnarray}
and for the states for which $\langle\theta_{0}|f\rangle =0$,
the second term on the right-hand side vanishes, giving the
value demanded by~\citeAY{GW70}. A detailed comparison of the
Garrison-Wong and Pegg-Barnett formalisms was given
by~\citeAY{BP92} and by ~\citeAY{GMT92}. The difference
between the two formalisms is, in mathematical sense, the
difference between the weak and strong limits for the phase
operators that is taken when
$\sigma\rightarrow\infty$~\cite{VP93}.

%%%%%%%%%%%%%%%%%%%%%%%%%%%%%%%%%%%%%%%%%%%%%%%%%%%%%%%%%%%%%%%%%%%%%%%%
\section*{Appendix B. States for the Pegg-Barnett Phase Formalism}
%%%%%%%%%%%%%%%%%%%%%%%%%%%%%%%%%%%%%%%%%%%%%%%%%%%%%%%%%%%%%%%%%%%%%%%%
\def\theequation{B.\arabic{equation}} \setcounter{equation}{0}

The Pegg-Barnett optical phase operator~\mref{N12} is
constructed in a finite ($\sigma+1$)-dimensional Hilbert
space \Hs spanned by the number states $|0\rangle,
|1\rangle,\ldots,|\sigma\rangle$. Hence, all other
quantities, such as states, operators or probability
distributions, analyzed within the Pegg-Barnett formalism,
should also be defined in the same ($\sigma+1$)-dimensional
state space \Hs. \citeAY{BWKL92} emphasized that it is not
strictly correct to apply the definition~\mref{N23} of the
finite-dimensional phase distribution
%----------------------------------------------------------------------
\begin{eqnarray}
\label{N225} P(\theta_m) &=&
|_{(\sigma)}\langle\theta_m|f\rangle|^2 \ \ \ \ \ \ {\rm
(wrong)},
\end{eqnarray}
for the infinite-dimensional state
%----------------------------------------------------------------------
\begin{eqnarray}
  |f\rangle \;\equiv\; |f\rangle_{(\infty)} &=& \sum_{n=0}^{\infty}
  c_n |n\rangle.
\label{N226}
\end{eqnarray}
The problem of the precise definition of states in \Hs can be
overcome by assuming that $\sigma$ is large enough so that
the differences between the states in the finite, \Hs, and
infinite-dimensional, $\cal H$, spaces can be arbitrarily
small in the sense of the Cauchy condition~\cite{PB89}:
%----------------------------------------------------------------------
\begin{eqnarray}
  {\forall\atop\epsilon} \ \ \ {\exists\atop\sigma} \ \ \ \ \
  1-\sum_{n=0}^{\sigma} |c_n|^2 &<& \epsilon.
\label{N227}
\end{eqnarray}

The precise finite-dimensional phase distribution reads as
follows~\cite{BWKL92}:
%----------------------------------------------------------------------
\begin{eqnarray}
\label{N228} P(\theta_m) &=&
|_{(\sigma)}\langle\theta_m|f\rangle_{(\sigma)}|^2
\end{eqnarray}
for the ($\sigma+1$)-dimensional state
%----------------------------------------------------------------------
\begin{eqnarray}
  |f\rangle_{(\sigma)} &=& \sum_{n=0}^{\sigma} c^{(\sigma)}_n
  |n\rangle,
\label{N229}
\end{eqnarray}
which is properly normalized,
%----------------------------------------------------------------------
\begin{eqnarray}
  _{(\sigma)}\langle f|f\rangle_{(\sigma)} &=& \sum_{n=0}^{\sigma}
  |c_n^{(\sigma)}|^2 \;=\; 1
\label{N230}
\end{eqnarray}
for arbitrary $\sigma$. The main problem resides in the
construction of the normalized ($\sigma+1$)-dimensional
states $|f\rangle_{(\sigma)}$.  We restrict our attention to
finite-dimensional coherent states only.  However, other
finite-dimensional states of the electromagnetic field can be
defined in a similar manner; e.g., squeezed
states~\cite{BWKL92}, even and odd coherent
states~\cite{ZK94}, phase coherent states (\citeA{KC94b}
\citeY{KC94b,KC94c},~\citeAY{Gan94}) and displaced phase
states~\cite{Gan94}.

There exist several generalizations of coherent states
comprising the finite-dimensional case (see \citeAY{ZFG90}
and references therein). It is possible to define coherent
states using the concept of Lie group representations (see,
e.g., \citeAY{PHJ94}), or to postulate the validity of some
properties of the infinite-dimensional Hilbert-space coherent
states for the finite-dimensional coherent states. We present
two definitions of the latter case. Firstly, the coherent
states \cs in $(\sigma +1)$-dimensional Hilbert space of a
harmonic oscillator can be defined in the Glauber sense by
the action of an analogue of the Glauber displacement
operator ${\hat D}^{(\sigma
  )}(\alpha)$ on the vacuum state $|0\rangle$~\cite{BWKL92}:
%----------------------------------------------------------------------
\begin{eqnarray}
\label{N231} |\alpha\rangle_{(\sigma)} &=& {\hat D}^{(\sigma
)}(\alpha)|0\rangle \;\equiv\; \exp(\alpha \hat a\plus -
\alpha ^{*} \hat a) |0\rangle.
\end{eqnarray}
The operator ${\hat D}^{(\sigma )}(\alpha )$ is given in
terms of the modified annihilation operator:
%----------------------------------------------------------------------
\begin{eqnarray}
\label{N232}
\hat{a} &=& \exp({\rm i} \hat{\Phi}_\theta) \sqrt{\hat{N}} \nonumber\\
&=& |0\rangle\langle 1|+ \sqrt{2}|1\rangle\langle 2|+ \cdots
+\sqrt{\sigma}|\sigma-1\rangle\langle \sigma|
\end{eqnarray}
and modified creation operator $\hat{a}\plus$.  The coherent
states \cs are close {\em analogues} of Glauber's (i.e.,
infinite-dimensional) coherent states $\ket{\alpha}$. They
were introduced and discussed by \citeAY{BWKL92}, and their
analytical Fock expansion was found by \citeAY{MPT94} in the
form [eq.~\mref{N229}] $|f\rangle=|\alpha\rangle$, with the
superposition coefficients:
%----------------------------------------------------------------------
\begin{eqnarray}
\label{N233} c_n^{(\sigma)} &=& \frac{\exp[\I
n(\theta-\pi/2)]} {\sqrt{n!}} \frac{\sigma!}{\sigma+1}\,
\sum_{k=0}^{\sigma} \e^{\I x_k|\alpha|}\, {\rm He}_n(x_k)
{\rm He}_{\sigma}^{-2}(x_k).
\end{eqnarray}
Here, $x_{l} \equiv x_{l}^{(\sigma+1)}$ are the roots of the
modified Hermite polynomial of order $(\sigma+1)$,
$\mbox{He}_{\sigma+1}(x_{l}) = 0$, $\mbox{He}_{n}(x) \equiv
2^{-n/2}\mbox{H}_{n}(x/\sqrt{2})$, and
$\alpha=|\alpha|\exp(\I\theta)$.

\citeA{KWZ93} \citeY{KWZ93,KWZ94} defined the normalized
finite-dimensional coherent states in another manner by
truncating the Fock-basis expansion of the Glauber
infinite-dimensional coherent states or, equivalently, by the
action of the formally designed ``displacement'' operator
$\exp(\ba{\hat
  a}\plus)\exp(-\ba{\hat a})$ on the vacuum state.  This approach is
close to that of~\citeAY{VP90b} in the construction of a
finite-dimensional Wigner function for coherent states.  The
states \tcs can be defined as follows~\cite{KWZ93}:
%----------------------------------------------------------------------
\begin{eqnarray}
\label{N234} \ket{\ba}_{(\sigma)} &=& {\cal N}^{(\sigma)}
\exp(\ba{\hat a}\plus) \ket{0} \;=\; \sum_{n=0}^{\sigma}
c_n^{(\sigma)} \ket{n},
\end{eqnarray}
where
%----------------------------------------------------------------------
\begin{eqnarray}
\label{N235} c_{n}^{(\sigma)} &=& {\cal N}^{(\sigma)}
\frac{\ba^{n}}{\sqrt{n!}}.
\end{eqnarray}
and the normalization constant is~\cite{OMB96}:
%----------------------------------------------------------------------
\begin{eqnarray}
\label{N236} {\cal N}^{(\sigma)} &=& \left\{
(-1)^{\sigma}{\rm
    L}^{-\sigma-1}_{\sigma}(|\ba|^2) \right\}^{-1/2}
\end{eqnarray}
in terms of generalized Laguerre polynomials
L$^n_{\sigma}(x)$.

The differences between the finite-dimensional coherent
states~\mref{N231} and~\mref{N234} were discussed in detail
by~\citeAY{OMB96} using the finite-dimensional Wigner
function~\cite{Woo87,VP90b} and in terms of the Stokes
parameters.

In the limit $\sigma\rightarrow\infty$, the coherent states
\cs and \tcs go over into ($\alpha=\ba$):
%----------------------------------------------------------------------
\begin{eqnarray}
  \lim\limits_{\sigma\rightarrow\infty} |\alpha\rangle_{(\sigma)} &=&
  \lim\limits_{\sigma\rightarrow\infty} |\ba\rangle_{(\sigma)} \;=\;
  |\alpha\rangle,
\label{N237}
\end{eqnarray}
as was shown analytically by~\citeAY{OMB96}. However, the
states \cs and \tcs are essentially different, particularly
for $|\alpha|^2,|\ba|^2\ge\sigma$, from the ordinary
(infinite-dimensional) Glauber coherent states
$|\alpha\rangle$ as revealed by their photon-number,
squeezing and phase properties (\citeAY{BWKL92},
\citeA{KWZ93} \citeY{KWZ93,KWZ94},~\citeAY{MPT94}). Let us
only mention that the well-known property of the ordinary
coherent state $|\alpha\rangle$ for the mean photon-number is
not fulfilled in the case of the finite-dimensional coherent
states:
%----------------------------------------------------------------------
\begin{eqnarray}
  \left.
\begin{array}{ll}
  _{(\sigma)}\langle\alpha|\hat{n}|\alpha\rangle_{(\sigma)}\\
  _{(\sigma)}\langle\ba|\hat{n}|\ba\rangle_{(\sigma)}
\end{array}
\right\} &\neq& \langle\alpha|\hat{n}|\alpha\rangle \;=\;
|\alpha|^2. \label{N238}
\end{eqnarray}

The finite-dimensional states discussed here are not only
mathematical structures. A framework for their physical
interpretation is provided by cavity quantum electrodynamics
and atomic physics. Moreover, they can be generated, e.g., in
a single-mode resonator.  Several methods have been proposed
for the preparation of an arbitrary field state
(e.g.,~\citeAY{VAS93},~\citeAY{GSMKK94} and references
therein), which can be readily applied to prepare these
finite-dimensional states.  Recently, \citeAY{LT94} have
presented a scheme of field generation in a cavity containing
a nonlinear Kerr medium,kicked periodically with classical
pulses. The field generated in this process is actually the
finite-dimensional coherent state \cs in Hilbert space \Hs of
arbitrary dimension.

% \small \bibliographystyle{progopt} \bibliography{phase}

\begin{thebibliography}{}

\bibitem[\protect\citeauthoryear{{Adam, Janszky and Vinogradov}}{Adam
et~al.}{1991}]{AJV91} {Adam, P., J.~Janszky and A.V.
Vinogradov}, 1991, Amplitude squeezed and number-phase
intelligent states via coherent state superposition, {Phys.
Lett. A\/}~{\bf 160}, 506.

\bibitem[\protect\citeauthoryear{{Agarwal}}{Agarwal}{1986}]{Aga86}
{Agarwal, G.S.}, 1986, Generation of pair coherent states and
squeezing via the competition of four-wave mixing and
amplified spontaneous emission, {Phys. Rev. Lett.\/}~{\bf
57}, 827.

\bibitem[\protect\citeauthoryear{{Agarwal}}{Agarwal}{1988}]{Aga88}
{Agarwal, G.S.}, 1988, Nonclassical statistics of fields in
pair coherent states, {J. Opt. Soc. Am. B\/}~{\bf 5}, 1940.

\bibitem[\protect\citeauthoryear{{Agarwal}}{Agarwal}{1993}]{Aga93}
{Agarwal, G.S.}, 1993, Eigenstates of the phase operator \`a
la {D}irac for a two-mode field, {Opt. Commun.\/}~{\bf 100},
479.

\bibitem[\protect\citeauthoryear{{Agarwal, Chaturvedi, Tara and
Srinivasan}}{Agarwal et~al.}{1992}]{ACTS92} {Agarwal, G.S.,
S.~Chaturvedi, K.~Tara and V.~Srinivasan}, 1992, Classical
phase changes in nonlinear processes and their quantum
counterparts, {Phys. Rev. A\/}~{\bf 45}, 4904.

\bibitem[\protect\citeauthoryear{{Agarwal, Graf, Orszag, Scully and
Walther}}{Agarwal et~al.}{1994}]{AGOSW94} {Agarwal, G.S.,
M.~Graf, M.~Orszag, M.O. Scully and H.~Walther}, 1994, State
preparation via quantum coherence and continuous measurement,
{Phys. Rev. A\/}~{\bf 49}, 4077.

\bibitem[\protect\citeauthoryear{{Agarwal and Puri}}{Agarwal and
Puri}{1989}]{AP89a} {Agarwal, G.S., and R.R. Puri}, 1989,
Quantum theory of propagation of elliptically polarized light
through a {K}err medium, {Phys. Rev. A\/}~{\bf 40}, 5179.

\bibitem[\protect\citeauthoryear{{Agarwal, Scully and
Walther}}{Agarwal et~al.}{1993}]{ASW93} {Agarwal, G.S., M.O.
Scully and H.~Walther}, 1993, Phase narrowing a coherent
state via repeated measures: {O}nly the no counts count,
{Phys. Scripta\/}~{\bf T 48}, 128.

\bibitem[\protect\citeauthoryear{{Averbukh and Perelman}}{Averbukh
and Perelman}{1989}]{AP89} {Averbukh, J.S. and N.F.
Perelman}, 1989, Fractional revivals: {U}niversality in the
long-term evolution of quantum wave packets beyond the
correspondence principle dynamics, {Phys. Lett. A\/}~{\bf
139}, 449.

\bibitem[\protect\citeauthoryear{{Ban}}{Ban}{1991a}]{Ban91a}
{Ban, M.}, 1991a, Number-phase quantization in ultra-small
tunnel junctions, {Phys. Lett. A\/}~{\bf 152}, 223.

\bibitem[\protect\citeauthoryear{{Ban}}{Ban}{1991b}]{Ban91b}
{Ban, M.}, 1991b, Phase operator and its eigenstate in
{L}iouville space, {Phys. Lett. A\/}~{\bf 155}, 397.

\bibitem[\protect\citeauthoryear{{Ban}}{Ban}{1991c}]{Ban91c}
{Ban, M.}, 1991c, Relative number state representation and
phase operator for Phys.l systems, {J. Math. Phys.\/}~{\bf
32}, 3077.

\bibitem[\protect\citeauthoryear{{Ban}}{Ban}{1992}]{Ban92}
{Ban, M.}, 1992, Phase operator formalism for a two-mode
photon system, {J. Opt. Soc. Am. B\/}~{\bf 9}, 1189.

\bibitem[\protect\citeauthoryear{{Ban}}{Ban}{1993}]{Ban93a}
{Ban, M.}, 1993, Phase operator in quantum optics, {Phys.
Lett. A\/}~{\bf 176}, 47.

\bibitem[\protect\citeauthoryear{{Bandilla}}{Bandilla}{1991}]{Ban91}
{Bandilla, A.}, 1991, The broadening of the phase
distribution due to linear amplification, {Opt.
Commun.\/}~{\bf 80}, 267.

\bibitem[\protect\citeauthoryear{{Bandilla}}{Bandilla}{1992}]{Ban92c}
{Bandilla, A.}, 1992, How to realize phase optimized quantum
states, {Opt. Commun.\/}~{\bf 94}, 273.

\bibitem[\protect\citeauthoryear{{Bandilla}}{Bandilla}{1993}]{Ban93}
{Bandilla, A.}, 1993, Strong local oscillator limit of the
operational approach for quantum phase measurement, {Phys.
Scripta\/}~{\bf T 48}, 49.

\bibitem[\protect\citeauthoryear{{Bandilla and Paul}}{Bandilla and
Paul}{1969}]{BP69} {Bandilla, A., and H.~Paul}, 1969,
Laser-verst\"arker und {P}hasenunsch\"arfe, {Ann.
Physik\/}~{\bf 23}, 323.

\bibitem[\protect\citeauthoryear{{Bandilla and Ritze}}{Bandilla and
Ritze}{1993}]{BR93} {Bandilla, A., and H.H. Ritze}, 1993,
Realistic phase distributions derived from the {W}igner
function, {Quantum Opt.\/}~{\bf 5}, 213.

\bibitem[\protect\citeauthoryear{{Barnett and Dalton}}{Barnett and
Dalton}{1993}]{BD93} {Barnett, S.M., and B.J. Dalton}, 1993,
Conceptions of quantum optical phase, {Phys. Scripta\/}~{\bf
T 48}, 13.

\bibitem[\protect\citeauthoryear{{Barnett and Pegg}}{Barnett and
Pegg}{1986}]{BP86} {Barnett, S.M., and D.T. Pegg}, 1986,
Phase in quantum optics, {J. Phys. A\/}~{\bf 19}, 3849.

\bibitem[\protect\citeauthoryear{{Barnett and Pegg}}{Barnett and
Pegg}{1989}]{BP89} {Barnett, S.M., and D.T. Pegg}, 1989, On
the {H}ermitian optical phase operator, {J. Mod. Opt.\/}~{\bf
36}, 7.

\bibitem[\protect\citeauthoryear{{Barnett and Pegg}}{Barnett and
Pegg}{1990}]{BP90} {Barnett, S.M., and D.T. Pegg}, 1990,
Quantum theory of optical phase correlations, {Phys. Rev.
A\/}~{\bf 42}, 6713.

\bibitem[\protect\citeauthoryear{{Barnett and Pegg}}{Barnett and
Pegg}{1992}]{BP92} {Barnett, S.M., and D.T. Pegg}, 1992,
Limiting procedures for the optical phase operator, {J. Mod.
Opt.\/}~{\bf 39}, 2121.

\bibitem[\protect\citeauthoryear{{Barnett and Pegg}}{Barnett and
Pegg}{1993}]{BP93} {Barnett, S.M., and D.T. Pegg}, 1993,
Phase measurements, {Phys. Rev. A\/}~{\bf 47}, 4537.

\bibitem[\protect\citeauthoryear{{Barnett, Pegg and Vaccaro}}{Barnett
et~al.}{1990}]{BPV90} {Barnett, S.M., D.T. Pegg and J.A.
Vaccaro}, 1990, Applications of the optical phase operator,
in: Coherence and Quantum Optics VI, eds. J.H. Eberly,
L.~Mandel and E.~Wolf (Plenum Press, New York), p.\ ~89.

\bibitem[\protect\citeauthoryear{{Barnett, Stenholm and
Pegg}}{Barnett et~al.}{1989}]{BSP89} {Barnett, S.M.,
S.~Stenholm and D.T. Pegg}, 1989, A new approach to optical
phase diffusion, {Opt. Commun.\/}~{\bf 73}, 314.

\bibitem[\protect\citeauthoryear{{Beck, Smithey, Cooper and
Raymer}}{Beck et~al.}{1993}]{BSCR93} {Beck, M., D.T. Smithey,
J.~Cooper and M.G. Raymer}, 1993, Experimental determination
of number-phase uncertainty relations, {Opt. Lett.\/}~{\bf
18}, 1259.

\bibitem[\protect\citeauthoryear{{Beck, Smithey and Raymer}}{Beck
et~al.}{1993}]{BSR93} {Beck, M., D.T. Smithey and M.G.
Raymer}, 1993, Experimental determination of quantum phase
distributions using optical homodyne tomography, {Phys. Rev.
A\/}~{\bf 48}, R890.

\bibitem[\protect\citeauthoryear{{Belavkin and
Bendjaballah}}{Belavkin and Bendjaballah}{1994}]{BB94a}
{Belavkin, V.P., and C.~Bendjaballah}, 1994, Continuous
measurements of quantum phase, {Quantum Opt.\/}~{\bf 6}, 169.

\bibitem[\protect\citeauthoryear{{Bergou and Englert}}{Bergou and
Englert}{1991}]{BE91} {Bergou, J., and B.G. Englert}, 1991,
Operators of the phase: {F}undamentals, {Ann. Phys. (New
York)\/}~{\bf 209}, 479.

\bibitem[\protect\citeauthoryear{{Bia\l{}ynicka-Birula}}{Bia\l{}ynicka-Birula}{1968}]{Bia68}
{Bia\l{}ynicka-Birula, Z.}, 1968, Properties of generalized
coherent states, {Phys. Rev.\/}~{\bf 173}, 1207.

\bibitem[\protect\citeauthoryear{{Bia\l{}ynicka-Birula and
Bia\l{}ynicki-Birula}}{Bia\l{}ynicka-Birula and
Bia\l{}ynicki-Birula}{1994}]{BBB94} {Bia\l{}ynicka-Birula,
Z., and I.~Bia\l{}ynicki-Birula}, 1994, Reconstruction of the
wavefunction from the photon number and quantum phase
distributions, {J. Mod. Opt.\/}~{\bf 41}, 2203.

\bibitem[\protect\citeauthoryear{{Bia\l{}ynicka-Birula and
Bia\l{}ynicki-Birula}}{Bia\l{}ynicka-Birula and
Bia\l{}ynicki-Birula}{1995}]{BBB95} {Bia\l{}ynicka-Birula,
Z., and I.~Bia\l{}ynicki-Birula}, 1995, On the measurability
of the quantum phase distribution, {Appl. Phys. B\/}~{\bf
60}, 275.

\bibitem[\protect\citeauthoryear{{Bia\l{}ynicki-Birula, Freyberger and
Schleich}}{Bia\l{}ynicki-Birula et~al.}{1993}]{BFS93}
{Bia\l{}ynicki-Birula, I., M.~Freyberger and W.~Schleich},
1993, Various measures of quantum phase uncertainty: {A}
comparative study, {Phys. Scripta\/}~{\bf T 48}, 113.

\bibitem[\protect\citeauthoryear{{Braunstein}}{Braunstein}{1992}]{Bra92}
{Braunstein, S.L.}, 1992, Quantum limits on precision
measurement of phase, {Phys. Rev. Lett.\/}~{\bf 69}, 3598.

\bibitem[\protect\citeauthoryear{{Braunstein and Caves}}{Braunstein
and Caves}{1990}]{BC90} {Braunstein, S.L., and C.M. Caves},
1990, Phase and homodyne statistics of generalized squeezed
states, {Phys. Rev. A\/}~{\bf 42}, 4115.

\bibitem[\protect\citeauthoryear{{Braunstein, Lane and
Caves}}{Braunstein et~al.}{1992}]{BLC92} {Braunstein, S.L.,
A.S. Lane and C.M. Caves}, 1992, Maximum likelihood analysis
of multiple quantum phase measurements, {Phys. Rev.
Lett.\/}~{\bf 69}, 2153.

\bibitem[\protect\citeauthoryear{{Brif and Ben-Aryeh}}{Brif and
Ben-Aryeh}{1994a}]{BB94} {Brif, C., and Y.~Ben-Aryeh}, 1994a,
Antinormal ordering of {S}usskind-{G}logower quantum phase
operators, {Phys. Rev. A\/}~{\bf 50}, 2727.

\bibitem[\protect\citeauthoryear{{Brif and Ben-Aryeh}}{Brif and
Ben-Aryeh}{1994b}]{BB94b} {Brif, C., and Y.~Ben-Aryeh},
1994b, Phase-state representation in quantum optics, {Phys.
Rev. A\/}~{\bf 50}, 3505.

\bibitem[\protect\citeauthoryear{{Brune, Haroche, Raimond, Davidovich and Zagury}}{Brune et~al.}{1992}]{BHRDZ92} {Brune, M.,
S.~Haroche, J.M. Raimond, L.~Davidovich and N.~Zagury}, 1992,
Manipulation of photons in a cavity by dispersive atom-field
coupling: {Q}uantum-nondemolition measurements and generation
of ``{S}chr\"odinger cat'' states, {Phys. Rev. A\/}~{\bf 45},
5193.

\bibitem[\protect\citeauthoryear{{Burak and W\'odkiewicz}}{Burak and
W\'odkiewicz}{1992}]{BW92} {Burak, D., and K.~W\'odkiewicz},
1992, Phase properties of quantum states of light, {Phys.
Rev. A\/}~{\bf 46}, 2744.

\bibitem[\protect\citeauthoryear{{Bu\v{z}ek}}{Bu\v{z}ek}{1993}]{Buz93}
{Bu\v{z}ek, V.}, 1993, Phase properties of {s}chr\"odinger
cats, Habilitation Thesis (Slovak Academy of Sciences,
Bratislava).

\bibitem[\protect\citeauthoryear{{Bu\v{z}ek, Gantsog and
Kim}}{Bu\v{z}ek et~al.}{1993}]{BGK93} {Bu\v{z}ek, V.,
Ts.~Gantsog and M.S. Kim}, 1993, Phase properties of
{S}chr\"odinger cat states of light decaying in
phase-sensitive reservoirs, {Phys. Scripta\/}~{\bf T 48},
131.

\bibitem[\protect\citeauthoryear{{Bu\v{z}ek, Kim and
Gantsog}}{Bu\v{z}ek et~al.}{1993}]{BKG93} {Bu\v{z}ek, V.,
M.S. Kim and Ts.~Gantsog}, 1993, Quantum phase distributions
of amplified {S}chr\"odinger-cat states of light, {Phys. Rev.
A\/}~{\bf 48}, 3394.

\bibitem[\protect\citeauthoryear{{Bu\v{z}ek and Knight}}{Bu\v{z}ek
and Knight}{1995}]{BK95} {Bu\v{z}ek, V., and P.L. Knight},
1995, Quantum interference, superposition states of light and
nonclassical effects, in: {Progress in Optics\/}, Vol. XXXIV,
ed. E. Wolf (North-Holland, Amsterdam) p. 1.

\bibitem[\protect\citeauthoryear{{Bu\v{z}ek, Wilson-Gordon, Knight and Lai}}{Bu\v{z}ek et~al.}{1992}]{BWKL92} {Bu\v{z}ek, V.,
A.D. Wilson-Gordon, P.L. Knight and W.K. Lai}, 1992, Coherent
states in a finite-dimensional basis: {T}heir phase
properties and relationship to coherent states of light,
{Phys. Rev. A\/}~{\bf 45}, 8079.

\bibitem[\protect\citeauthoryear{{Cahill and Glauber}}{Cahill and
Glauber}{1969a}]{CG69a} {Cahill, K.E., and R.J. Glauber},
1969a, ~{O}rdered expansions in boson amplitude operators,
{Phys. Rev.\/}~{\bf 177}, 1857.

\bibitem[\protect\citeauthoryear{{Cahill and Glauber}}{Cahill and
Glauber}{1969b}]{CG69b} {Cahill, K.E., and R.J. Glauber},
1969b, Density operators and quasiprobability distributions,
{Phys. Rev.\/}~{\bf 177}, 1882.

\bibitem[\protect\citeauthoryear{{Carruthers and Nieto}}{Carruthers
and Nieto}{1968}]{CN68} {Carruthers, P., and M.~Nieto}, 1968,
Phase and angle variables in quantum mechanics, {Rev. Mod.
Phys.\/}~{\bf 40}, 411.

\bibitem[\protect\citeauthoryear{{Caves and Schumaker}}{Caves and
Schumaker}{1985}]{CS85} {Caves, C.M., and B.L. Schumaker},
1985, New formalism for two-photon optics: {I.} {Q}uadrature
phases and squeezed states, {Phys. Rev. A\/}~{\bf 31}, 3068.

\bibitem[\protect\citeauthoryear{{Chaichian and Ellinas}}{Chaichian
and Ellinas}{1990}]{CE90} {Chaichian, M., and D.~Ellinas},
1990, On the polar decomposition of the quantum ${SU}(2)$
algebra, {J. Phys. A\/}~{\bf 23}, L291.

\bibitem[\protect\citeauthoryear{{Chizhov, Gantsog and
Murzakhmetov}}{Chizhov et~al.}{1993}]{CGM93} {Chizhov, A.V.,
Ts.~Gantsog and B.K. Murzakhmetov}, 1993, Phase distributions
of squeezed number states and squeezed thermal states,
{Quantum Opt.\/}~{\bf 5}, 85.

\bibitem[\protect\citeauthoryear{{Chizhov and Murzakhmetov}}{Chizhov
and Murzakhmetov}{1993}]{CM93} {Chizhov, A.V., and B.K.
Murzakhmetov}, 1993, Photon statistics and phase properties
of two-mode squeezed number states, {Phys. Lett. A\/}~{\bf
176}, 33.

\bibitem[\protect\citeauthoryear{{Cibils, Cuche, Marvulle and
Wreszinski}}{Cibils et~al.}{1991}]{CCMW91} {Cibils, M.B.,
Y.~Cuche, V.~Marvulle and W.F. Wreszinski}, 1991, Connection
between the {P}egg-{B}arnett and the {B}ialynicki-{B}irula
phase operators, {Phys. Rev. A\/}~{\bf 43}, 4044.

\bibitem[\protect\citeauthoryear{{Cohen, Ben-Aryeh and Mann}}{Cohen
et~al.}{1992}]{CBM92} {Cohen, D., Y.~Ben-Aryeh and A.~Mann},
1992, Phase variance of squeezed states, {Opt.
Commun.\/}~{\bf 94}, 227.

\bibitem[\protect\citeauthoryear{{Collett}}{Collett}{1993a}]{Col93}
{Collett, M.J.}, 1993a, Generation of number-phase squeezed
states, {Phys. Rev. Lett.\/}~{\bf 70}, 3400.

\bibitem[\protect\citeauthoryear{{Collett}}{Collett}{1993b}]{Col93a}
{Collett, M.J.}, 1993b, Phase noise in a squeezed state,
{Phys. Scripta\/}~{\bf T 48}, 124.

\bibitem[\protect\citeauthoryear{{Daeubler, Miller, Risken and
Schoendorff}}{Daeubler et~al.}{1993}]{DMRS93} {Daeubler, B.,
C.~Miller, H.~Risken and L.~Schoendorff}, 1993, Quantum
states with minimum phase uncertainty for the {S}\"ussmann
measure, {Phys. Scripta\/}~{\bf T 48}, 119.

\bibitem[\protect\citeauthoryear{{Damaskinsky and
Yarunin}}{Damaskinsky and Yarunin}{1978}]{DY78} {Damaskinsky,
E.V., and V.S. Yarunin}, 1978, {H}ermitian phase operator and
{H}eisenberg representation of canonical commutation
relation, {Izv. Vyssh. Uchebn. Zaved. Tomsk Univ.\/}~{\bf 6},
59. In Russian.

\bibitem[\protect\citeauthoryear{{Daniel and Milburn}}{Daniel and
Milburn}{1989}]{DM89} {Daniel, D.J., and G.J. Milburn}, 1989,
Destruction of quantum coherence in a nonlinear oscillator
via attenuation and amplification, {Phys. Rev. A\/}~{\bf 39},
4628.

\bibitem[\protect\citeauthoryear{{d'Ariano and Paris}}{d'Ariano and
Paris}{1993}]{DP93} {d'Ariano, G.M., and M.G.A. Paris}, 1993,
Necessity of sine-cosine joint measurement, {Phys. Rev.
A\/}~{\bf 48}, R4039.

\bibitem[\protect\citeauthoryear{{d'Ariano and Paris}}{d'Ariano and
Paris}{1994}]{DP94} {d'Ariano, G.M., and M.G.A. Paris}, 1994,
Lower bounds on phase sensitivity in ideal and feasible
measurements, {Phys. Rev. A\/}~{\bf 49}, 3022.

\bibitem[\protect\citeauthoryear{{Das}}{Das}{1994}]{Das94}
{Das, H.K.}, 1994, Phase distribution in the {Y}urke-{S}toler
states, {Phys. Scripta\/}~{\bf 49}, 606.

\bibitem[\protect\citeauthoryear{{de~Oliveira, Kim, Knight and
Bu\v{z}ek}}{de~Oliveira et~al.}{1990}]{DKKB90} {de~Oliveira,
F. A.M., M.S. Kim, P.L. Knight and V.~Bu\v{z}ek}, 1990,
Properties of displaced number states, {Phys. Rev. A\/}~{\bf
41}, 2645.

\bibitem[\protect\citeauthoryear{{Dirac}}{Dirac}{1927}]{Dir27}
{Dirac, P.A.M.}, 1927, The quantum theory of the emission and
absorption of radiation, {Proc. R. Soc. London A\/}~{\bf
114}, 243.

\bibitem[\protect\citeauthoryear{{Dowling}}{Dowling}{1991}]{Dow91}
{Dowling, J.P.}, 1991, A quantum state of ultra-low phase
noise, {Opt. Commun.\/}~{\bf 86}, 119.

\bibitem[\protect\citeauthoryear{{Drobn\'y, Gantsog and
Jex}}{Drobn\'y et~al.}{1994}]{DGJ94} {Drobn\'y, G.,
Ts.~Gantsog and I.~Jex}, 1994, Phase properties of a field
mode interacting with $n$ two-level atoms, {Phys. Rev.
A\/}~{\bf 49}, 622.

\bibitem[\protect\citeauthoryear{{Drobn\'y and Jex}}{Drobn\'y and
Jex}{1992}]{DJ92} {Drobn\'y, G., and I.~Jex}, 1992, Phase
properties of field modes in the process of $k$th-harmonic
generation, {Phys. Lett. A\/}~{\bf 169}, 273.

\bibitem[\protect\citeauthoryear{{Dung, Huyen and Shumovsky}}{Dung
et~al.}{1992}]{DHS92} {Dung, H.T., N.D. Huyen and A.S.
Shumovsky}, 1992, Phase properties of a coherent field
interacting with two two-level atoms in a cavity, {Phys.
A\/}~{\bf 182}, 467.

\bibitem[\protect\citeauthoryear{{Dung and Shumovsky}}{Dung and
Shumovsky}{1992}]{DS92} {Dung, H.T., and A.S. Shumovsky},
1992, Quantum phase fluctuations in the {J}aynes-{C}ummings
model: {E}ffects of cavity damping, {Phys. Lett. A\/}~{\bf
169}, 379.

\bibitem[\protect\citeauthoryear{{Dung, Tana\'s and Shumovsky}}{Dung
et~al.}{1990}]{DTS90} {Dung, H.~T., R.~Tana\'s and A.S.
Shumovsky}, 1990, Collapses, revivals, and phase properties
of the field in {J}aynes-{C}ummings type models, {Opt.
Commun.\/}~{\bf 79}, 462.

\bibitem[\protect\citeauthoryear{{Dung, Tana\'s and Shumovsky}}{Dung
et~al.}{1991a}]{DTS91a} {Dung, H.T., R.~Tana\'s and A.S.
Shumovsky}, 1991a, Dynamic properties of the field phase in
the {J}aynes-{C}ummings model, {J. Mod. Opt.\/}~{\bf 38},
2069.

\bibitem[\protect\citeauthoryear{{Dung, Tana\'s and Shumovsky}}{Dung
et~al.}{1991b}]{DTS91b} {Dung, H.T., R.~Tana\'s and A.S.
Shumovsky}, 1991b, Phase properties of the field interacting
with a three-level atom. {II}. {T}wo-mode case, {Quantum
Opt.\/}~{\bf 3}, 255.

\bibitem[\protect\citeauthoryear{{Eberly, Narozhny and
Sanchez-Mondragon}}{Eberly et~al.}{1980}]{ENS80} {Eberly,
J.H., N.B. Narozhny and J.J. Sanchez-Mondragon}, 1980,
Periodic spontaneous collapse and revival in a simple quantum
model, {Phys. Rev. Lett.\/}~{\bf 44}, 1323.

\bibitem[\protect\citeauthoryear{{Eiselt and Risken}}{Eiselt and
Risken}{1989}]{ER89} {Eiselt, J., and H.~Risken}, 1989,
Calculation of quasiprobabilities for damped
{J}aynes-{C}ummings model, {Opt. Commun.\/}~{\bf 72}, 351.

\bibitem[\protect\citeauthoryear{{Eiselt and Risken}}{Eiselt and
Risken}{1991}]{ER91} {Eiselt, J., and H.~Risken}, 1991,
Quasiprobability distributions for the {J}aynes-{C}ummings
model with cavity damping, {Phys. Rev. A\/}~{\bf 43}, 346.

\bibitem[\protect\citeauthoryear{{Ellinas}}{Ellinas}{1991a}]{Ell91}
{Ellinas, D.}, 1991a, Phase operators via group contraction,
{J. Math. Phys.\/}~{\bf 32}, 135.

\bibitem[\protect\citeauthoryear{{Ellinas}}{Ellinas}{1991b}]{Ell91a}
{Ellinas, D.}, 1991b, Quantum phase angles and
${SU}(\infty)$, {J. Mod. Opt.\/}~{\bf 38}, 2393.

\bibitem[\protect\citeauthoryear{{Ellinas}}{Ellinas}{1992}]{Ell92}
{Ellinas, D.}, 1992, Quantum phase and a $q$-deformed quantum
oscillator, {Phys. Rev. A\/}~{\bf 45}, 3358.

\bibitem[\protect\citeauthoryear{{Englert and W\'odkiewicz}}{Englert
and W\'odkiewicz}{1995}]{EW95} {Englert, B.G., and
K.~W\'odkiewicz}, 1995, Intrinsic and operational observables
in quantum mechanics, {Phys. Rev. A\/}~{\bf 51}, R2661.

\bibitem[\protect\citeauthoryear{{Englert, W\'odkiewicz and
Riegler}}{Englert et~al.}{1995}]{EWR95} {Englert, B.G.,
K.~W\'odkiewicz and P.~Riegler}, 1995, Intrinsic phase
operator of the {N}oh-{F}oug\`eres-{M}andel experiments,
{Phys. Rev. A\/}~{\bf 52}, 1704.

\bibitem[\protect\citeauthoryear{{Fan}}{Fan}{1993}]{Fan93}
{Fan, A.F.}, 1993, Phase properties of a field in the
{J}aynes-{C}ummings model for nonresonant behavior, {Opt.
Commun.\/}~{\bf 98}, 340.

\bibitem[\protect\citeauthoryear{{Fan and Wang}}{Fan and
Wang}{1994}]{FW94} {Fan, A.F., and Z.W. Wang}, 1994, Phase,
coherence properties, and the numerical analysis of the field
in the nonresonant {J}aynes-{C}ummings model, {Phys. Rev.
A\/}~{\bf 49}, 1509.

\bibitem[\protect\citeauthoryear{{Fan and Zaidi}}{Fan and
Zaidi}{1988}]{FZ88} {Fan, H.Y., and H.R. Zaidi}, 1988, An
exact calculation of the expectation values of the phase
operators in squeezed states, {Opt. Commun.\/}~{\bf 68}, 143.

\bibitem[\protect\citeauthoryear{{Foug\'eres, Monken and
Mandel}}{Foug\'eres et~al.}{1994}]{FMM94} {Foug\'eres, A.,
C.H. Monken and L.~Mandel}, 1994, Measurements of the
probability distribution of the phase difference between two
quantum fields, {Opt. Lett.\/}~{\bf 19}, 1771.

\bibitem[\protect\citeauthoryear{{Franson}}{Franson}{1994}]{Fra94}
{Franson, J.D.}, 1994, Nonlocal reduction of the
wave-function by quantum phase measurements, {Phys. Rev.
A\/}~{\bf 49}, 3221.

\bibitem[\protect\citeauthoryear{{Freyberger and
Schleich}}{Freyberger and Schleich}{1993}]{FS93} {Freyberger,
M., and W.~Schleich}, 1993, Photon-counting, quantum phase,
and phase-space distributions, {Phys. Rev. A\/}~{\bf 47},
R30.

\bibitem[\protect\citeauthoryear{{Freyberger and
Schleich}}{Freyberger and Schleich}{1994}]{FS94} {Freyberger,
M., and W.~Schleich}, 1994, Phase uncertainties of a squeezed
state, {Phys. Rev. A\/}~{\bf 49}, 5056.

\bibitem[\protect\citeauthoryear{{Freyberger, Vogel and
Schleich}}{Freyberger et~al.}{1993a}]{FVS93} {Freyberger, M.,
K.~Vogel and W.~Schleich}, 1993a, From photon counts to
quantum phase, {Phys. Lett. A\/}~{\bf 176}, 41.

\bibitem[\protect\citeauthoryear{{Freyberger, Vogel and
Schleich}}{Freyberger et~al.}{1993b}]{FVS93a} {Freyberger,
M., K.~Vogel and W.~Schleich}, 1993b, Quantum phase from
photon counting and the ${Q}$-function, {Quantum Opt.\/}~{\bf
5}, 65.

\bibitem[\protect\citeauthoryear{{Galindo}}{Galindo}{1984a}]{Gal84a}
{Galindo, A.}, 1984a, Phase and number, {Lett. Math.
Phys.\/}~{\bf 8}, 495.

\bibitem[\protect\citeauthoryear{{Galindo}}{Galindo}{1984b}]{Gal84}
{Galindo, A.}, 1984b, A quantum phase, {Ann. Fis. A\/}~{\bf
81}, 191.

\bibitem[\protect\citeauthoryear{{Gangopadhyay}}{Gangopadhyay}{1994}]{Gan94} {Gangopadhyay,
G.}, 1994, Coherent phase state and displaced phase state in
a finite-dimensional basis and their light field limits, {J.
Mod. Opt.\/}~{\bf 41}, 525.

\bibitem[\protect\citeauthoryear{{Gantsog}}{Gantsog}{1992}]{Gan92}
{Gantsog, Ts.}, 1992, Collapses and revivals of phase
fluctuations in parametric down conversion with quantum pump,
{Phys. Lett. A\/}~{\bf 170}, 249.

\bibitem[\protect\citeauthoryear{{Gantsog, Miranowicz and
Tana\'s}}{Gantsog et~al.}{1992}]{GMT92} {Gantsog, Ts.,
A.~Miranowicz and R.~Tana\'s}, 1992, Phase properties of real
field states: {T}he {G}arrison-{W}ong versus {P}egg-{B}arnett
predictions, {Phys. Rev. A\/}~{\bf 46}, 2870.

\bibitem[\protect\citeauthoryear{{Gantsog and Tana\'s}}{Gantsog and
Tana\'s}{1991a}]{GT91a} {Gantsog, Ts., and R.~Tana\'s},
1991a, Discrete superpositions of coherent states and phase
properties of elliptically polarized light propagating in a
{K}err medium, {Quantum Opt.\/}~{\bf 3}, 33.

\bibitem[\protect\citeauthoryear{{Gantsog and Tana\'s}}{Gantsog and
Tana\'s}{1991b}]{GT91b} {Gantsog, Ts., and R.~Tana\'s},
1991b, Phase properties of a damped anharmonic oscillator,
{Phys. Rev. A\/}~{\bf 44}, 2086.

\bibitem[\protect\citeauthoryear{{Gantsog and Tana\'s}}{Gantsog and
Tana\'s}{1991c}]{GT91c} {Gantsog, Ts., and R.~Tana\'s},
1991c, Phase properties of elliptically polarized light
propagating in a {K}err medium, {J. Mod. Opt.\/}~{\bf 38},
1537.

\bibitem[\protect\citeauthoryear{{Gantsog and Tana\'s}}{Gantsog and
Tana\'s}{1991d}]{GT91d} {Gantsog, Ts., and R.~Tana\'s},
1991d, Phase properties of fractional coherent states, {Phys.
Lett. A\/}~{\bf 157}, 330.

\bibitem[\protect\citeauthoryear{{Gantsog and Tana\'s}}{Gantsog and
Tana\'s}{1991e}]{GT91e} {Gantsog, Ts., and R.~Tana\'s},
1991e, Phase properties of pair coherent states, {Opt.
Commun.\/}~{\bf 82}, 145.

\bibitem[\protect\citeauthoryear{{Gantsog and Tana\'s}}{Gantsog and
Tana\'s}{1991f}]{GT91f} {Gantsog, Ts., and R.~Tana\'s},
1991f, Phase properties of self-squeezed states generated by
the anharmonic oscillator, {J. Mod. Opt.\/}~{\bf 38}, 1021.

\bibitem[\protect\citeauthoryear{{Gantsog and Tana\'s}}{Gantsog and
Tana\'s}{1991g}]{GT91g} {Gantsog, Ts., and R.~Tana\'s},
1991g, Phase properties of the two-mode squeezed vacuum
states, {Phys. Lett. A\/}~{\bf 152}, 251.

\bibitem[\protect\citeauthoryear{{Gantsog, Tana\'s and
Zawodny}}{Gantsog et~al.}{1991a}]{GTZ91b} {Gantsog, Ts.,
R.~Tana\'s and R.~Zawodny}, 1991a, Quantum phase fluctuations
in the second-harmonic generation, {Phys. Lett. A\/}~{\bf
155}, 1.

\bibitem[\protect\citeauthoryear{{Gantsog, Tana\'s and
Zawodny}}{Gantsog et~al.}{1991b}]{GTZ91a} {Gantsog, Ts.,
R.~Tana\'s and R.~Zawodny}, 1991b, Quantum phase fluctuations
in parametric down conversion with quantum pump, {Opt.
Commun.\/}~{\bf 82}, 345.

\bibitem[\protect\citeauthoryear{{Gantsog, Tana\'s and
Zawodny}}{Gantsog et~al.}{1993}]{GTZ93} {Gantsog, Ts.,
R.~Tana\'s and R.~Zawodny}, 1993, Collapses and revivals of
quantum phase fluctuations in the down conversion with
quantum pump, {Acta Phys. Slov.\/}~{\bf 43}, 74.

\bibitem[\protect\citeauthoryear{{Garraway and Knight}}{Garraway and
Knight}{1992}]{GK92} {Garraway, B.M., and P.L. Knight}, 1992,
Quantum phase distributions and quasidistributions, {Phys.
Rev. A\/}~{\bf 46}, R5346.

\bibitem[\protect\citeauthoryear{{Garraway and Knight}}{Garraway and
Knight}{1993}]{GK93} {Garraway, B.M., and P.L. Knight}, 1993,
Quantum superpositions, phase distributions and
quasi-probabilities, {Phys. Scripta\/}~{\bf T 48}, 66.

\bibitem[\protect\citeauthoryear{{Garraway, Sherman, Moya-Cessa, Knight
and Kurizki}}{Garraway et~al.}{1994}]{GSMKK94} {Garraway,
B.M., B.~Sherman, H.~Moya-Cessa, P.L. Knight and G.~Kurizki},
1994, Generation and detection of nonclassical field states
by conditional measurements following two-photon resonant
interactions, {Phys. Rev. A\/}~{\bf 49}, 535.

\bibitem[\protect\citeauthoryear{{Garrison and Wong}}{Garrison and
Wong}{1970}]{GW70} {Garrison, J.C., and J.~Wong}, 1970,
Canonically conjugate pairs, uncertainty relations, and phase
operators, {J. Math. Phys.\/}~{\bf 11}, 2242.

\bibitem[\protect\citeauthoryear{{Gennaro, Leonardi, Lillo, Vaglica and Vetri}}{Gennaro et~al.}{1994}]{GLLVV94} {Gennaro, G.,
C.~Leonardi, F.~Lillo, A.~Vaglica and G.~Vetri}, 1994,
Internal coherence and quantum phase difference between two
electromagnetic fields, {Opt. Commun.\/}~{\bf 112}, 67.

\bibitem[\protect\citeauthoryear{{Gerhardt, B\"uchler and
Litfin}}{Gerhardt et~al.}{1974}]{GBL74} {Gerhardt, H.,
U.~B\"uchler and G.~Litfin}, 1974, Phase measurement of a
microscopic radiation field, {Phys. Lett. A\/}~{\bf 49}, 119.

\bibitem[\protect\citeauthoryear{{Gerhardt, Welling and
Fr\"olich}}{Gerhardt et~al.}{1973}]{GWF73} {Gerhardt, H.,
H.~Welling and D.~Fr\"olich}, 1973, Ideal laser amplifier as
a phase measuring system of a microscopic radiation field,
{Appl. Phys.\/}~{\bf 2}, 91.

\bibitem[\protect\citeauthoryear{{Gerry}}{Gerry}{1987}]{Ger87}
{Gerry, C.C.}, 1987, On the phase fluctuations of
coherent-light interacting with an anharmonic oscillator,
{Opt. Commun.\/}~{\bf 63}, 278.

\bibitem[\protect\citeauthoryear{{Gerry}}{Gerry}{1990}]{Ger90}
{Gerry, C.C.}, 1990, Phase fluctuations of coherent light in
an anharmonic oscillator using the {H}ermitian phase
operator, {Opt. Commun.\/}~{\bf 75}, 168.

\bibitem[\protect\citeauthoryear{{Gerry}}{Gerry}{1995}]{Ger95}
{Gerry, C.C.}, 1995, Two-mode squeezed pair coherent states,
{J. Mod. Opt.\/}~{\bf 42}, 585.

\bibitem[\protect\citeauthoryear{{Gerry and Urbanski}}{Gerry and
Urbanski}{1990}]{GU90} {Gerry, C.C., and K.E. Urbanski},
1990, {H}ermitian phase-difference operator analysis of
microscopic radiation-field measurements, {Phys. Rev.
A\/}~{\bf 42}, 662.

\bibitem[\protect\citeauthoryear{{Goldhirsh}}{Goldhirsh}{1980}]{Gol80}
{Goldhirsh, I.}, 1980, Phase operator and phase fluctuations
of spins, {J. Phys. A\/}~{\bf 13}, 3479.

\bibitem[\protect\citeauthoryear{{Gou}}{Gou}{1993}]{Gou93}
{Gou, S.C.}, 1993, Characteristic oscillations of phase
properties for pair coherent states in the two-mode
{J}aynes-{C}ummings model dynamics, {Phys. Rev. A\/}~{\bf
48}, 3233.

\bibitem[\protect\citeauthoryear{{Gr{\o}nbech-Jensen, Christiansen and Ramanujam}}{Gr{\o}nbech-Jensen et~al.}{1989}]{GCR89}
{Gr{\o}nbech-Jensen, N., P.L. Christiansen and P.S.
Ramanujam}, 1989, Phase properties of squeezed states, {J.
Opt. Soc. Am. B\/}~{\bf 6}, 2423.

\bibitem[\protect\citeauthoryear{{{Hach~III} and Gerry}}{{Hach~III}
and Gerry}{1993}]{HG93} {{Hach~III}, E.E., and C.C. Gerry},
1993, Phase properties of macroscopic superposition states,
{Quantum Opt.\/}~{\bf 5}, 327.

\bibitem[\protect\citeauthoryear{{Hall}}{Hall}{1991}]{Hal91}
{Hall, M.J.W.}, 1991, The quantum description of optical
phase, {Quantum Opt.\/}~{\bf 3}, 7.

\bibitem[\protect\citeauthoryear{{Hall}}{Hall}{1993}]{Hal93}
{Hall, M.J.W.}, 1993, Phase resolution and coherent phase
states, {J. Mod. Opt.\/}~{\bf 40}, 809.

\bibitem[\protect\citeauthoryear{{Hall and Fuss}}{Hall and
Fuss}{1991}]{HF91} {Hall, M.J.W., and I.G. Fuss}, 1991,
Quantum phase detection and digital communication, {Quantum
Opt.\/}~{\bf 3}, 147.

\bibitem[\protect\citeauthoryear{{Helstrom}}{Helstrom}{1976}]{Hel76}
{Helstrom, C.W.}, 1976, Quantum Detection and Estimation
Theory, (Academic Press, New York).

\bibitem[\protect\citeauthoryear{{Herzog, Paul and Richter}}{Herzog
et~al.}{1993}]{HPR93} {Herzog, U., H.~Paul and T.~Richter},
1993, {W}igner function for a phase state, {Phys.
Scripta\/}~{\bf T 48}, 61.

\bibitem[\protect\citeauthoryear{{Hillery, Freyberger and
Schleich}}{Hillery et~al.}{1995}]{HFS95} {Hillery, M.,
M.~Freyberger and W.~Schleich}, 1995, Phase distributions and
large-amplitude states, {Phys. Rev. A\/}~{\bf 51}, 1792.

\bibitem[\protect\citeauthoryear{{Hradil}}{Hradil}{1990}]{Hra90}
{Hradil, Z.}, 1990, Phase in quantum optics and number phase
minimum uncertainty states, {Phys. Lett. A\/}~{\bf 146}, 1.

\bibitem[\protect\citeauthoryear{{Hradil}}{Hradil}{1992a}]{Hra92}
{Hradil, Z.}, 1992a, Performance-measures of quantum-phase
measurement, {Phys. Rev. A\/}~{\bf 46}, R2217.

\bibitem[\protect\citeauthoryear{{Hradil}}{Hradil}{1992b}]{Hra92a}
{Hradil, Z.}, 1992b, Phase measurement in quantum optics,
{Quantum Opt.\/}~{\bf 4}, 93.

\bibitem[\protect\citeauthoryear{{Hradil}}{Hradil}{1993a}]{Hra93a}
{Hradil, Z.}, 1993a, Operational approach to the phase of a
quantum-field: {C}omment, {Phys. Rev. A\/}~{\bf 47}, 4532.

\bibitem[\protect\citeauthoryear{{Hradil}}{Hradil}{1993b}]{Hra93b}
{Hradil, Z.}, 1993b, Relation between ideal and feasible
phase concepts, {Phys. Rev. A\/}~{\bf 47}, 2376.

\bibitem[\protect\citeauthoryear{{Hradil}}{Hradil}{1995}]{Hra95}
{Hradil, Z.}, 1995, Estimation of counted quantum phase,
{Phys. Rev. A\/}~{\bf 51}, 1870.

\bibitem[\protect\citeauthoryear{{Hradil and Bajer}}{Hradil and
Bajer}{1993}]{HB93} {Hradil, Z., and J.~Bajer}, 1993, Further
investigations of the operationally defined quantum phase:
{C}omment, {Phys. Rev. A\/}~{\bf 48}, 1717.

\bibitem[\protect\citeauthoryear{{Hradil and Shapiro}}{Hradil and
Shapiro}{1992}]{HS92} {Hradil, Z., and J.H. Shapiro}, 1992,
Quantum phase measurements with infinite peak-likelihood and
zero phase information, {Quantum Opt.\/}~{\bf 4}, 31.

\bibitem[\protect\citeauthoryear{{Janszky, Adam, Bertolotti and
Sibilia}}{Janszky et~al.}{1992}]{JABS92} {Janszky, J.,
P.~Adam, M.~Bertolotti and C.~Sibilia}, 1992,
Amplitude-squeezed and number-phase intelligent states in a
directional coupler, {Quantum Opt.\/}~{\bf 4}, 163.

\bibitem[\protect\citeauthoryear{{Jaynes and Cummings}}{Jaynes and
Cummings}{1963}]{JC63} {Jaynes, E.T., and F.W. Cummings},
1963, Comparison of quantum and semiclassical radiation
theories with application to the beam maser, {Proc.
IEEE.\/}~{\bf 51}, 89.

\bibitem[\protect\citeauthoryear{{Jex and Drobn\'y}}{Jex and
Drobn\'y}{1993}]{JD93} {Jex, I., and G.~Drobn\'y}, 1993,
Phase properties and entanglement of the field modes in a
two-mode coupler with intensity-dependent coupling, {Phys.
Rev. A\/}~{\bf 47}, 3251.

\bibitem[\protect\citeauthoryear{{Jex, Drobn\'y and Matsuoka}}{Jex
et~al.}{1992}]{JDM92} {Jex, I., G.~Drobn\'y and M.~Matsuoka},
1992, Quantum phase properties of the process of
$\kappa$-photon down conversion with quantized pump, {Opt.
Commun.\/}~{\bf 94}, 619.

\bibitem[\protect\citeauthoryear{{Jex, Matsuoka and Koashi}}{Jex
et~al.}{1993}]{JMK93} {Jex, I., M.~Matsuoka and M.~Koashi},
1993, Phase of the field in the interaction with two
two-level atoms, {Quantum Opt.\/}~{\bf 5}, 275.

\bibitem[\protect\citeauthoryear{{Jones}}{Jones}{1993}]{Jon93}
{Jones, K.R.W.}, 1993, Information theory and optimal phase
measurement, {Phys. Scripta\/}~{\bf T 48}, 100.

\bibitem[\protect\citeauthoryear{{Judge}}{Judge}{1964}]{Jud64}
{Judge, D.}, 1964, On the uncertainty relation for angle
variables, {Nuovo Cimento\/}~{\bf 31}, 332.

\bibitem[\protect\citeauthoryear{{Khan and Chaudhry}}{Khan and
Chaudhry}{1994}]{KC94a} {Khan, M.A., and M.A. Chaudhry},
1994, Applicability of {V}ogel-{S}chleich phase theory to
quantum phase measurements, {Optik\/}~{\bf 96}, 49.

\bibitem[\protect\citeauthoryear{{Kielich, Kozierowski and
Tana\'s}}{Kielich et~al.}{1985}]{KKT85} {Kielich, S.,
M.~Kozierowski and R.~Tana\'s}, 1985, Photon antibunching and
squeezing: {T}wo nontrivial effects of the nonlinear
interaction of laser light with matter, {Opt. Acta\/}~{\bf
32}, 1023.

\bibitem[\protect\citeauthoryear{{Kitagawa and Yamamoto}}{Kitagawa
and Yamamoto}{1986}]{KY86} {Kitagawa, H., and Y.~Yamamoto},
1986, Number-phase minimum-uncertainty state with reduced
number uncertainty in a {K}err nonlinear interferometer,
{Phys. Rev. A\/}~{\bf 34}, 3974.

\bibitem[\protect\citeauthoryear{{Kozierowski and
Tana\'s}}{Kozierowski and Tana\'s}{1977}]{KT77} {Kozierowski,
M., and R.~Tana\'s}, 1977, Quantum fluctuations in
second-harmonic light generation, {Opt. Commun.\/}~{\bf 21},
229.

\bibitem[\protect\citeauthoryear{{Kuang and Chen}}{Kuang and
Chen}{1994a}]{KC94b} {Kuang, L.M., and X.~Chen}, 1994a,
Coherent-state formalism for the {P}egg-{B}arnett {H}ermitian
phase theory, {Phys. Lett. A\/}~{\bf 186}, 8.

\bibitem[\protect\citeauthoryear{{Kuang and Chen}}{Kuang and
Chen}{1994b}]{KC94c} {Kuang, L.M., and X.~Chen}, 1994b,
Phase-coherent states and their squeezing properties, {Phys.
Rev. A\/}~{\bf 50}, 4228.

\bibitem[\protect\citeauthoryear{{Kuang, Wang and Zhou}}{Kuang
et~al.}{1993}]{KWZ93} {Kuang, L.M., F.B. Wang and Y.G. Zhou},
1993, Dynamics of a harmonic oscillator in a
finite-dimensional {H}ilbert space, {Phys. Lett. A\/}~{\bf
183}, 1.

\bibitem[\protect\citeauthoryear{{Kuang, Wang and Zhou}}{Kuang
et~al.}{1994}]{KWZ94} {Kuang, L.M., F.B. Wang and Y.G. Zhou},
1994, Coherent states of a harmonic oscillator in a
finite-dimensional {H}ilbert space and their squeezing
properties, {J. Mod. Opt.\/}~{\bf 41}, 1307.

\bibitem[\protect\citeauthoryear{{Lakshmi and Swain}}{Lakshmi and
Swain}{1990}]{LS90} {Lakshmi, P.A., and S.~Swain}, 1990,
Phase in the correlated-emission laser, {Phys. Rev. A\/}~{\bf
42}, 5632.

\bibitem[\protect\citeauthoryear{{Lane, Braunstein and Caves}}{Lane
et~al.}{1993}]{LBC93} {Lane, A.S., S.L. Braunstein and C.M.
Caves}, 1993, Maximum-likelihood statistics of multiple
quantum phase measurements, {Phys. Rev. A\/}~{\bf 47}, 1667.

\bibitem[\protect\citeauthoryear{{Leonhardt and Paul}}{Leonhardt and
Paul}{1993a}]{LP93b} {Leonhardt, U., and H.~Paul}, 1993a,
Phase measurement and ${Q}$-function, {Phys. Rev. A\/}~{\bf
47}, R2460.

\bibitem[\protect\citeauthoryear{{Leonhardt and Paul}}{Leonhardt and
Paul}{1993b}]{LP93c} {Leonhardt, U., and H.~Paul}, 1993b,
Realistic measurement of phase, {Phys. Scripta\/}~{\bf T 48},
45.

\bibitem[\protect\citeauthoryear{{Leonhardt, Vaccaro, B\"ohmer and
Paul}}{Leonhardt et~al.}{1995}]{LVBP95} {Leonhardt, U., J.A.
Vaccaro, B.~B\"ohmer and H.~Paul}, 1995, Canonical and
measured phase distributions, {Phys. Rev. A\/}~{\bf 51}, 84.

\bibitem[\protect\citeauthoryear{{Leo\'nski and Tana\'s}}{Leo\'nski
and Tana\'s}{1994}]{LT94} {Leo\'nski, W., and R.~Tana\'s},
1994, Possibility of producing the one-photon state in a
kicked cavity with a nonlinear {K}err medium, {Phys. Rev.
A\/}~{\bf 49}, R20.

\bibitem[\protect\citeauthoryear{{L\'evy-Leblond}}
{L\'evy-Leblond}{1973}]{Lev73} {L\'evy-Leblond, J.M.}, 1
973, Azimuthal quantization of
angular momentum, {Rev. Mex. Fis.\/}~{\bf 22}, 15.

\bibitem[\protect\citeauthoryear{{L\'evy-Leblond}}
{L\'evy-Leblond}{1976}]{Lev76} {L\'evy-Leblond, J.M.}, 1976,
Who is afraid of nonhermitian operators? {A} quantum
description of angle and phase, {Ann. Phys.\/}~{\bf 101},
319.

\bibitem[\protect\citeauthoryear{{L\'evy-Leblond}}
{L\'evy-Leblond}{1977}]{Lev77} {L\'evy-Leblond, J.M.}, 1977,
On the theoretical analysis of phase measurement for
microscopic radiation fields, {Phys. Lett. A\/}~{\bf 64},
159.

\bibitem[\protect\citeauthoryear{{Loudon}}{Loudon}{1973}]{Lou73}
{Loudon, R.}, 1973, {The Quantum Theory of Light\/}, 1st Ed.
(Oxford University Press, Oxford).

\bibitem[\protect\citeauthoryear{{Loudon and Knight}}{Loudon and
Knight}{1987}]{LK87} {Loudon, R., and P.L. Knight}, 1987,
Squeezed light, {J. Mod. Opt.\/}~{\bf 34}, 709.

\bibitem[\protect\citeauthoryear{{Louisell}}{Louisell}{1963}]{Lou63}
{Louisell, W.H.}, 1963, Amplitude and phase uncertainty
relations, {Phys. Lett.\/}~{\bf 7}, 60.

\bibitem[\protect\citeauthoryear{{Luis and S\'anchez-Soto}}{Luis and
S\'anchez-Soto}{1993a}]{LS93a} {Luis, A., and L.L.
S\'anchez-Soto}, 1993a, Alternative derivation of the
{P}egg-{B}arnett phase operator, {Phys. Rev. A\/}~{\bf 47},
1492.

\bibitem[\protect\citeauthoryear{{Luis and S\'anchez-Soto}}{Luis and
S\'anchez-Soto}{1993b}]{LS93b} {Luis, A., and L.L.
S\'anchez-Soto}, 1993b, Canonical-transformations to action
and phase-angle variables and phase operators, {Phys. Rev.
A\/}~{\bf 48}, 752.

\bibitem[\protect\citeauthoryear{{Luis and S\'anchez-Soto}}{Luis and
S\'anchez-Soto}{1993c}]{LS93c} {Luis, A., and L.L.
S\'anchez-Soto}, 1993c, Phase-difference operator, {Phys.
Rev. A\/}~{\bf 48}, 4702.

\bibitem[\protect\citeauthoryear{{Luis and S\'anchez-Soto}}{Luis and
S\'anchez-Soto}{1994}]{LS94} {Luis, A., and L.L.
S\'anchez-Soto}, 1994, Reply to ``{C}omment on
`{P}hase-difference operator'~'', {Phys. Rev. A\/}~{\bf 51},
861.

\bibitem[\protect\citeauthoryear{{Luis, S\'anchez-Soto and
Tana\'s}}{Luis et~al.}{1995}]{LST95} {Luis, A., L.L.
S\'anchez-Soto and R.~Tana\'s}, 1995, Phase properties of
light propagating in a {K}err medium: {S}tokes parameters
versus {P}egg-{B}arnett predictions, {Phys. Rev. A\/}~{\bf
51}, 1634.

\bibitem[\protect\citeauthoryear{{Luk\v{s} and
Pe\v{r}inov\'a}}{Luk\v{s} and Pe\v{r}inov\'a}{1990}]{LP90}
{Luk\v{s}, A., and V.~Pe\v{r}inov\'a}, 1990, Compatibility of
the cosine and sine operators, {Phys. Rev. A\/}~{\bf 42},
5805.

\bibitem[\protect\citeauthoryear{{Luk\v{s} and
Pe\v{r}inov\'a}}{Luk\v{s} and Pe\v{r}inov\'a}{1991}]{LP91}
{Luk\v{s}, A., and V.~Pe\v{r}inov\'a}, 1991, Extended number
state basis and number-phase intelligent states of
light-field: {I.} {M}apping and operator ordering approach to
quantum phase problem, {Czech. J. Phys. A\/}~{\bf 41}, 1205.

\bibitem[\protect\citeauthoryear{{Luk\v{s} and
Pe\v{r}inov\'a}}{Luk\v{s} and Pe\v{r}inov\'a}{1992}]{LP92}
{Luk\v{s}, A., and V.~Pe\v{r}inov\'a}, 1992, Number-phase
uncertainty products and minimizing states, {Phys. Rev.
A\/}~{\bf 45}, 6710.

\bibitem[\protect\citeauthoryear{{Luk\v{s} and
Pe\v{r}inov\'a}}{Luk\v{s} and Pe\v{r}inov\'a}{1993a}]{LP93}
{Luk\v{s}, A., and V.~Pe\v{r}inov\'a}, 1993a, Finite photon
number and discrete quantum phase, {Quantum Opt.\/}~{\bf 5},
287.

\bibitem[\protect\citeauthoryear{{Luk\v{s} and
Pe\v{r}inov\'a}}{Luk\v{s} and Pe\v{r}inov\'a}{1993b}]{LP93a}
{Luk\v{s}, A., and V.~Pe\v{r}inov\'a}, 1993b, Ordering of
ladder operators, the {W}igner function for number and phase,
and the enlarged {H}ilbert space, {Phys. Scripta\/}~{\bf T
48}, 94.

\bibitem[\protect\citeauthoryear{{Luk\v{s} and
Pe\v{r}inov\'a}}{Luk\v{s} and Pe\v{r}inov\'a}{1994}]{LP94}
{Luk\v{s}, A., and V.~Pe\v{r}inov\'a}, 1994, Presumable
solutions of quantum phase problem and their laws, {Quantum
Opt.\/}~{\bf 6}, 125.

\bibitem[\protect\citeauthoryear{{Luk\v{s}, Pe\v{r}inov\'a and
K\v{r}epelka}}{Luk\v{s} et~al.}{1992a}]{LPK92} {Luk\v{s}, A.,
V.~Pe\v{r}inov\'a and J.~K\v{r}epelka}, 1992a, Extended
number state basis and number-phase intelligent states of
light-field: {II.} {T}he reduced number-sine intelligent and
{J}ackiw states, {Czech. J. Phys. A\/}~{\bf 42}, 59.

\bibitem[\protect\citeauthoryear{{Luk\v{s}, Pe\v{r}inov\'a and
K\v{r}epelka}}{Luk\v{s} et~al.}{1992b}]{LPK92a} {Luk\v{s},
A., V.~Pe\v{r}inov\'a and J.~K\v{r}epelka}, 1992b, Special
states of the plane rotator relevant to the light-field,
{Phys. Rev. A\/}~{\bf 46}, 489.

\bibitem[\protect\citeauthoryear{{Luk\v{s}, Pe\v{r}inov\'a and
K\v{r}epelka}}{Luk\v{s} et~al.}{1994}]{LPK94} {Luk\v{s}, A.,
V.~Pe\v{r}inov\'a and J.~K\v{r}epelka}, 1994, Rotation angle,
phases of oscillators with definite circular polarizations,
and the composite ideal phase operator, {Phys. Rev. A\/}~{\bf
50}, 818.

\bibitem[\protect\citeauthoryear{{Lynch}}{Lynch}{1987}]{Lyn87}
{Lynch, R.}, 1987, Phase fluctuations in a squeezed state
using measured phase operators, {J. Opt. Soc. Am. B\/}~{\bf
4}, 1723.

\bibitem[\protect\citeauthoryear{{Lynch}}{Lynch}{1990}]{Lyn90}
{Lynch, R.}, 1990, Fluctuation of the {B}arnett-{P}egg phase
operator in a coherent state, {Phys. Rev. A\/}~{\bf 41},
2841.

\bibitem[\protect\citeauthoryear{{Lynch}}{Lynch}{1993}]{Lyn93}
{Lynch, R.}, 1993, Comparison of {V}ogel-{S}chleich phase
theory with quantum phase measurements, {Phys. Rev. A\/}~{\bf
47}, 1576.

\bibitem[\protect\citeauthoryear{{Lynch}}{Lynch}{1995}]{Lyn95}
{Lynch, R.}, 1995, The quantum phase problem: {A} critical
review, {Phys. Rep.\/}~{\bf 256}, 367.

\bibitem[\protect\citeauthoryear{{Mandel}}{Mandel}{1982}]{Man82}
{Mandel, L.}, 1982, Squeezing and photon antibunching in
harmonic generation, {Opt. Commun.\/}~{\bf 42}, 437.

\bibitem[\protect\citeauthoryear{{Matthys and Jaynes}}{Matthys and
Jaynes}{1980}]{MJ80} {Matthys, D.R., and E.T. Jaynes}, 1980,
Phase-sensitive optical amplifier, {J. Opt. Soc. Am.\/}~{\bf
70}, 263.

\bibitem[\protect\citeauthoryear{{Meng and Chai}}{Meng and
Chai}{1991}]{MC91} {Meng, H.X., and C.L. Chai}, 1991, Phase
properties of coherent light in the {J}aynes-{C}ummings
model, {Phys. Lett. A\/}~{\bf 155}, 500.

\bibitem[\protect\citeauthoryear{{Meng, Chai and Zhang}}{Meng
et~al.}{1992}]{MCZ92} {Meng, H.X., C.L. Chai and Z.M. Zhang},
1992, Phase dynamics of coherent light in the $m$-photon
{J}aynes-{C}ummings model, {Phys. Rev. A\/}~{\bf 45}, 2131.

\bibitem[\protect\citeauthoryear{{Meng, Guo and Xing}}{Meng
et~al.}{1994}]{MGX94} {Meng, H.X., A.Q. Guo and C.Z. Xing},
1994, Dependence of phase dynamics of light interacting with
two-level atoms on the initial-state of atoms, {Phys. Lett.
A\/}~{\bf 190}, 455.

\bibitem[\protect\citeauthoryear{{Meystre, Slosser and
Wilkens}}{Meystre et~al.}{1991}]{MSW91} {Meystre, P.,
J.~Slosser and M.~Wilkens}, 1991, Cotangent states of the
electromagnetic field: {S}queezing and phase properties,
{Phys. Rev. A\/}~{\bf 43}, 4959.

\bibitem[\protect\citeauthoryear{{Milburn}}{Milburn}{1986}]{Mil86}
{Milburn, G.J.}, 1986, Quantum and classical {L}iouville
dynamics of the anharmonic oscillator, {Phys. Rev. A\/}~{\bf
33}, 674.

\bibitem[\protect\citeauthoryear{{Milburn and Holmes}}{Milburn and
Holmes}{1986}]{MH86} {Milburn, G.J., and C.A. Holmes}, 1986,
Dissipative quantum and classical {L}iouville mechanics of
the anharmonic oscillator, {Phys. Rev. Lett.\/}~{\bf 56},
2237.

\bibitem[\protect\citeauthoryear{{Miranowicz}}{Miranowicz}{1994}]{Mir94}
{Miranowicz, A.}, 1994, Superposition states, phase
distributions and quasiprobabilities of quantum optical
fields, Ph.\ D. Thesis (A. Mickiewicz University, Pozna\'n).

\bibitem[\protect\citeauthoryear{{Miranowicz, Pi\c{a}tek and
Tana\'s}}{Miranowicz et~al.}{1994}]{MPT94} {Miranowicz, A.,
K.~Pi\c{a}tek and R.~Tana\'s}, 1994, Coherent states in a
finite dimensional {H}ilbert space, {Phys. Rev. A\/}~{\bf
50}, 3423.

\bibitem[\protect\citeauthoryear{{Miranowicz, Tana\'s and
Kielich}}{Miranowicz et~al.}{1990}]{MTK90} {Miranowicz, A.,
R.~Tana\'s and S.~Kielich}, 1990, Generation of discrete
superpositions of coherent states in the anharmonic
oscillator model, {Quantum Opt.\/}~{\bf 2}, 253.

\bibitem[\protect\citeauthoryear{{Narozhny, Sanchez-Mondragon and
Eberly}}{Narozhny et~al.}{1981}]{NSE81} {Narozhny, N.B., J.J.
Sanchez-Mondragon and J.H. Eberly}, 1981, Coherence versus
incoherence: {C}ollapse and revival in a single quantum
model, {Phys. Rev.\/}~{\bf 23}, 236.

\bibitem[\protect\citeauthoryear{{Nath and Kumar}}{Nath and
Kumar}{1989}]{NK89} {Nath, R., and P.~Kumar}, 1989, Phase
states and their statistical properties, {J. Mod.
Opt.\/}~{\bf 36}, 1615.

\bibitem[\protect\citeauthoryear{{Nath and Kumar}}{Nath and
Kumar}{1990}]{NK90} {Nath, R., and P.~Kumar}, 1990,
Higher-order coherence functions of phase states, {Opt.
Commun.\/}~{\bf 76}, 51.

\bibitem[\protect\citeauthoryear{{Nath and Kumar}}{Nath and
Kumar}{1991a}]{NK91a} {Nath, R., and P.~Kumar}, 1991a, Phase
properties of squeezed number states, {J. Mod. Opt.\/}~{\bf
38}, 1655.

\bibitem[\protect\citeauthoryear{{Nath and Kumar}}{Nath and
Kumar}{1991b}]{NK91b} {Nath, R., and P.~Kumar}, 1991b,
Quasi-photon phase states, {J. Mod. Opt.\/}~{\bf 38}, 263.

\bibitem[\protect\citeauthoryear{{Newton}}{Newton}{1980}]{New80}
{Newton, R.G.}, 1980, Quantum action-angle variables for
harmonic oscillators, {Ann. Phys. (New York)\/}~{\bf 124},
327.

\bibitem[\protect\citeauthoryear{{Nieto}}{Nieto}{1977}]{Nie77}
{Nieto, M.M.}, 1977, Phase-difference operator analysis of
microscopic radiation field measurements, {Phys. Lett.
A\/}~{\bf 60}, 401.

\bibitem[\protect\citeauthoryear{{Nieto}}{Nieto}{1993}]{Nie93}
{Nieto, M.M.}, 1993, Quantum phase and quantum phase
operators: {S}ome physics and some history, {Phys.
Scripta\/}~{\bf T 48}, 5.

\bibitem[\protect\citeauthoryear{{Nikitin and Masalov}}{Nikitin and
Masalov}{1991}]{NM91} {Nikitin, S.P., and A.V. Masalov},
1991, Quantum state evolution of the fundamental mode in the
process of second-harmonic generation, {Quantum Opt.\/}~{\bf
3}, 105.

\bibitem[\protect\citeauthoryear{{Noh, Foug\`eres and Mandel}}{Noh
et~al.}{1991}]{NFM91} {Noh, J.W., A.~Foug\`eres and
L.~Mandel}, 1991, Measurement of the quantum phase by photon
counting, {Phys. Rev. Lett.\/}~{\bf 67}, 1426.

\bibitem[\protect\citeauthoryear{{Noh, Foug\`{e}res and Mandel}}{Noh
et~al.}{1992a}]{NFM92} {Noh, J.W., A.~Foug\`{e}res and
L.~Mandel}, 1992a, Further investigations of the
operationally defined quantum phase, {Phys. Rev. A\/}~{\bf
46}, 2840.

\bibitem[\protect\citeauthoryear{{Noh, Foug\`{e}res and Mandel}}{Noh
et~al.}{1992b}]{NFM92a} {Noh, J.W., A.~Foug\`{e}res and
L.~Mandel}, 1992b, Operational approach to the phase of a
quantum field, {Phys. Rev. A\/}~{\bf 45}, 424.

\bibitem[\protect\citeauthoryear{{Noh, Foug\`{e}res and Mandel}}{Noh
et~al.}{1993a}]{NFM93} {Noh, J.W., A.~Foug\`{e}res and
L.~Mandel}, 1993a, Further investigations of the
operationally defined quantum phase: {R}eply, {Phys. Rev.
A\/}~{\bf 48}, 1719.

\bibitem[\protect\citeauthoryear{{Noh, Foug\`{e}res and Mandel}}{Noh
et~al.}{1993b}]{NFM93a} {Noh, J.W., A.~Foug\`{e}res and
L.~Mandel}, 1993b, Measurements of the
probability-distribution of the operationally defined quantum
phase difference, {Phys. Rev. Lett.\/}~{\bf 71}, 2579.

\bibitem[\protect\citeauthoryear{{Noh, Foug\`{e}res and Mandel}}{Noh
et~al.}{1993c}]{NFM93b} {Noh, J.W., A.~Foug\`{e}res and
L.~Mandel}, 1993c, Operational approach to phase operators
based on classical optics, {Phys. Scripta\/}~{\bf T 48}, 29.

\bibitem[\protect\citeauthoryear{{Noh, Foug\`{e}res and Mandel}}{Noh
et~al.}{1993d}]{NFM93c} {Noh, J.W., A.~Foug\`{e}res and
L.~Mandel}, 1993d, Operational approach to the phase of a
quantum field: {R}eply, {Phys. Rev. A\/}~{\bf 47}, 4535.

\bibitem[\protect\citeauthoryear{{Noh, Foug\`{e}res and Mandel}}{Noh
et~al.}{1993e}]{NFM93d} {Noh, J.W., A.~Foug\`{e}res and
L.~Mandel}, 1993e, Phase measurements: {R}eply, {Phys. Rev.
A\/}~{\bf 47}, 4537.

\bibitem[\protect\citeauthoryear{{Opatrn\'y}}{Opatrn\'y}{1994}]{Opa94}
{Opatrn\'y, T.}, 1994, Mean value and uncertainty of optical
phase: {A} simple mechanical analogy, {J. Phys. A\/}~{\bf
27}, 7201.

\bibitem[\protect\citeauthoryear{{Opatrn\'y, Miranowicz and
Bajer}}{Opatrn\'y et~al.}{1996}]{OMB96} {Opatrn\'y, T.,
A.~Miranowicz and J.~Bajer}, 1996, Coherent states in
finite-dimensional {H}ilbert space and their {W}igner
representation, {J. Mod. Opt.\/}~{\bf 43}, 417.

\bibitem[\protect\citeauthoryear{{Orszag and Saavedra}}{Orszag and
Saavedra}{1991a}]{OS91a} {Orszag, M., and C.~Saavedra},
1991a, Phase-difference fluctuations of the quantum-beat
laser, {Phys. Rev. A\/}~{\bf 43}, 2557.

\bibitem[\protect\citeauthoryear{{Orszag and Saavedra}}{Orszag and
Saavedra}{1991b}]{OS91b} {Orszag, M., and C.~Saavedra},
1991b, Phase fluctuations in a laser with atomic memory
effects, {Phys. Rev. A\/}~{\bf 43}, 554.

\bibitem[\protect\citeauthoryear{{Paprzycka and Tana\'s}}{Paprzycka
and Tana\'s}{1992}]{PT92} {Paprzycka, M., and R.~Tana\'s},
1992, Discrete superpositions of coherent states and phase
properties of the $m$-photon anharmonic oscillator, {Quantum
Opt.\/}~{\bf 4}, 331.

\bibitem[\protect\citeauthoryear{{Paul}}{Paul}{1974}]{Pau74}
{Paul, H.}, 1974, Phase of a microscopic electromagnetic
field and its measurement, {Fortschr. Phys.\/}~{\bf 22}, 657.

\bibitem[\protect\citeauthoryear{{Paul}}{Paul}{1991}]{Pau91}
{Paul, H.}, 1991, Phase destruction in quantum non-demolition
measurement of the photon number, {Quantum Opt.\/}~{\bf 3},
169.

\bibitem[\protect\citeauthoryear{{Paul and Leonhardt}}{Paul and
Leonhardt}{1994}]{PL94} {Paul, H., and U.~Leonhardt}, 1994,
Realistic measurement of phase, {Acta Phys. Pol. A\/}~{\bf
86}, 213.

\bibitem[\protect\citeauthoryear{{Pegg and Barnett}}{Pegg and
Barnett}{1988}]{PB88} {Pegg, D.T., and S.M. Barnett}, 1988,
Unitary phase operator in quantum mechanics, {Europhys.
Lett.\/}~{\bf 6}, 483.

\bibitem[\protect\citeauthoryear{{Pegg and Barnett}}{Pegg and
Barnett}{1989}]{PB89} {Pegg, D.T., and S.M. Barnett}, 1989,
Phase properties of the quantized single-mode electromagnetic
field, {Phys. Rev. A\/}~{\bf 39}, 1665.

\bibitem[\protect\citeauthoryear{{Pegg, Barnett and Vaccaro}}{Pegg
et~al.}{1989}]{PBV89} {Pegg, D.T., S.M. Barnett and J.A.
Vaccaro}, 1989, Phase in quantum electrodynamics, in: Quantum
Optics V, eds. J.D. Harvey and D.F. Walls (Springer, Berlin),
p.\  122.

\bibitem[\protect\citeauthoryear{{Pegg, Vaccaro and Barnett}}{Pegg
et~al.}{1990}]{PVB90} {Pegg, D.T., J.A. Vaccaro and S.M.
Barnett}, 1990, Quantum-optical phase and canonical
conjugation, {J. Mod. Opt.\/}~{\bf 37}, 1703.

\bibitem[\protect\citeauthoryear{{Peng and Li}}{Peng and
Li}{1992}]{PL92} {Peng, J.S., and G.X. Li}, 1992, Phase
fluctuations in the {J}aynes-{C}ummings model with and
without the rotating-wave approximation, {Phys. Rev.
A\/}~{\bf 45}, 3289.

\bibitem[\protect\citeauthoryear{{Peng, Li and Zhou}}{Peng
et~al.}{1992}]{PLZ92} {Peng, J.S., G.~X. Li and P.~Zhou},
1992, Phase properties and atomic coherent trapping in the
system of a three-level atom interacting with a bimodal
field, {Phys. Rev. A\/}~{\bf 46}, 1516.

\bibitem[\protect\citeauthoryear{{Pe\v{r}ina}}{Pe\v{r}ina}{1991}]{Per91}
{Pe\v{r}ina, J.}, 1991, {Quantum Statistics of Linear and
Nonlinear Optical Phenomena\/}, 2nd Ed. (Kluwer Academic
Publisheres, Dortrecht).

\bibitem[\protect\citeauthoryear{{Pe\v{r}ina, Hradil and
Jur\v{c}o}}{Pe\v{r}ina et~al.}{1994}]{PHJ94} {Pe\v{r}ina, J.,
Z.~Hradil and B.~Jur\v{c}o}, 1994, Quantum Optics and
Fundamentals of Physics, vol.~63 of Fundamental Theories of
Physics (Kluwer Academic Publisheres, Dortrecht).

\bibitem[\protect\citeauthoryear{{Pe\v{r}inov\'a and
Luk\v{s}}}{Pe\v{r}inov\'a and Luk\v{s}}{1988}]{PL88}
{Pe\v{r}inov\'a, V., and A.~Luk\v{s}}, 1988, Third-order
nonlinear dissipative oscillator with initial squeezed state,
{J. Mod. Opt.\/}~{\bf 35}, 1513.

\bibitem[\protect\citeauthoryear{{Pe\v{r}inov\'a and
Luk\v{s}}}{Pe\v{r}inov\'a and Luk\v{s}}{1990}]{PL90}
{Pe\v{r}inov\'a, V., and A.~Luk\v{s}}, 1990, Exact quantum
statistics of a nonlinear dissipative oscillator evolving
from an arbitrary state, {Phys. Rev. A\/}~{\bf 41}, 414.

\bibitem[\protect\citeauthoryear{{Pe\v{r}inov\'a, Luk\v{s} and
K\'arsk\'a}}{Pe\v{r}inov\'a et~al.}{1990}]{PLK90}
{Pe\v{r}inov\'a, V., A.~Luk\v{s} and M.~K\'arsk\'a}, 1990,
Third-order nonlinear dissipative oscillator with initial
Gaussian pure and mixed states, {J. Mod. Opt.\/}~{\bf 37},
1055.

\bibitem[\protect\citeauthoryear{{Phoenix and Knight}}{Phoenix and
Knight}{1990}]{PK90} {Phoenix, S.J.D., and P.L. Knight},
1990, Periodicity, phase, and entropy in models of two-photon
resonance, {J. Opt. Soc. Am. B\/}~{\bf 7}, 116.

\bibitem[\protect\citeauthoryear{{Popov and Yarunin}}{Popov and
Yarunin}{1973}]{PY73} {Popov, V.N., and V.S. Yarunin}, 1973,
On the problem of phase operator for linear harmonic
oscillator, {Vest. Leningr. Univ.\/}~{\bf 22}, 7. In
{R}ussian.

\bibitem[\protect\citeauthoryear{{Popov and Yarunin}}{Popov and
Yarunin}{1992}]{PY92} {Popov, V.N., and V.S. Yarunin}, 1992,
Quantum and quasi-classical states of the photon phase
operator, {J. Mod. Opt.\/}~{\bf 39}, 1525.

\bibitem[\protect\citeauthoryear{{Rempe, Walther and Klein}}{Rempe
et~al.}{1987}]{RWK87} {Rempe, G., H.~Walther and N.~Klein},
1987, Observation of quantum collapse and revival in a
one-atom maser, {Phys. Rev. Lett.\/}~{\bf 58}, 353.

\bibitem[\protect\citeauthoryear{{Riegler and W\'odkiewicz}}{Riegler
and W\'odkiewicz}{1994}]{RW94} {Riegler, P., and
K.~W\'odkiewicz}, 1994, Phase-space representation of
operational phase operators, {Phys. Rev. A\/}~{\bf 49}, 1387.

\bibitem[\protect\citeauthoryear{{Ritze}}{Ritze}{1980}]{Rit80}
{Ritze, H.H.}, 1980, Photon statistics of self-induced
gyrotropic birefringence, {Z. Phys. B\/}~{\bf 39}, 353.

\bibitem[\protect\citeauthoryear{{Ritze}}{Ritze}{1992}]{Rit92}
{Ritze, H.H.}, 1992, A proposal for the measurement of
extremely small phase fluctuations, {Opt. Commun.\/}~{\bf
92}, 127.

\bibitem[\protect\citeauthoryear{{Ritze and Bandilla}}{Ritze and
Bandilla}{1979}]{RB79} {Ritze, H.H., and A.~Bandilla}, 1979,
Quantum effects of a nonlinear interferometer with a {K}err
cell, {Opt. Commun.\/}~{\bf 29}, 126.

\bibitem[\protect\citeauthoryear{{S\'anchez-Soto and
Luis}}{S\'anchez-Soto and Luis}{1994}]{SL94} {S\'anchez-Soto,
L.L., and A.~Luis}, 1994, Quantum {S}tokes parameters and
phase difference operator, {Opt. Commun.\/}~{\bf 105}, 84.

\bibitem[\protect\citeauthoryear{{Sanders}}{Sanders}{1992}]{San92}
{Sanders, B.C.}, 1992, Superpositions of distinct phase
states by a nonlinear evolution, {Phys. Rev. A\/}~{\bf 45},
7746.

\bibitem[\protect\citeauthoryear{{Sanders, Barnett and
Knight}}{Sanders et~al.}{1986}]{SBK86} {Sanders, B.C., S.M.
Barnett and P.L. Knight}, 1986, Phase variables and squeezed
states, {Opt. Commun.\/}~{\bf 58}, 290.

\bibitem[\protect\citeauthoryear{{Santhanam}}{Santhanam}{1976}]{San76}
{Santhanam, T.S.}, 1976, Canonical commutation relation for
operators with bounded spectrum, {Phys. Lett. A\/}~{\bf 56},
345.

\bibitem[\protect\citeauthoryear{{Santhanam}}{Santhanam}{1977a}]{San77}
{Santhanam, T.S.}, 1977a, Does {W}eyl's commutation relation
imply a generalized statistics?, {Nuovo Cimento Lett.\/}~{\bf
20}, 13.

\bibitem[\protect\citeauthoryear{{Santhanam}}{Santhanam}{1977b}]{San77a}
{Santhanam, T.S.}, 1977b, Quantum mechanics in discrete space
and angular momentum, {Found. Phys.\/}~{\bf 7}, 121.

\bibitem[\protect\citeauthoryear{{Santhanam and Sinha}}{Santhanam and
Sinha}{1978}]{SS78} {Santhanam, T.S., and K.B. Sinha}, 1978,
Quantum mechanics in finite dimensions, {Aust. J.
Phys.\/}~{\bf 31}, 233.

\bibitem[\protect\citeauthoryear{{Santhanam and Tekumalla}}{Santhanam
and Tekumalla}{1976}]{ST76} {Santhanam, T.S., and A.R.
Tekumalla}, 1976, Quantum mechanics in finite dimensions,
{Found. Phys.\/}~{\bf 6}, 583.

\bibitem[\protect\citeauthoryear{{Schaufler, Freyberger and
Schleich}}{Schaufler et~al.}{1994}]{SFS94} {Schaufler, S.,
M.~Freyberger and W.~P. Schleich}, 1994, The birth of a
phase-cat, {J. Mod. Opt.\/}~{\bf 41}, 1765.

\bibitem[\protect\citeauthoryear{{Schieve and McGowan}}{Schieve and
McGowan}{1993}]{SM93} {Schieve, W.C., and R.R. McGowan},
1993, Phase distribution and linewidth in the micromaser,
{Phys. Rev. A\/}~{\bf 48}, 2315.

\bibitem[\protect\citeauthoryear{{Schleich, Horowicz and
Varro}}{Schleich et~al.}{1989a}]{SHV89a} {Schleich, W., R.J.
Horowicz and S.~Varro}, 1989a, A bifurcation in squeezed
state physics: But how? or area-of-overlap in phase space as
a guide to the phase distribution and the action-angle
{W}igner function of a squeezed state, in: Quantum Optics V,
eds. J.D. Harvey and D.F. Walls (Springer, Berlin), p.\ 133.

\bibitem[\protect\citeauthoryear{{Schleich, Horowicz and
Varro}}{Schleich et~al.}{1989b}]{SHV89} {Schleich, W., R.J.
Horowicz and S.~Varro}, 1989b, Bifurcation in the phase
probability distribution of a highly squeezed state, {Phys.
Rev. A\/}~{\bf 40}, 7405.

\bibitem[\protect\citeauthoryear{{Schleich and Wheeler}}{Schleich and
Wheeler}{1987}]{SW87} {Schleich, W., and J.A. Wheeler}, 1987,
Oscillations in photon distribution of squeezed states and
interference in phase space, {Nature\/}~{\bf 326}, 574.

\bibitem[\protect\citeauthoryear{{Schleich, Dowling and
Horowicz}}{Schleich et~al.}{1991}]{SDH91} {Schleich, W.P.,
J.P. Dowling and R.J. Horowicz}, 1991, Exponential decrease
in phase uncertainty, {Phys. Rev. A\/}~{\bf 44}, 3365.

\bibitem[\protect\citeauthoryear{{Schleich, Dowling, Horowicz and
Varro}}{Schleich et~al.}{1990}]{SDHV90} {Schleich, W.P., J.P.
Dowling, R.J. Horowicz and S.~Varro}, 1990, Asymptology in
quantum optics, in: New Frontiers in Quantum Electrodynamics
and Quantum Optics, ed. A.O. Barut (Plenum Press, New York),
p.\ ~31.

\bibitem[\protect\citeauthoryear{{Schumaker and Caves}}{Schumaker and
Caves}{1985}]{SC85} {Schumaker, B.L., and C.M. Caves}, 1985,
New formalism for two-photon optics: {II.} {M}athematical
foundation and compact notation, {Phys. Rev. A\/}~{\bf 31},
3093.

\bibitem[\protect\citeauthoryear{{Shapiro}}{Shapiro}{1993}]{Sha93}
{Shapiro, J.H.}, 1993, Phase conjugate quantum communication
with zero error probability at finite average photon number,
{Phys. Scripta\/}~{\bf T 48}, 105.

\bibitem[\protect\citeauthoryear{{Shapiro and Shepard}}{Shapiro and
Shepard}{1991}]{SS91} {Shapiro, J.H., and S.R. Shepard},
1991, Quantum phase measurement: {A} system theory
perspective, {Phys. Rev. A\/}~{\bf 43}, 3795.

\bibitem[\protect\citeauthoryear{{Shapiro, Shepard and Wong}}{Shapiro
et~al.}{1989}]{SSW89} {Shapiro, J.H., S.R. Shepard and W.C.
Wong}, 1989, Ultimate quantum limits on phase measurement,
{Phys. Rev. Lett.\/}~{\bf 62}, 2377.

\bibitem[\protect\citeauthoryear{{Shore and Knight}}{Shore and
Knight}{1993}]{SK93} {Shore, B.W., and P.L. Knight}, 1993,
The {J}aynes-{C}ummings model, {J. Mod. Opt.\/}~{\bf 40},
1195.

\bibitem[\protect\citeauthoryear{{Smith, Dubin and Hennings}}{Smith
et~al.}{1992}]{SDH92} {Smith, T.B., D.A. Dubin and M.A.
Hennings}, 1992, The {W}eyl quantization of phase-angle, {J.
Mod. Opt.\/}~{\bf 39}, 1603.

\bibitem[\protect\citeauthoryear{{Smithey, Beck, Cooper and
Raymer}}{Smithey et~al.}{1993}]{SBCR93} {Smithey, D.T.,
M.~Beck, J.~Cooper and M.G. Raymer}, 1993, Measurement of
number-phase uncertainty relations of optical fields, {Phys.
Rev. A\/}~{\bf 48}, 3159.

\bibitem[\protect\citeauthoryear{{Smithey, Beck, Cooper, Raymer and
Faridani}}{Smithey et~al.}{1993}]{SBCRF93} {Smithey, D.T.,
M.~Beck, J.~Cooper, M.G. Raymer and M.B.A.~Faridani}, 1993,
Complete experimental characterization of the quantum state
of a light mode via the {W}igner function and the density
matrix: {A}pplication to quantum phase distributions of
vacuum and squeezed-vacuum states, {Phys. Scripta\/}~{\bf T
48}, 35.

\bibitem[\protect\citeauthoryear{{Smithey, Faridani,  and
Raymer}}{Smithey et~al.}{1993}]{SBFR93} {Smithey, D.T.,
M.B.A. Faridani,  and M.G. Raymer}, 1993, Measurement of the
{W}igner distribution and the density-matrix of a light mode
using optical homodyne tomography: {A}pplication to squeezed
states and the vacuum, {Phys. Rev. Lett.\/}~{\bf 70}, 1244.

\bibitem[\protect\citeauthoryear{{Stenholm}}{Stenholm}{1993}]{Ste93}
{Stenholm, S.}, 1993, Some formal properties of operator
polar decomposition, {Phys. Scripta\/}~{\bf T 48}, 77.

\bibitem[\protect\citeauthoryear{{Stoler}}{Stoler}{1971}]{Sto71}
{Stoler, D.}, 1971, Generalized coherent states, {Phys. Rev.
D\/}~{\bf 4}, 2309.

\bibitem[\protect\citeauthoryear{{Summy and Pegg}}{Summy and
Pegg}{1990}]{SP90} {Summy, G.S., and D.T. Pegg}, 1990, Phase
optimized quantum states of light, {Opt. Commun.\/}~{\bf 77},
75.

\bibitem[\protect\citeauthoryear{{Susskind and Glogower}}{Susskind
and Glogower}{1964}]{SG64} {Susskind, L., and J.~Glogower},
1964, Quantum mechanical phase and time operator,
{Physics\/}~{\bf 1}, 49.

\bibitem[\protect\citeauthoryear{{Tana\'s}}{Tana\'s}{1984}]{Tan84}
{Tana\'s, R.}, 1984, Squeezed states of an anharmonic
oscillator, in: Coherence and Quantum Optics V (Plenum Press,
New York), p. 645.

\bibitem[\protect\citeauthoryear{{Tana\'s}}{Tana\'s}{1991}]{Tan91b}
{Tana\'s, R.}, 1991, Quantum phase correlations in nonlinear
optical processes, {J. Sov. Las. Res.\/}~{\bf 12}, 395.

\bibitem[\protect\citeauthoryear{{Tana\'s and Gantsog}}{Tana\'s and
Gantsog}{1991}]{TG91} {Tana\'s, R., and Ts.~Gantsog}, 1991,
Phase properties of elliptically polarized light propagating
in a {K}err medium with dissipation, {J. Opt. Soc. Am.
B\/}~{\bf 8}, 2505.

\bibitem[\protect\citeauthoryear{{Tana\'s and Gantsog}}{Tana\'s and
Gantsog}{1992a}]{TG92a} {Tana\'s, R., and Ts.~Gantsog},
1992a, Number and phase quantum fluctuations in the down
conversion with quantum pump, {Quantum Opt.\/}~{\bf 4}, 245.

\bibitem[\protect\citeauthoryear{{Tana\'s and Gantsog}}{Tana\'s and
Gantsog}{1992b}]{TG92b} {Tana\'s, R., and Ts.~Gantsog},
1992b, Phase properties of fields generated in a multiphoton
down converter, {Phys. Rev. A\/}~{\bf 45}, 5031.

\bibitem[\protect\citeauthoryear{{Tana\'s, Gantsog, Miranowicz and
Kielich}}{Tana\'s et~al.}{1991}]{TGMK91} {Tana\'s, R.,
Ts.~Gantsog, A.~Miranowicz and S.~Kielich}, 1991,
Quasi-probability distribution ${Q}(\alpha,\alpha^{*})$
versus phase distribution ${P}(\theta)$ in a description of
superpositions of coherent states, {J. Opt. Soc. Am.
B\/}~{\bf 8}, 1576.

\bibitem[\protect\citeauthoryear{{Tana\'s, Gantsog and
Zawodny}}{Tana\'s et~al.}{1991a}]{TGZ91a} {Tana\'s, R.,
Ts.~Gantsog and R.~Zawodny}, 1991a, Number and phase quantum
fluctuations in the second-harmonic generation, {Quantum
Opt.\/}~{\bf 3}, 221.

\bibitem[\protect\citeauthoryear{{Tana\'s, Gantsog and
Zawodny}}{Tana\'s et~al.}{1991b}]{TGZ91b} {Tana\'s, R.,
Ts.~Gantsog and R.~Zawodny}, 1991b, Phase properties of
second harmonics generated by different initial fields, {Opt.
Commun.\/}~{\bf 83}, 278.

\bibitem[\protect\citeauthoryear{{Tana\'s and Kielich}}{Tana\'s and
Kielich}{1979}]{TK79} {Tana\'s, R., and S.~Kielich}, 1979,
Polarization dependence of photon antibunching phenomena
involving light propagation in isotropic media, {Opt.
Commun.\/}~{\bf 30}, 443.

\bibitem[\protect\citeauthoryear{{Tana\'s and Kielich}}{Tana\'s and
Kielich}{1983}]{TK83} {Tana\'s, R., and S.~Kielich}, 1983,
Self-squeezing of light propagating through nonlinear
optically isotropic media, {Opt. Commun.\/}~{\bf 45}, 351.

\bibitem[\protect\citeauthoryear{{Tana\'s and Kielich}}{Tana\'s and
Kielich}{1984}]{TK84} {Tana\'s, R., and S.~Kielich}, 1984, On
the possibility of almost complete self-squeezing of strong
electromagnetic fields, {Opt. Acta\/}~{\bf 31}, 81.

\bibitem[\protect\citeauthoryear{{Tana\'s and Kielich}}{Tana\'s and
Kielich}{1990}]{TK90} {Tana\'s, R., and S.~Kielich}, 1990,
Quantum fluctuations in the {S}tokes parameters of light
propagating in a {K}err medium, {J. Mod. Opt.\/}~{\bf 37},
1935.

\bibitem[\protect\citeauthoryear{{Tana\'s, Miranowicz and
Gantsog}}{Tana\'s et~al.}{1993}]{TMG93} {Tana\'s, R.,
A.~Miranowicz and Ts.~Gantsog}, 1993, Phase distributions of
real field states, {Phys. Scripta\/}~{\bf T 48}, 53.

\bibitem[\protect\citeauthoryear{{Tana\'s, Murzakhmetov, Gantsog and
Chizhov}}{Tana\'s et~al.}{1992}]{TMGC92} {Tana\'s, R., B.K.
Murzakhmetov, Ts.~Gantsog and A.V. Chizhov}, 1992, Phase
properties of displaced number states, {Quantum Opt.\/}~{\bf
4}, 1.

\bibitem[\protect\citeauthoryear{{Tara, Agarwal and Chaturvedi}}{Tara
et~al.}{1993}]{TAC93} {Tara, K., G.S. Agarwal and
S.~Chaturvedi}, 1993, Production of {S}chr\"odinger
macroscopic quantum-superposition states in a {K}err medium,
{Phys. Rev. A\/}~{\bf 47}, 5024.

\bibitem[\protect\citeauthoryear{{Titulaer and Glauber}}{Titulaer and
Glauber}{1966}]{TG66} {Titulaer, U.M., and R.J. Glauber},
1966, Density operators for coherent fields, {Phys.
Rev.\/}~{\bf 145}, 1041.

\bibitem[\protect\citeauthoryear{{Tombesi and Mecozzi}}{Tombesi and
Mecozzi}{1987}]{TM87} {Tombesi, P., and A.~Mecozzi}, 1987,
Generation of macroscopically distinguishable quantum states
and detection by the squeezed-vacuum technique, {J. Opt. Soc.
Am. B\/}~{\bf 4}, 1700.

\bibitem[\protect\citeauthoryear{{Tsui and Reid}}{Tsui and
Reid}{1992}]{TR92} {Tsui, Y.K., and M.F. Reid}, 1992, Unitary
and {H}ermitian phase operators for the electromagnetic
field, {Phys. Rev. A\/}~{\bf 46}, 549.

\bibitem[\protect\citeauthoryear{{Tu and Gong}}{Tu and
Gong}{1993}]{TG93} {Tu, H.T., and C.D. Gong}, 1993,
Properties of the measured phase operators in the squeezed
number states, {J. Mod. Opt.\/}~{\bf 40}, 57.

\bibitem[\protect\citeauthoryear{{Turski}}{Turski}{1972}]{Tur72}
{Turski, L.A.}, 1972, The velocity operator for many-boson
sytems, {Phys.\/}~{\bf 57}, 432.

\bibitem[\protect\citeauthoryear{{Vaccaro, Barnett and Pegg}}{Vaccaro
et~al.}{1992}]{VBP92} {Vaccaro, J.A., S.M. Barnett and D.T.
Pegg}, 1992, Phase fluctuations and squeezing, {J. Mod.
Opt.\/}~{\bf 39}, 603.

\bibitem[\protect\citeauthoryear{{Vaccaro and Ben-Aryeh}}{Vaccaro and
Ben-Aryeh}{1995}]{VB95} {Vaccaro, J.A., and Y.~Ben-Aryeh},
1995, Antinormally ordering of phase operators and the
algebra of weak limits, {Opt. Commun.\/}~{\bf 113}, 427.

\bibitem[\protect\citeauthoryear{{Vaccaro and Pegg}}{Vaccaro and
Pegg}{1989}]{VP89} {Vaccaro, J.A., and D.T. Pegg}, 1989,
Phase properties of squeezed states of light, {Opt.
Commun.\/}~{\bf 70}, 529.

\bibitem[\protect\citeauthoryear{{Vaccaro and Pegg}}{Vaccaro and
Pegg}{1990a}]{VP90a} {Vaccaro, J.A., and D.T. Pegg}, 1990a,
Phys.l number-phase intelligent and minimum-uncertainty
states of light, {J. Mod. Opt.\/}~{\bf 37}, 17.

\bibitem[\protect\citeauthoryear{{Vaccaro and Pegg}}{Vaccaro and
Pegg}{1990b}]{VP90b} {Vaccaro, J.A., and D.T. Pegg}, 1990b,
{W}igner function for number and phase, {Phys. Rev. A\/}~{\bf
41}, 5156.

\bibitem[\protect\citeauthoryear{{Vaccaro and Pegg}}{Vaccaro and
Pegg}{1993}]{VP93} {Vaccaro, J.A., and D.T. Pegg}, 1993,
Consistency of quantum descriptions of phase, {Phys.
Scripta\/}~{\bf T 48}, 22.

\bibitem[\protect\citeauthoryear{{Vaccaro and Pegg}}{Vaccaro and
Pegg}{1994a}]{VP94a} {Vaccaro, J.A., and D.T. Pegg}, 1994a,
Nondiffusive phase dynamics from linear amplifiers and
attenuators in the weak-field regime, {J. Mod. Opt.\/}~{\bf
41}, 1079.

\bibitem[\protect\citeauthoryear{{Vaccaro and Pegg}}{Vaccaro and
Pegg}{1994b}]{VP94c} {Vaccaro, J.A., and D.T. Pegg}, 1994b,
On measuring extremely small phase fluctuations, {Opt.
Commun.\/}~{\bf 105}, 335.

\bibitem[\protect\citeauthoryear{{Vaccaro and Pegg}}{Vaccaro and
Pegg}{1994c}]{VP94b} {Vaccaro, J.A., and D.T. Pegg}, 1994c,
Phase properties of optical linear amplifiers, {Phys. Rev.
A\/}~{\bf 49}, 4985.

\bibitem[\protect\citeauthoryear{{Vogel, Akulin and Schleich}}{Vogel
et~al.}{1993}]{VAS93} {Vogel, K., V.M. Akulin and W.P.
Schleich}, 1993, Quantum state engineering of the radiation
field, {Phys. Rev. Lett.\/}~{\bf 71}, 1816.

\bibitem[\protect\citeauthoryear{{Vourdas}}{Vourdas}{1990}]{Vou90}
{Vourdas, A.}, 1990, ${SU}(2)$ and ${SU}(1,1)$ phase states,
{Phys. Rev. A\/}~{\bf 41}, 1653.

\bibitem[\protect\citeauthoryear{{Vourdas}}{Vourdas}{1992}]{Vou92}
{Vourdas, A.}, 1992, Analytic representations in the unit
disk and applications to phase states and squeezing, {Phys.
Rev. A\/}~{\bf 45}, 1943.

\bibitem[\protect\citeauthoryear{{Vourdas}}{Vourdas}{1993}]{Vou93}
{Vourdas, A.}, 1993, Phase states: {A}n analytic approach in
the unit disc, {Phys. Scripta\/}~{\bf T 48}, 84.

\bibitem[\protect\citeauthoryear{{Wagner, Brecha, Schenzle and
Walther}}{Wagner et~al.}{1992}]{WBSW92} {Wagner, C., R.J.
Brecha, A.~Schenzle and H.~Walther}, 1992, Phase diffusion
and continuous quantum measurements in the micromaser, {Phys.
Rev. A\/}~{\bf 46}, R5350.

\bibitem[\protect\citeauthoryear{{Wagner, Brecha, Schenzle and
Walther}}{Wagner et~al.}{1993}]{WBSW93} {Wagner, C., R.J.
Brecha, A.~Schenzle and H.~Walther}, 1993, Phase diffusion,
entangled states, and quantum measurements in the micromaser,
{Phys. Rev. A\/}~{\bf 47}, 5068.

\bibitem[\protect\citeauthoryear{{Walker and Carroll}}{Walker and
Carroll}{1984}]{WC84} {Walker, N.G., and J.E. Carroll}, 1984,
Simultaneous phase and amplitude measurements on optical
signals using a multiport junction, {J. Math. Phys.\/}~{\bf
20}, 981.

\bibitem[\protect\citeauthoryear{{Walker and Carroll}}{Walker and
Carroll}{1986}]{WC86} {Walker, N.G., and J.E. Carroll}, 1986,
Multiport homodyne detection near the quantum noise limit,
{Opt. Quantum Electron.\/}~{\bf 18}, 355.

\bibitem[\protect\citeauthoryear{{Walls and Barakat}}{Walls and
Barakat}{1970}]{WB70} {Walls, D.F., and R.~Barakat}, 1970,
Quantum-mechanical amplification and frequency conversion
with a trilinear {H}amiltonian, {Phys. Rev. A\/}~{\bf 1},
446.

\bibitem[\protect\citeauthoryear{{Wilson-Gordon, Bu\v{z}ek and
Knight}}{Wilson-Gordon et~al.}{1991}]{WBK91} {Wilson-Gordon,
A.D., V.~Bu\v{z}ek and P.L. Knight}, 1991, Statistical and
phase properties of displaced {K}err states, {Phys. Rev.
A\/}~{\bf 44}, 7647.

\bibitem[\protect\citeauthoryear{{Wootters}}{Wootters}{1987}]{Woo87}
{Wootters, W.K.}, 1987, A {W}igner-function formulation of
finite-state quantum mechanics, {Ann. Phys.\/}~{\bf 176}, 1.

\bibitem[\protect\citeauthoryear{{Wu, Kimble, Hall and Wu}}{Wu
et~al.}{1986}]{WKH86} {Wu, L., H.J. Kimble, J.L. Hall and
H.~Wu}, 1986, Generation of squeezed states by parametric
down conversion, {Phys. Rev. Lett.\/}~{\bf 57}, 2520.

\bibitem[\protect\citeauthoryear{{Yao}}{Yao}{1987}]{Yao87}
{Yao, D.M.}, 1987, Phase properties of squeezed states of
light, {Phys. Lett. A\/}~{\bf 122}, 77.

\bibitem[\protect\citeauthoryear{{Yoo and Eberly}}{Yoo and
Eberly}{1985}]{YE85} {Yoo, H.I., and J.H. Eberly}, 1985,
Dynamical theory of an atom with two or three levels
interacting with quantized cavity fields, {Phys. Rep.\/}~{\bf
118}, 239.

\bibitem[\protect\citeauthoryear{{Yurke and Stoler}}{Yurke and
Stoler}{1986}]{YS86} {Yurke, B., and D.~Stoler}, 1986,
Generating quantum mechanical superpositions of
macroscopically distinguishable states via amplitude
dispersion, {Phys. Rev. Lett.\/}~{\bf 57}, 13.

\bibitem[\protect\citeauthoryear{{Yurke and Stoler}}{Yurke and
Stoler}{1988}]{YS88} {Yurke, B., and D.~Stoler}, 1988, The
dynamical generation of {S}chr\"odinger cats and their
detection, {Phys. B\/}~{\bf 151}, 298.

\bibitem[\protect\citeauthoryear{{Zhang, Feng and Gilmore}}{Zhang
et~al.}{1990}]{ZFG90} {Zhang, W.M., D.H. Feng and
R.~Gilmore}, 1990, Coherent states: Theory and some
applications, {Rev. Mod. Phys.\/}~{\bf 62}, 867.

\bibitem[\protect\citeauthoryear{{Zhu and Kuang}}{Zhu and
Kuang}{1994}]{ZK94} {Zhu, J.Y., and L.M. Kuang}, 1994, Even
and odd coherent states of a harmonic oscillator in a
finite-dimensional {H}ilbert space and their squeezing
properties, {Phys. Lett. A\/}~{\bf 193}, 227.

\end{thebibliography}
%aaa

\end{document}